\documentclass[11pt]{article}
\usepackage[letterpaper, margin=1in]{geometry}
\usepackage{graphicx}
\usepackage{braket}
\usepackage{amsmath}
\usepackage{wasysym}
\usepackage{hyperref}
\usepackage{enumitem}
\usepackage{titlesec}
\usepackage[numbers,sort&compress]{natbib}
\usepackage{caption}
\usepackage{amssymb} 
\usepackage{subcaption}
\usepackage{float}
\usepackage{booktabs}
\usepackage{multirow}
\usepackage{siunitx}
\usepackage{authblk}

\titleformat{\section}[block]{\large\bfseries}{\thesection.}{1em}{}
\makeatletter
\patchcmd{\@maketitle}{\vskip 2em}{\vskip -3em}{}{}
\makeatother

\title{\Large\bfseries Threshold Behavior of ZX and ZY Surface Codes Under Circuit-Level Biased and Crosstalk Noise}

\author[1\thanks{Work done while employed at Brigham Young University, Provo, UT, USA as a Fletcher Research Intern. Corresponding author}]{Pritesh Thakur}
\author[2]{Jean-Fran\c{c}ois Van Huele}
\affil[1]{Brandeis University, Waltham, MA, USA, pritesht@brandeis.edu}
\affil[2]{Brigham Young University, Provo, UT, USA, vanhuele@byu.edu}

\date{\vspace{-4em}}

\begin{document}

\maketitle


\begin{abstract}

Studying the threshold behavior of surface codes under biased noise models is an active area of research. Previously Tuckett et al. (2018) \cite{Tuckett2018Ultrahigh}, using an optimal tensor-network decoder, demonstrated that replacing $Z$-type stabilizers with $Y$-type stabilizers significantly improves the surface code threshold under code-capacity level dephasing noise. In this work, we construct and study a $ZY$ surface code by replacing the $X$-type stabilizers with $Y$-type stabilizers. We compare it with the standard $ZX$ surface code under circuit-level Pauli-$X$ biased noise, with and without an additional gate-based $XX$ crosstalk noise. We find that for the $ZX$ surface code, the $X$-memory threshold increases monotonically with bias while the $Z$-memory threshold decreases and saturates. For the $ZY$ surface code, the $Y$-memory threshold is nearly constant across all bias values. The $Z$-memory thresholds of the $ZX$ and $ZY$ codes are consistent within the uncertainty. Adding $XX$ crosstalk reduces the $Z$-memory threshold beyond the fitting uncertainty while leaving the $X$ memory threshold largely unaffected. The choice of CNOT ordering redistributes threshold performance between the two logical memories. Our work extends prior observations from code-capacity level noise to circuit-level noise. It also indicates the need for decoders capable of jointly reasoning over correlated syndrome information so that tailored stabilizer structures could be fully utilized for quantum error correction.

\end{abstract}

\section{Introduction}
Richard Feynman emphasized the necessity of building fundamentally different computing machines that function on the laws of quantum mechanics to simulate quantum phenomena \cite{feynman1982}. Many physical systems are being researched as potential candidates for useful qubits for large-scale quantum computing, including trapped ion qubits, superconducting qubits, neutral-atom qubits, spin qubits, topological qubits, and others \cite{Bruzewicz_2019, Kjaergaard_2020, Henriet2020quantumcomputing, burkard2023semiconductor, Nayak_2008}. None of these qubit candidates are perfect, as they suffer errors through different physical noise mechanisms, such as magnetic field fluctuations, pulse shape imperfections, flux noise, cosmic rays, and so on \cite{Bruzewicz_2019, Kjaergaard_2020, Henriet2020quantumcomputing, burkard2023semiconductor, Vepsalainen2020, McEwen2022}. In the current Noisy Intermediate-Scale Quantum (NISQ) era, these types of noise challenge the construction of useful quantum computers \cite{preskill2018quantum}. Since there are physical limitations to how noise resistant the hardware can be, error correcting codes need to be built to detect and correct errors. Quantum error correction uses multiple qubits redundantly to create logical qubits that are robust against physical noise, increasing the accuracy of computations \cite{nielsen2010}. 

Surface codes are among the most promising candidates for fault-tolerant quantum error correction \cite{fowler2012, dennis2002}. However, unbiased physical noise is rare on realistic hardware. For example, superconducting qubits, trapped ions, and semiconductor spin qubits can fall into dephasing biased noise, i.e. Pauli-$Z$ errors dominate over $X$ and $Y$ errors~\cite{Aliferis_2009,Seis_2023,burkard2023semiconductor}. This asymmetry motivates the study of surface codes whose stabilizer structure is tailored to the dominant error. Ref.~\cite{Tuckett2018Ultrahigh} demonstrated that replacing $Z$-type stabilizers of the standard surface code with $Y$-type stabilizers results in significantly improved thresholds under pure dephasing noise. However, that result was obtained under code-capacity level noise using tensor-network decoders, which are computationally expensive and not practical for real-time decoding. Unlike code-capacity noise, in circuit-level noise every gate, measurement, and reset operation is fully faulty, and the error correction must be performed with a practical matching decoder. It is an open question whether any distinct behavior introduced by $Y$-type stabilizers can also be seen under circuit-level noise and practical matching decoder.

In this work, to address this question, we construct a $ZY$ surface code in which we replace the $X$-type stabilizers with $Y$-type stabilizers, and study it under Pauli-$X$ (bit-flip) biased noise. We adopt the $X$-biased convention rather than the more common dephasing convention as a matter of relabeling. At the code-capacity level the two constructions are equivalent, since a transversal Hadamard relabels $X$ to $Z$ and vice versa. It maps a dephasing-biased noise channel onto an $X$-biased one, leaves the standard surface code invariant, and because a $Y$-type stabilizer is invariant under this relabeling, it maps our $ZY$ code onto the $XY$ code of Ref.~\cite{Tuckett2018Ultrahigh}. Thus, our comparison of $ZY$ and $ZX$ surface codes under $X$-biased noise is a mirror image of the comparison of the $XY$ code and standard surface code under dephasing. Crucially, this equivalence holds only at code-capacity level. Under circuit-level noise it is broken by the fixed CNOT schedule and by the fixed basis in which ancillas are prepared and measured, so circuit-level behavior cannot be inferred from code-capacity results. We compare the threshold behavior of the $ZY$ surface code with the standard $ZX$ surface code under two circuit-level noise models: Pauli-$X$ biased noise and a combined noise model that additionally includes gate-based $XX$ crosstalk noise. The use of crosstalk noise was motivated by residual cross-resonance interactions in superconducting systems and Mølmer–Sørensen gate crosstalk trapped-ion systems \cite{zhao2023, Fang2022Crosstalk}. We use PyMatching v2.3.1 \cite{Higgott2025sparseblossom} to decode the surface codes. We simulate both the rotated and unrotated surface codes using two different CNOT orders for the stabilizer measurement. We sweep over nine values of the bias parameter ($\eta$) that span from unbiased depolarizing noise to pure $X$-biased noise.

This paper is organized as follows. In section \ref{sec:surface_code}, we give a background description of surface codes and introduce the construction of our $ZY$ surface codes. In section~\ref{sec:memory experiment}, we give a background of memory experiments. In section \ref{sec:cnot_order}, we introduce the different orders of performing CNOT operations for syndrome extraction and mention the CNOT orders we used. In section \ref{sec:methods}, we describe the noise model we implemented, our simulation pipeline and plotting logic. In sections \ref{sec:results} and \ref{sec:discussion}, we present the results that we obtained from our experiments, and finally in section \ref{sec:outlook}, we point towards future directions that can be pursued to extend this work.

\subsection{Surface Code}
\label{sec:surface_code}
The surface code is a topological stabilizer quantum error correction code originally introduced by Kitaev as toric codes in which qubits are distributed on the surface of a toroidal geometry \cite{kitaev2003}. Later, it was realized that the complicated toroidal geometry was not necessary and planar surface codes were created in which the qubits are arranged in a two-dimensional lattice \cite{fowler2012}. The planar codes were further modified to create rotated surface codes that require half the number of physical qubits compared to the unrotated planar surface codes for the same distance $d$ \cite{Bombin_2007}. The distance $d$ of a surface code is defined as the minimum number of physical-qubit errors required to cause an undetectable logical error. In this work, we have used both rotated and unrotated geometries. In surface codes, we have two types of qubits: data qubits (to store the information) and measurement qubits (to perform parity check). In the standard surface codes, there are two types of measurement qubits: measure-X qubits and measure-Z qubits. 

\begin{figure}[h!] 
    \centering
    \includegraphics[width=0.8\linewidth]{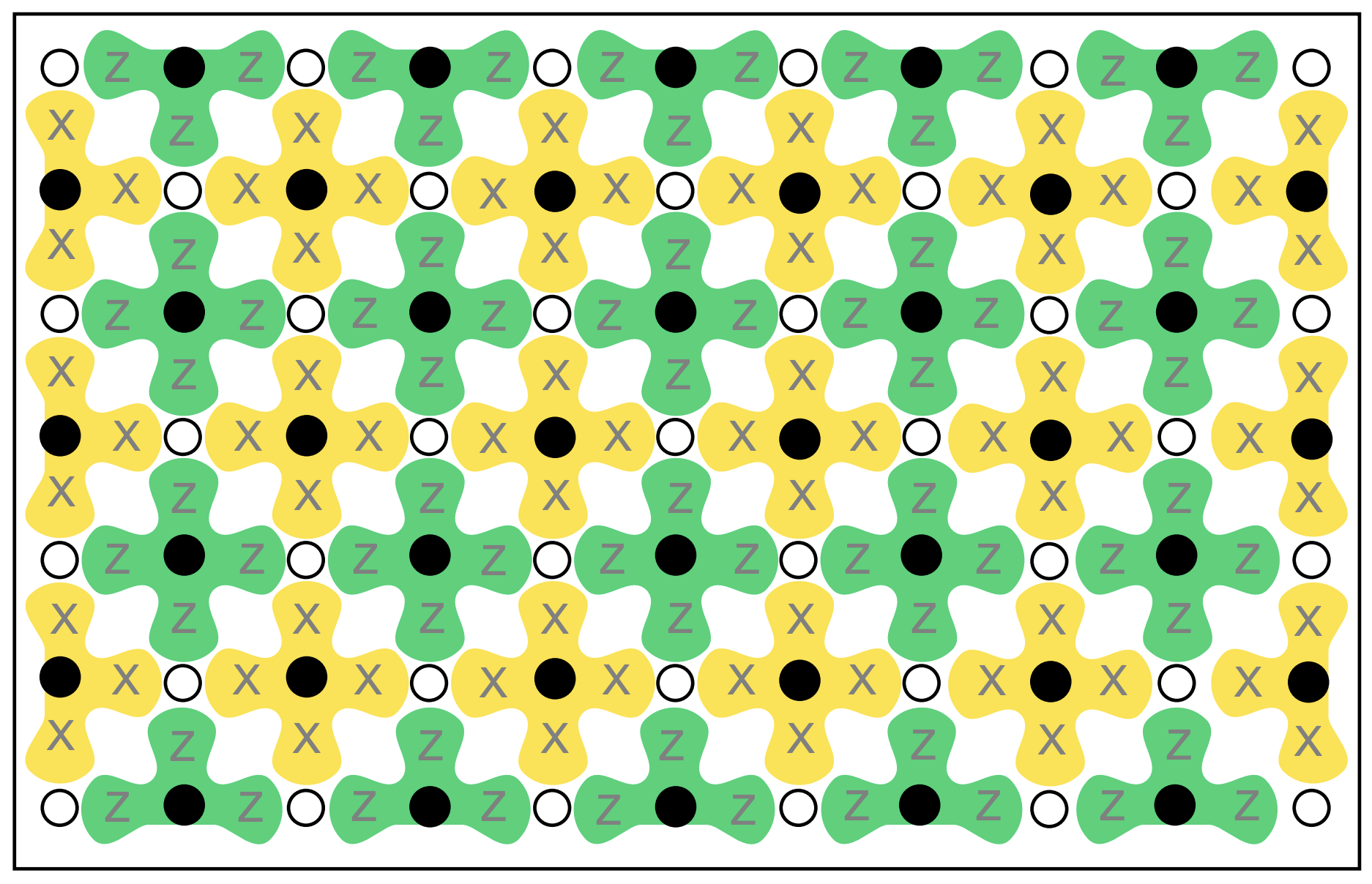}
    \caption{A two-dimensional implementation of unrotated surface code with Z-type (green) and X-type (yellow) stabilizers. Each filled circle ($\bullet$) denotes a measurement qubit, and each open circle ($\circ$) denotes a data qubit. Each data qubit is surrounded by four measurement qubits and each measurement qubit is connected to four data qubits to perform syndrome extraction using the circuits shown in Fig.~\ref{fig:stabilizers}.}
    \label{fig:surface_code}
\end{figure}

As it can be seen in Fig.~\ref{fig:surface_code}, taking a surface code as an array of physical qubits arranged in a $2D$ grid, the data qubits are placed on the edges $(e)$ of the grid, the $X-$stabilizer measurement qubits are placed on the vertices $(v)$ of the grid, and the $Z-$stabilizer measurement qubits are placed in the plaquettes $(p)$ of the grid. In this setup, the $X-$stabilizer generators can be defined as the product of Pauli $X-$operators acting on edges $(e)$ surrounding the vertices $v$, i.e.,

\begin{equation}
    A_v = \prod_{e\in v} X_e
\end{equation}
and the $Z-$stabilizer generators can be defined as the product of Pauli $Z-$ operators acting on the edges $(e)$ enclosing the plaquette $(p)$, i.e.,

\begin{equation}
    B_p = \prod_{e\in p} Z_e.    
\end{equation}

In this research, we implemented Pauli-$X$ biased noise. For $X$ error in the standard surface code, the stabilizers consisting of products of $X$ around the data qubits provide no useful syndrome information because the error commutes with the stabilizer. Since Pauli-$Y$ anti-commutes with Pauli-$Z$ (as $X$ does), and each vertex and plaquette operator share an even number of data qubits, the new $Y$-type stabilizers commute with the retained $Z$-type stabilizers. So, the surface code formed after replacing the $X$-type stabilizers with $Y$-type stabilizers still forms a valid surface code. Thus, in the surface code we replaced the measure$-X$ stabilizers from the vertex with measure$-Y$ stabilizers, which can be formally defined as

\begin{equation}
    C_v = \prod_{e\in v} Y_e.    
\end{equation}

\begin{figure}[htbp]
    \centering

    \begin{subfigure}{0.48\textwidth}
        \centering
        \includegraphics[width=\linewidth]{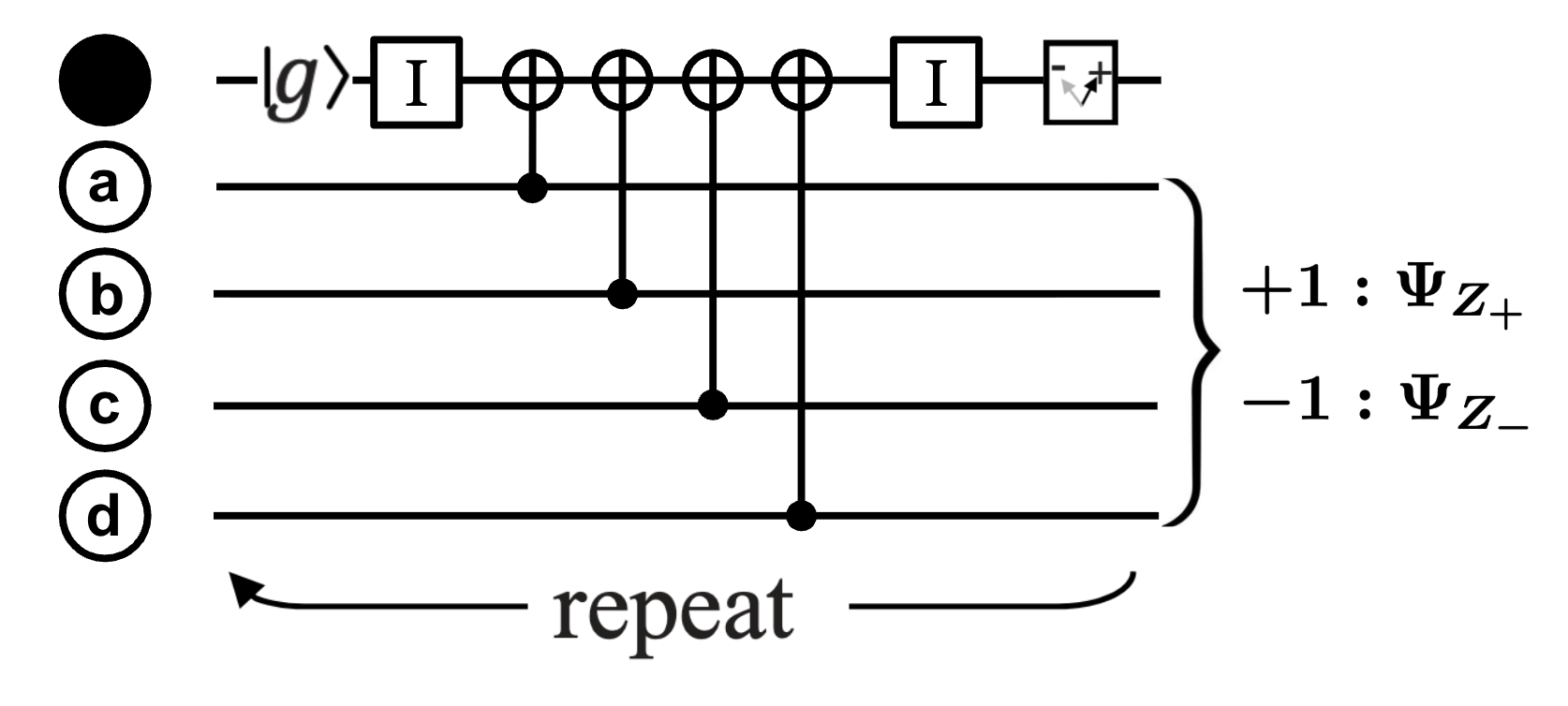}
        \caption{}
        \label{fig:z_stabilizer}
    \end{subfigure}
    \hfill
    \begin{subfigure}{0.48\textwidth}
        \centering
        \includegraphics[width=\linewidth]{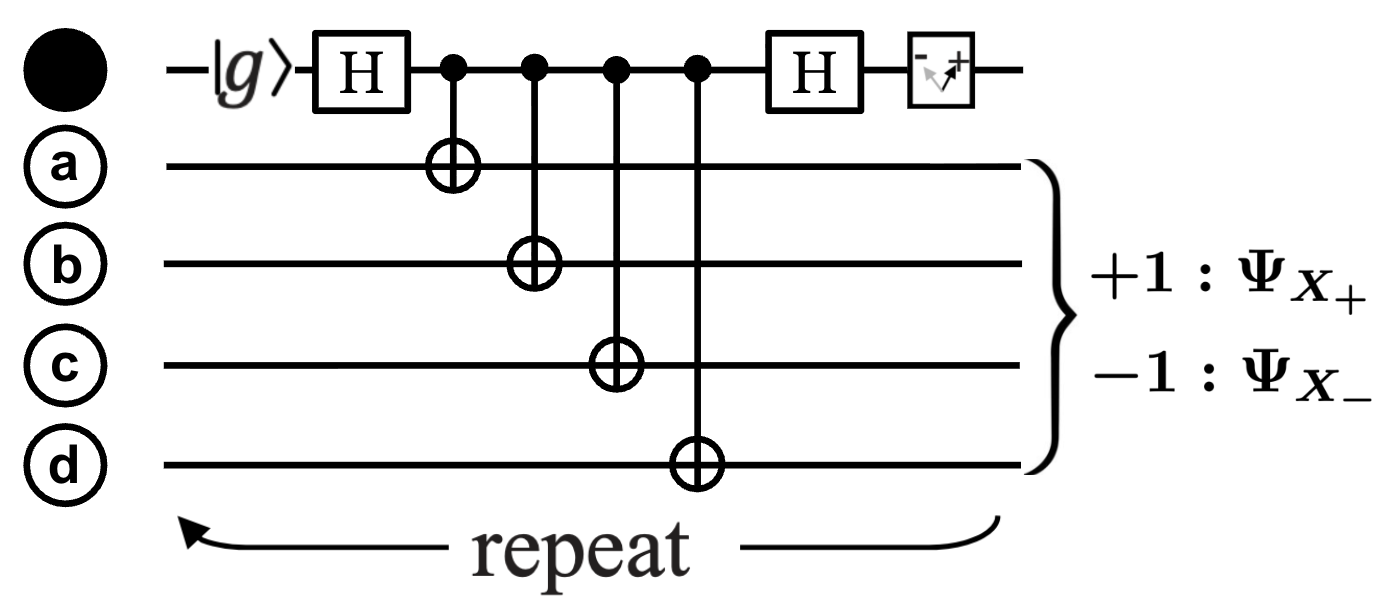}
        \caption{}
        \label{fig:x_stabilizer}
    \end{subfigure}

    \vspace{1em}

    \begin{subfigure}{0.48\textwidth}
        \centering
        \includegraphics[width=\linewidth]{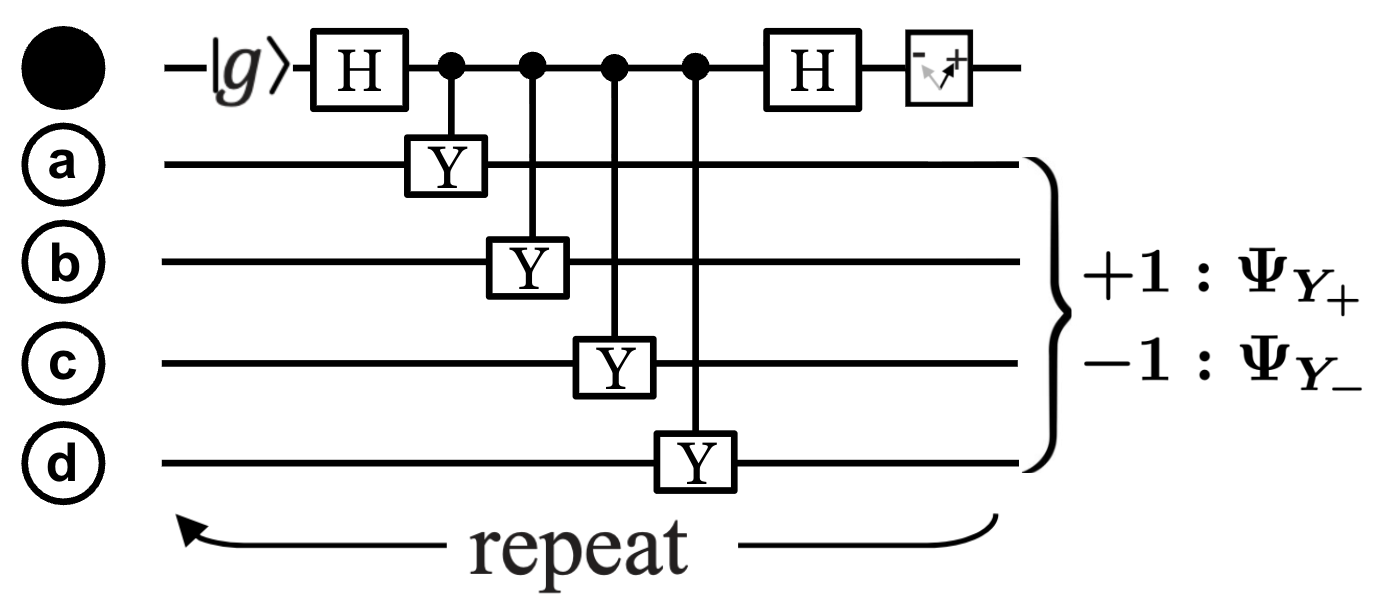}
        \caption{}
        \label{fig:y_stabilizer}
    \end{subfigure}

    \caption{Stabilizer measurement circuits of the Z stabilizer (a), X stabilizer (b), and Y stabilizer (c). Each filled circle ($\bullet$) is the measurement qubit, that acts as an ancilla and performs repeated rounds of controlled operation on four data qubits to give parity information. That parity information helps us find out if an error occurred in the surface code. In the circuits, $\ket{g}$ represents ground state initialization in the computational basis, and I on the ancilla of (a) denotes idle step when the ancilla for $X$ checks and $Y$ checks undergo H gate. The box with arrow at the end of each circuit denotes measurement in computational basis, and $\Psi$ denotes the possible outcomes from the circuit.}
    \label{fig:stabilizers}
\end{figure}

Additionally, since Pauli-$Y$ anti-commutes with both Pauli-$X$ and Pauli-$Z$, it provides an additional bit of syndrome information about the noise biased in $X$. Ref.~\cite{Tuckett2018Ultrahigh} claims to have exploited similar extra bit of syndrome information in their $XY$ surface code for pure dephasing noise. Using tensor-network decoders \cite{Bravyi2014tensorNetwork} would near-optimally utilize this additional bit of information to perform error correction. But in this research, we have implemented a very pessimistic noise model, assuming that each quantum operation is faulty, including syndrome extraction. Under circuit-level noise and noisy syndrome extraction, the decoding problem becomes three-dimensional, two spatial dimensions and one time dimension. So, the maximum-likelihood decoding via tensor networks would require $3D$ tensor-network contraction \cite{Piveteau2024TNBeyond2D}. Though \cite{Piveteau2024TNBeyond2D} present a robust tensor network decoder for such $3D$ decoding problem, they are still only useful when decoding is to be performed offline, and such decoders are computationally expensive. So, in this work, we used minimum-weight perfect matching algorithm PyMatching v2.3.1, which claims to support correlated matching and is still a more practical decoder \cite{Higgott2025sparseblossom}. Developing efficient and practical fault-tolerant decoders with the best achievable thresholds still remains a significant challenge in this field.

\subsection{Memory Experiment}
\label{sec:memory experiment}
The main working mechanism of quantum error correction codes is to perform repeated stabilizer measurements on the data qubits to get the parity information without disturbing their encoded state \cite{nielsen2010, fowler2012}. Standard surface codes have both measure-$Z$ qubits and measure-$X$ qubits that work together to protect the system from bit-flip and phase-flip errors respectively \cite{fowler2012}. A $Z (X)$-memory experiment involves initializing the logical qubits at the eigenstate of $Z_L(X_L)$ i.e., $\ket{0_L} (\ket{+_L})$ state, which is done by initializing the data qubits in the eigenstate of $Z(X)$ i.e., $\ket0 (\ket+)$ state and performing the first round of stabilizer measurements to determine the original parity information of the system \cite{fowler2012, Acharya2023Suppressing}. Then after taking the system through $3d$ (where, $d$ is the code distance) rounds of syndrome extraction, the logical qubit is measured by measuring the data qubits in $Z (X)$ basis, and then checking the parity under $Z_L (X_L)$ \cite{Acharya2023Suppressing}. Then the collected parity information is decoded to determine if a logical error has occurred \cite{Higgott2025sparseblossom}. A logical error is recorded when the decoder's inferred correction and the actual error chain differ by a nontrivial logical operator so that the recovered logical observable disagrees with the initialized logical state \cite{fowler2012, dennis2002}. The logical error rate $(p_L)$ is calculated per $d$ rounds of stabilizer measurement. Then by performing Monte Carlo simulations for different physical error rates ($p$) and distance ($d$), we plot $p_L$ against $p$ \cite{gidney2021stim}. The physical error rate below which increasing the code distance suppresses logical errors is called the threshold $p_{th}$ of the surface code \cite{fowler2012, dennis2002}. In this work, we introduce the $Y$-memory experiment, where the data qubits are initialized in the eigenstate of $Y$ i.e., $\ket{i}$ to prepare the logical qubit in the eigenstate of $Y_L$ i.e., $\ket{i_L}$, and then parity is checked under $Y_L$.

\subsection{CNOT Order and Hook Errors}
\label{sec:cnot_order}

For parity check, each measurement qubit measures four data qubits. Thus, each data qubit is measured by four measure qubits (ancillas) \cite{fowler2012}. In practical implementations, syndrome extraction is performed using ancilla qubits that interact with neighboring data qubits through two-qubit gates \cite{fowler2012}. While performing these two-qubit CNOT operations (Controlled-$Y$ operations for $Y$-type stabilizers), we should make sure that the order of the CNOT operations do not make syndrome information non-deterministic, and that we can parallelly perform the two-qubit gate operation without introducing unnecessary error in the system and without propagating error from one pair to another \cite{lowDistance-cnot, orourke2024comparethepair}. Any CNOT order that does not introduce any such additional errors and does not propagate single qubit errors to other qubits is a valid CNOT order \cite{orourke2024comparethepair}. The copying of a single physical error to data qubits that align with a logical operator is called hook error \cite{dennis2002}. The goal is to use a valid CNOT order that avoids hook errors.

\begin{figure}[htbp]
\centering
\subcaptionbox{\label{fig:cnot_order_circuits}}{%
    \includegraphics[height=0.3\textheight]{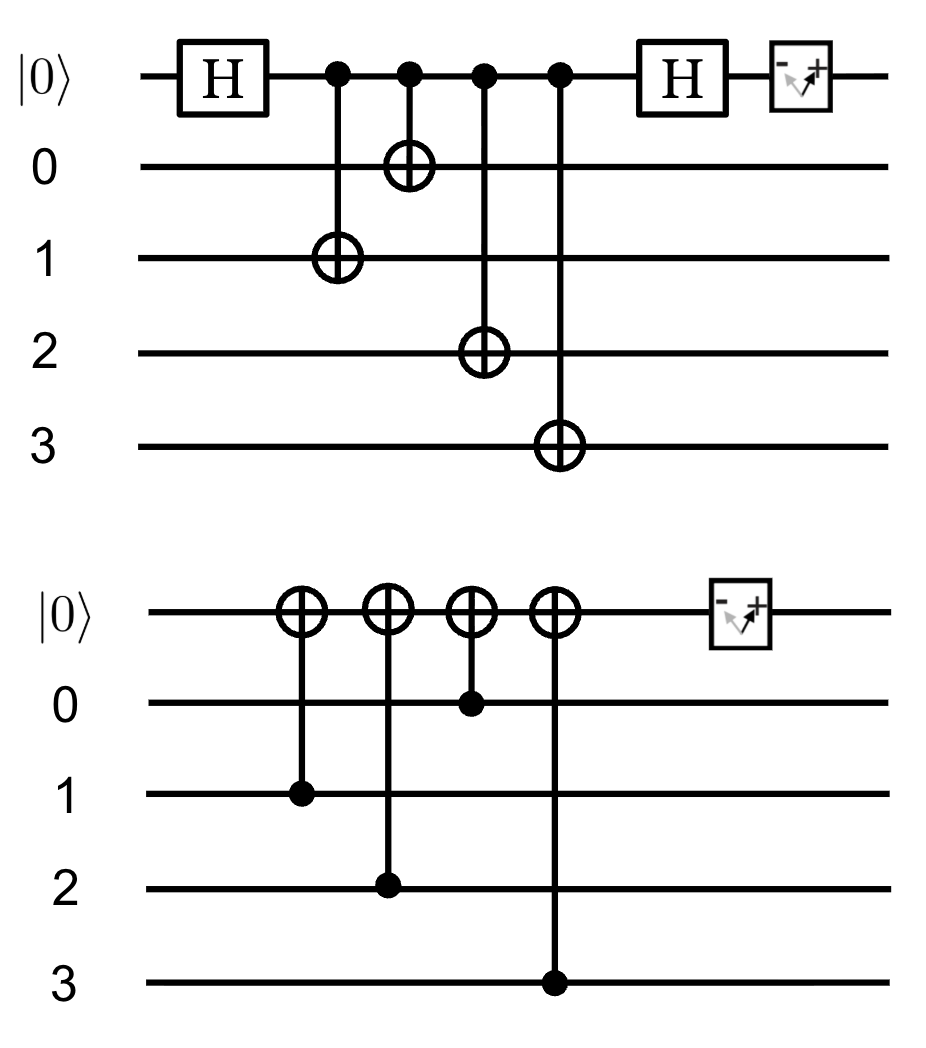}}%
\subcaptionbox{\label{fig:cnot_order_unrot}}{%
    \includegraphics[height=0.3\textheight]{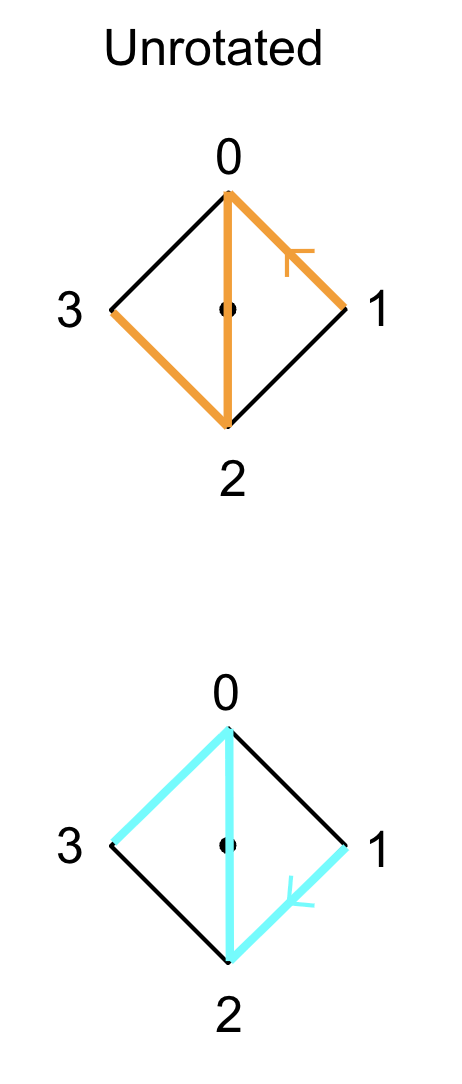}}%
\subcaptionbox{\label{fig:cnot_order_rot}}{%
    \includegraphics[height=0.3\textheight]{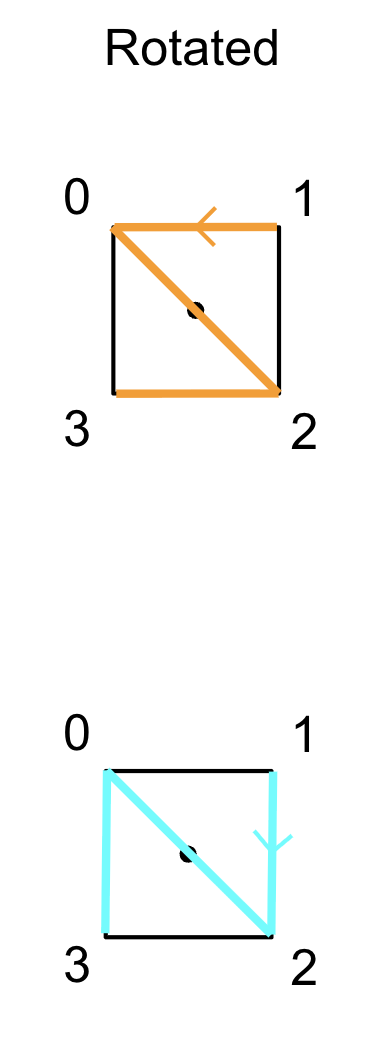}}%
\caption{(a) Representations of syndrome extraction circuits for $X$-type stabilizer (upper diagram) and $Z$-type stabilizer (lower diagram). The upper most qubit in both circuits in (a) are measurement qubits that act as ancilla qubits, and the following four qubits are data qubits whose parity is measured by the circuit. (b) and (c) denote the order of performing CNOT gate in unrotated and rotated codes respectively. This figure shows 10231203 CNOT order, where the first four digits denote the order of CNOT operations in X-type stabilizer measurement circuit and the later four digits represent the order of CNOT operations in $Z$-type stabilizer measurement circuit. We call this CNOT order the $X$-aligned CNOT order in this paper because in this order the second and third qubits are aligned along the $X$-logical operator in unrotated surface code. The $Z$-aligned CNOT order can similarly be inferred referring to this diagram.}
\label{fig:cnot_order}
\end{figure}

There are many valid CNOT orders, but they can be divided into two categories \cite{orourke2024comparethepair}. Let us take the unrotated surface code. First, consider a case in which the second and third qubits are vertically aligned; we will call it $X$-aligned CNOT order. This order is shown in Fig.~\ref{fig:cnot_order}. Second, consider the case in which the second and third qubits are horizontally aligned. We call it $Z$-aligned CNOT order. Here, we want to emphasize that the terminology $X$-aligned and $Z$-aligned do not imply CNOT schedule that is hook-aligned with either of the logical observables. These labels were given by looking at the alignment of the second and third qubits in unrotated surface codes. The CNOT orders used in this paper avoid hook errors for the rotated surface codes.

Among the experiments performed by \cite{orourke2024comparethepair} on many CNOT orders, $10231203$ is claimed to be the $X$-aligned CNOT order that has the worst threshold performance. So, we chose this CNOT order for this research, aligning with our maximally pessimistic philosophy. Additionally, for the $Z$-aligned case, we chose $01320312$ CNOT order because all the Z-aligned CNOT orders show similar behavior \cite{orourke2024comparethepair}. Here, in the eight digit number used to define the CNOT order, the first four digits tell the order of CNOT performed for the $X$-type stabilizer measurement and the last four digits are used to tell the order of CNOT performed for $Z$-type stabilizer measurement. To better understand what these numbers mean, refer to Fig.~\ref{fig:cnot_order}. We present results obtained by performing memory experiments on both of these CNOT orders.

We want to emphasize that the "$X$-aligned schedule" and "$Z$-aligned schedule" in this paper refer to the second and third qubit in the stabilizer check process being aligned along the respective logical operators. This does not refer to the hook-direction alignment of the checks in rotated surface code. We have chosen the schedule for the rotated surface code such that the hook error is avoided.


\section{Methods}
\label{sec:methods}

\subsection{Error Model}
\label{sec:error model}
In general, noise models can be categorized as independent noise models and correlated noise models. In the independent noise model we assume that the noise channel affects individual qubits independently, irrespective of change in any nearby qubits.

In this research, we simulate circuit-level noise that includes independent Pauli noise and \textit{crosstalk} noise. Formally, the standard depolarizing channel for single-qubit state denoted by the density operator $\rho$ is given by \cite{nielsen2010}

\begin{equation}
    \mathcal{E}(\rho) = (1-p)\rho + \frac{p}{3}(X\rho X+ Y\rho Y+ Z\rho Z).    
\end{equation}

More generally, it can be written as

\begin{equation}
    \mathcal{E}(\rho) = (1-p)\rho + \frac{p}{3} \sum_i \sigma^i \rho \sigma^i,    
\end{equation}
where, $i \in \{X, Y, Z \}$. The action of this channel on the qubit state $\ket\psi$ can be described as: with probability $(1-p)$, nothing happens to the qubit state, and with equal probability of $p/3$ each of the operators $X, Y, \text{and} \ Z$ is applied to the qubit state. Here, $p$ is the physical error rate.

The single-qubit Pauli noise channel can be redefined to work with varying probability of each error operator as
\begin{equation}
    \mathcal{E}_1(\rho) = (1-p)\rho + p_x \ X\rho X+ p_y \ Y\rho Y+ p_z \ Z\rho Z.
    \label{single-qubit-biased-error-channel}
\end{equation}
such that with probability $(1-p)$ nothing happens to the qubit, bit-flip error ($X$ operator) is applied to the qubit with probability $p_x$, phase-flip error ($Z$ operator) is applied to the qubit with probability $p_z$, and bit-phase flip error ($Y$ operator) is applied with probability $p_y$. For the Pauli-X biased noise model, we define a bias parameter $\eta$ in terms of error probabilities $p_x, \ p_y, \ \text{and} \ p_z$ as
\begin{equation}
    \eta = \frac{p_x}{p_y+p_z}.    
\end{equation}

The bias towards a certain noise in a system varies significantly based on the qubit processor and its complete system being considered \cite{Aliferis_2009,Seis_2023}. For simplicity, we consider $p_y = p_z$ \cite{Tuckett2018Ultrahigh}. Thus, when $\eta = \frac{1}{2}$, we have $p_x = p_y = p_z$, which is the standard depolarizing noise channel, i.e., all the error operators are applied with equal probability, and $\eta = \infty$ signifies a pure $X$-biased noise such that $p_x = p$, and $p_y = p_z = 0$. Rearranging the equations, we can get the relation for $p_x, \ p_y, \ \text{and} \ p_z$ in terms of $p$ and $\eta$ as these are the two parameters that we vary in our simulations to simulate the noise under different physical error rates and different bias. The individual probabilities in terms of bias parameter $\eta$ and physical error rate $p$ are
\begin{equation}
\begin{aligned}
    p_x &= \frac{\eta}{\eta + 1} p \hspace{2em}
\end{aligned}
\end{equation}

\begin{equation}
\begin{aligned}
    p_y &= p_z = \frac{1}{2(\eta + 1)} p.
\end{aligned}
\end{equation}
The bias is set up such that $p_x + p_y+p_z=p$. This ensures that the four channel probabilities, including the identity, sum to one. Additionally, it ensures that as the bias parameter increases, $p_x$ increases and, $p_y$ and $p_z$ decrease. 

Similarly, after every two-qubit gate, we apply an independent two-qubit Pauli channel: 
\begin{equation}
\mathcal{E}_2(\rho)
=
\left(1 - \sum_{(i,j)\neq(I,I)} p_{ij}\right)\rho
+
\sum_{(i,j)\neq(I,I)} p_{ij}
(\sigma_i \otimes \sigma_j)\,
\rho\,
(\sigma_i \otimes \sigma_j),
\label{two-qubit-pauli-error-channel}
\end{equation}
where $i,j \in \{I, X, Y, Z\}$. In the absence of correlated noise, the probabilities factorize as $p_{ij} = p_i p_j$, corresponding to independent biased Pauli errors acting on each qubit.


In addition to biased Pauli noise, we include correlated $XX$ crosstalk noise in our circuit-level error model. The unwanted coupling between target and spectator qubits due to control signal spillover is a well-documented error source in multiple quantum platforms, including cross-resonance interactions in superconducting qubits \cite{zhao2023}, M{\o}lmer--S{\o}rensen gate crosstalk in trapped-ion systems \cite{Fang2022Crosstalk, parrado2021}, and Rydberg addressing leakage in neutral-atom systems \cite{Warttmann_2026}. Trapped-ion two-qubit gates are commonly realized as M\o{}lmer--S\o{}rensen (MS) gates. Here, a bichromatic laser drive entangles a pair of ions through an effective $\hat{\sigma}_\phi^{(i)}\hat{\sigma}_\phi^{(j)}$ interaction, where $\hat{\sigma}_\phi = \hat{\sigma}_x\cos\phi + \hat{\sigma}_y\sin\phi$ is the spin operator along the equatorial-plane axis set by the laser phase $\phi$. For $\phi = 0$ this reduces to the $XX$-type coupling considered here. Because the addressing beam intended for a target ion $i$ leaks partially onto its neighbors, a nearby spectator ion $j$ is weakly co-driven, producing a residual two-body $\hat{\sigma}_x^{(i)}\hat{\sigma}_\phi^{(j)}$ interaction between target and spectator. The strength of this leakage is quantified by the addressing crosstalk ratio $\epsilon_{ij} = \Omega_j/\Omega_i$, where $\Omega_i$ is the Rabi frequency of the intended drive on the target ion and $\Omega_j$ is the parasitic Rabi frequency induced on the spectator. Measured values of $\epsilon_{ij}$ are typically $1$--$3\%$ for nearest neighbors \cite{Fang2022Crosstalk}. We note that in physical systems this crosstalk is predominantly coherent, and active suppression techniques exist \cite{Fang2022Crosstalk}.

In our simulations, we model residual crosstalk as a stochastic Pauli $X \otimes X$ channel applied after each two-qubit gate. Among the four categories of crosstalk: gate-based and always-on, each arising between data-data or data-ancilla qubit pairs, we inject gate-based data-ancilla crosstalk, as it has been identified as the most detrimental to error correction performance by \cite{zhou2025}.

We set the crosstalk probability to $p_{\times} = 5 \times 10^{-4}$. We choose this value to be small relative to the two-qubit gate error rates in our parameter sweep but large enough to produce observable effects on logical error rates. This allows us to study whether even weak residual correlated errors influence threshold behavior in surface codes under biased noise. We emphasize that $p_{\times}$ is treated as a fixed parameter independent of the single-qubit physical error rate $p$, reflecting the fact that crosstalk arises from a distinct physical mechanism (e.g., optical addressing imperfections) rather than from the same source of gate infidelity parameterized by $p$.


Thus, we have applied circuit-level $XX$ crosstalk noise in our surface codes. Formally, an $N$-qubit crosstalk noise is defined as
\begin{equation}
    \mathcal{E}_{N_i} (\rho) = (1-p_{\times}) \rho + p_{\times} \ \sigma_i^{\otimes N} \rho \ \sigma_i^{\otimes N},
\end{equation}
where, $i \in \{X, Y, Z\}$. As given by the following operator, we apply this $XX$ crosstalk noise in the qubit pair after each two-qubit gate
\begin{equation}
    \mathcal{E}_{2_X} (\rho) = (1-p_{\times}) \rho + p_{\times} \ XX \rho XX.
\end{equation}

Here, we make a maximally pessimistic assumption about occurrence of errors in syndrome extraction. Refer to Fig.~\ref{fig:cnot_order_circuits} for syndrome extraction steps. Below are all the errors we introduce in each step of syndrome extraction process:

\begin{itemize}
    \item \textit{Initialization error:}
    In each memory experiment, data qubits are initialized in the corresponding eigenbasis. After initialization, a basis-dependent Pauli error that flips the prepared eigenstate is applied to the data qubits. Ancilla qubits are initialized in the computational ($Z$) basis, and a bit-flip ($X$) error is applied after initialization to model imperfect ancilla preparation.

    \item \textit{Single-qubit gate and idling error:}
    After each single-qubit gate operation, a biased single-qubit Pauli channel $\mathcal{E}_1$  (Eq.~\ref{single-qubit-biased-error-channel}) is applied to the participating qubits. During the same time step, qubits on which no gate operation is performed are treated as idle qubits and the same biased single-qubit Pauli channel is applied to those qubits to model idling noise. Here, we use set the idling error rate $(p_{idle})$ to be same as the physical error rate $(p)$.

    \item \textit{Two-qubit gate error:}
    After each two-qubit gate operation, an independent two-qubit Pauli channel $\mathcal{E}_2$ (Eq.~\ref{two-qubit-pauli-error-channel}) is applied to the participating qubit pairs. In circuit instances that include correlated noise, an additional pairwise correlated $X \otimes X$ error is applied after each two-qubit gate with probability $p_{\times}$. Idle qubits during the same time step are subjected to the single-qubit Pauli channel $\mathcal{E}_1$.

    \item \textit{Ancilla measurement, readout, and reset error:}
    Ancilla qubits are measured in the computational ($Z$) basis for syndrome extraction. A bit-flip ($X$) error is applied immediately before measurement to model readout error, while a bit-flip error applied immediately after the reset operation to model imperfect reset and state preparation for the subsequent round.

    \item \textit{Final data-qubit readout error:}
    At the end of the memory experiment, data qubits are measured in the corresponding basis. Immediately prior to this final measurement, a basis-dependent Pauli error is applied to model measurement error by flipping the measurement outcome.

\end{itemize}


\subsection{Simulation}
\label{sec:simulation}

We simulated $X$ and $Z$ memory experiments on the $ZX$ surface code and $Y$ and $Z$ memory experiments on $ZY$ surface code. We simulated both the rotated and unrotated geometries of these surface code structures under circuit-level noise. For the $ZX$ surface code, we used two CNOT orders: the $X$-aligned CNOT order $10231203$ and the $Z$-aligned CNOT order $01320312$ as described in section \ref{sec:cnot_order}. For the $ZY$ surface code, we used only the $X$-aligned CNOT order $10231203$. Doing the $Z$-aligned CNOT order for the $ZY$ surface code is the beyond the scope of this work. All simulations were performed for surface codes of distances $d \in \{5, 6, 7, \ldots, 17\}$ and bias parameters $\eta \in \{0.5, 1, 3, 10, 30, 100, 300, 1000, \infty\}$. For each $(d, \eta)$ combination, we swept the physical error rate $p$ over $40$ values spanning from $5 \times 10^{-4}$ to $4.6 \times 10^{-1}$ with uneven gaps. Each circuit comprised $3d$ rounds of stabilizer measurement.

To generate the Stim \cite{gidney2021stim} circuits for stabilizer measurement, we extended the Python circuit-generation code \cite{higgott2024} to support $ZY$ stabilizer configurations, $Y$-stabilizer measurement circuits, and customizable CNOT orderings. Monte Carlo sampling was performed using Sinter, Stim's built-in sampling and statistics collection framework \cite{gidney2021stim, sinter}. For decoding, we used PyMatching v2.3.1 \cite{Higgott2025sparseblossom}. All the codes and the plots for this research can be found in this github repository: \textit{https://github.com/Pritesh402/zxzy-qec}.


\subsection{Threshold Extraction and Analysis}
\label{sec:plotting}

To study the error-correcting performance of each surface code variant, we plot the logical error rate $p_L$ as a function of the physical error rate $p$ for each code distance $d$ and bias parameter $\eta$. The threshold $p_{\mathrm{th}}$ is the critical physical error rate below which increasing the code distance suppresses logical errors, and above which increasing the code distance provides no error-correcting advantage. Although each circuit runs $3d$ rounds of stabilizer measurement (Sec.~\ref{sec:simulation}), we report $p_L$ per $d$ rounds, following \cite{orourke2024comparethepair}. Running $3d$ rounds rather than $d$ reduces time-boundary edge effects \cite{orourke2024comparethepair, Acharya2023Suppressing}. The raw per-shot logical error rate is
\begin{equation}
    p_L' = \frac{k}{n},
\end{equation}
where $k$ is the number of observed logical failures and $n$ is the total number of Monte Carlo trials. To perform meaningful comparison across code distances, we convert $p_L'$ to a logical error rate per $d$ rounds. For this conversion, we model the $3d$-round experiment as three consecutive and independent blocks of $d$ rounds. In each block, the logical state either remains correct (with probability $1 - p_L$) or is flipped by a logical error (with probability $p_L$). Since a logical error acts as a flip, the final logical state is incorrect only if an odd number of the three blocks produce a flip. The resulting per-shot error probability is therefore
\begin{equation}
    p_{L}' = \frac{1 - (1 - 2\,p_L)^{3}}{2}.
    \label{eq:shot_to_piece}
\end{equation}
Inverting this relation gives the reported logical error rate per $d$ rounds:
\begin{equation}
    p_L = \frac{1 - (1 - 2\,p_L')^{1/3}}{2}.
    \label{eq:piece_error_rate}
\end{equation}
For small $p_L'$, this reduces to $p_L \approx p_L'/3$, but the exact nonlinear inversion is used throughout this work. We estimate the uncertainties on $p_L$ by first computing the analytic root mean square error (RMSE) of the maximum likelihood estimator for the binomial parameter \cite{orourke_code}
\begin{equation}
    \mathrm{RMSE}_{\mathrm{shot}} = \sqrt{\frac{6k^2 - nk(k+6) + n^2(k+2)}{n^2(n+2)(n+3)}},
    \label{eq:rmse_mle}
\end{equation}
and then propagating this uncertainty through the nonlinear conversion in Eq.~\eqref{eq:piece_error_rate} via the delta method:
\begin{equation}
    \mathrm{RMSE}_{p_L} = \mathrm{RMSE}_{\mathrm{shot}} \cdot \frac{1}{3}\left(1 - 2\,p_L'\right)^{-2/3}.
    \label{eq:rmse_propagated}
\end{equation}
We plot the threshold curves on logarithmic axes as $\log_{10}(p_L)$ versus $\log_{10}(p)$ for each distance $d$, separately for each memory experiment type ($X$, $Y$, or $Z$).

To extract threshold estimates, we follow the critical exponent scaling method of \cite{Wang2003ConfinementHiggs, Tuckett2018Ultrahigh}. Near the threshold, the logical error rate is expected to obey a scaling ansatz of the form $p_L = f\!\big((p - p_{\mathrm{th}})\,d^{1/\nu}\big)$, where $\nu$ is the critical exponent. We fit our data in log-space to a truncated quadratic expansion of this scaling function:
\begin{equation}
    \log_{10}(p_L) = A + B\,(p - p_{\mathrm{th}})\,d^{1/\nu} + C\,(p - p_{\mathrm{th}})^2\,d^{2/\nu},
    \label{eq:log_scaling_fit}
\end{equation}
where $A$, $B$, $C$, $p_{\mathrm{th}}$, and $\nu$ are free parameters determined by weighted nonlinear least-squares fitting. The weights are derived from the propagated uncertainties in log-space,
\begin{equation}
    \sigma_{\log} = \frac{\mathrm{RMSE}_{p_L}}{p_L \cdot \ln 10},
    \label{eq:sigma_log}
\end{equation}
and the per-point RMSE is floored at $5\%$ of the $p_L$ to prevent individual data points from dominating the fit. For each memory experiment type and bias parameter $\eta$, the fit is performed over a data window centered on the expected threshold region. We determine the expected threshold region used here through visual inspection of the logical error rate plot.

\section{Results}
\label{sec:results}

Here, we report the results obtained by simulating different noise models on different surface code structures. As discussed in section \ref{sec:error model}, we simulated two noise models, one with only Pauli-$X$ biased noise, and another with an added $XX$ crosstalk noise.

We simulated these noise models on both rotated and unrotated variants of the $ZX$ and $ZY$ surface codes. For the $ZX$ surface code, we simulated two different CNOT orders, which as discussed in section \ref{sec:cnot_order}, we call $X$-aligned CNOT order and $Z$-aligned CNOT order. For these simulations, surface codes of distances $d=5$ to $d=17$ were used, with the bias parameters $\eta = \{0.5,1,3,10,30,100,300,1000, \infty \}$. 

In the following parts, we present all the results for each surface code type. Since there was a clear split in the results obtained from different memory experiments, we report different threshold values for each memory experiment type, $X, Y, \ \text{and} \ Z$, as $p_{\mathrm{th}}^{X}$, $p_{\mathrm{th}}^{Y}$ and $p_{\mathrm{th}}^{Z}$ respectively. The uncertainties from the scaling fits $(\sigma_{f})$ are not uniform across
all threshold values. For the $Z$ and $Y$ memory experiments, whose thresholds vary only weakly with bias, we obtain $\sigma_{f} \approx 0.002 \times 10^{-2}$ across the full range of $\eta$. For the $X$ memory experiments, whose thresholds increase substantially with bias, $\sigma_{f}$ increases accordingly, from $\approx 0.002 \times 10^{-2}$ at low bias to $\approx 0.02 \times 10^{-2}$ at high bias $(\eta \geq 30)$. Therefore, we quote a separate fit uncertainty for each reported threshold value rather than a single global value. We want to emphasize that the quoted uncertainties are statistical fit uncertainties only. The range of $p$ and $p_L$ is same for all the plots shown here. Every numerical value reported in this section carries its own fitting uncertainty $(\sigma_{f})$. The difference between two threshold values is assessed against the combined fitting uncertainty $\sqrt{\sigma_{f,1}^{2} + \sigma_{f,2}^{2}}$, evaluated from the individual uncertainties of the two values being compared.

\begin{table}[htbp]
\centering
\caption{Summary of surface-code configurations by figure and table.}
\label{tab:config-summary}
\begin{tabular}{cccccc}
\toprule
Table & Figure & Code type & Geometry & CNOT type & Noise \\
\midrule
\multirow{2}{*}{T.\ref{tab:zx_xbiased}} & Fig.\ref{fig:zx_rot_bi_selected} & \multirow{2}{*}{ZX} & Rotated & \multirow{2}{*}{X-aligned} & \multirow{2}{*}{Pauli-X biased} \\
                     & Fig.\ref{fig:zx_unrot_bi_selected} & & Unrotated & & \\
\midrule
\multirow{2}{*}{T.\ref{tab:zx_crosstalk}} & Fig.\ref{fig:zx_rot_wc_selected} & \multirow{2}{*}{ZX} & Rotated & \multirow{2}{*}{X-aligned} & \multirow{2}{*}{Pauli-X biased $+$ crosstalk} \\
                     & Fig.\ref{fig:zx_unrot_wc_selected} & & Unrotated & & \\
\midrule
\multirow{2}{*}{T.\ref{tab:zx_biased_zalign}} & Fig.\ref{fig:zx_zl_rot_bi} & \multirow{2}{*}{ZX} & Rotated & \multirow{2}{*}{Z-aligned} & \multirow{2}{*}{Pauli-X biased} \\
                     & Fig.\ref{fig:zx_zl_unrot_bi} & & Unrotated & & \\
\midrule
\multirow{2}{*}{T.\ref{tab:zx_crosstalk_zalign}} & Fig.\ref{fig:zx_zl_rot_wc} & \multirow{2}{*}{ZX} & Rotated & \multirow{2}{*}{Z-aligned} & \multirow{2}{*}{Pauli-X biased $+$ crosstalk} \\
                     & Fig.\ref{fig:zx_zl_unrot_wc} & & Unrotated & & \\
\midrule
\multirow{2}{*}{T.\ref{tab:zy_xbiased}} & Fig.\ref{fig:zy_rot_bi} & \multirow{2}{*}{ZY} & Rotated & \multirow{2}{*}{X-aligned} & \multirow{2}{*}{Pauli-X biased} \\
                     & Fig.\ref{fig:zy_unrot_bi} & & Unrotated & & \\
\midrule
\multirow{2}{*}{T.\ref{tab:zy_xbiased_crosstalk}} & Fig.\ref{fig:zy_rot_wc} & \multirow{2}{*}{ZY} & Rotated & \multirow{2}{*}{X-aligned} & \multirow{2}{*}{Pauli-X biased $+$ crosstalk} \\
                     & Fig.\ref{fig:zy_unrot_wc} & & Unrotated & & \\
\bottomrule
\end{tabular}
\end{table}

\subsection{ZX Surface Code with X-aligned CNOT Order}
\label{sec:zx_x-align_result}

\begin{figure}[!htbp]
    \centering
    
    \begin{subfigure}[t]{0.32\textwidth}
        \centering
        \includegraphics[width=\linewidth]{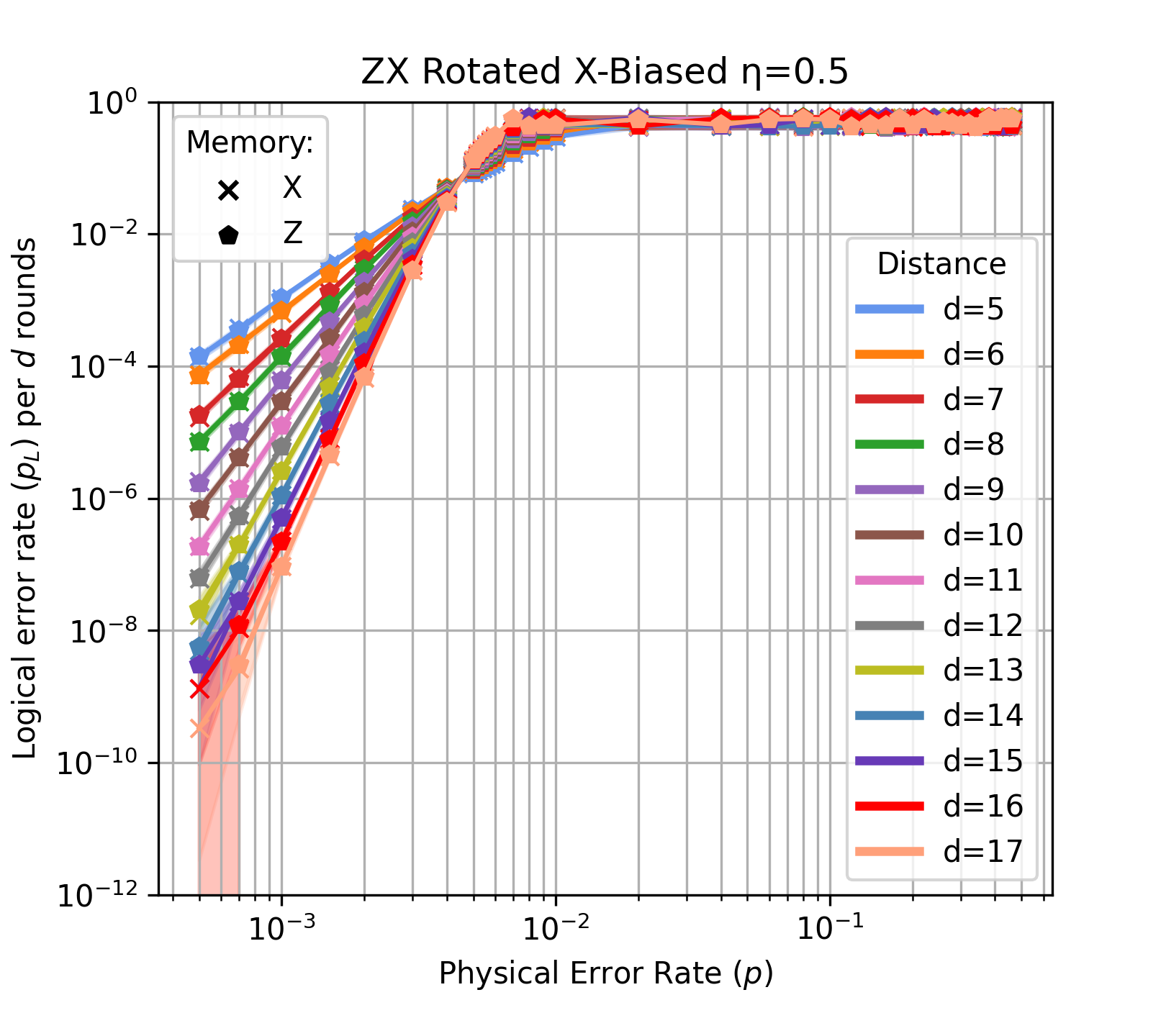}
        \caption{}
        \label{fig:4a}
    \end{subfigure}
    \hfill
    \begin{subfigure}[t]{0.32\textwidth}
        \centering
        \includegraphics[width=\linewidth]{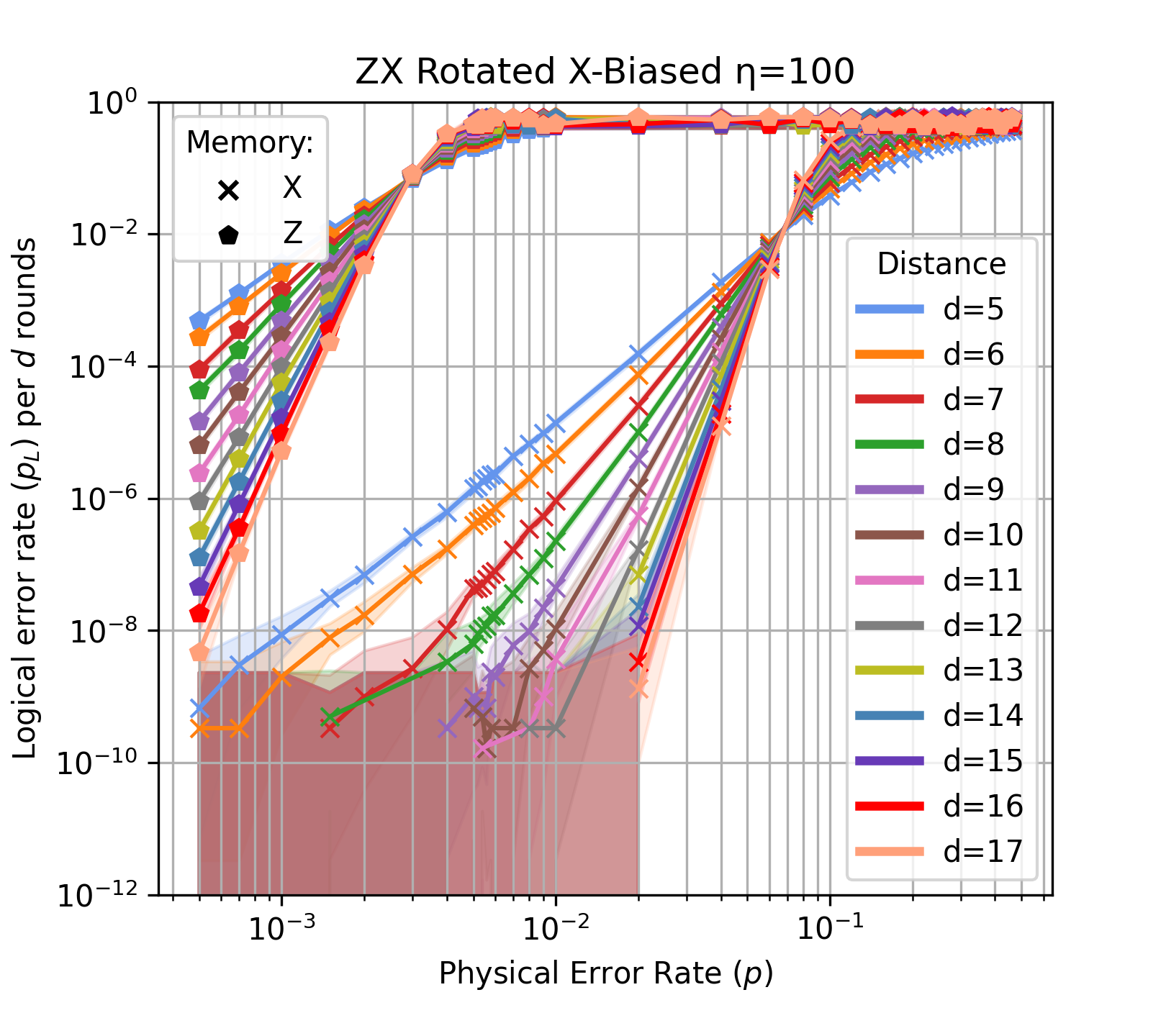}
        \caption{}
        \label{fig:4b}
    \end{subfigure}
    \hfill
    \begin{subfigure}[t]{0.32\textwidth}
        \centering
        \includegraphics[width=\linewidth]{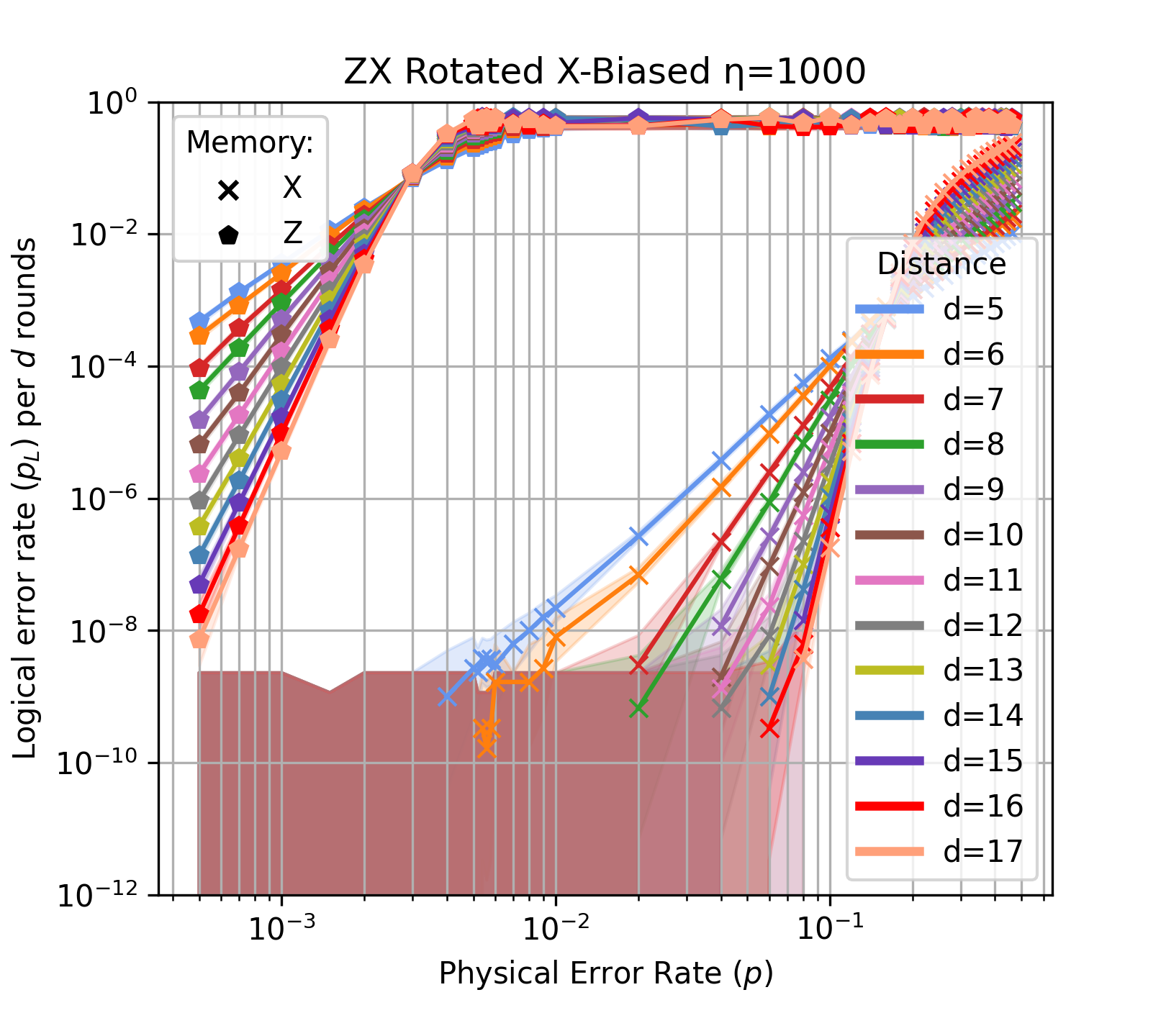}
        \caption{}
        \label{fig:4c}
    \end{subfigure}

    \caption{Logical error rate per $d$ rounds of syndrome extraction $p_L$ versus physical error rate $p$ simulated on rotated $ZX$ surface code with $X$-aligned CNOT order under Pauli-$X$ biased noise at $\eta = \{0.5, 100, 1000\}$. }
    
    \label{fig:zx_rot_bi_selected}
\end{figure}

\begin{figure}[!htbp]
    \centering
    
    \begin{subfigure}[t]{0.32\textwidth}
        \centering
        \includegraphics[width=\linewidth]{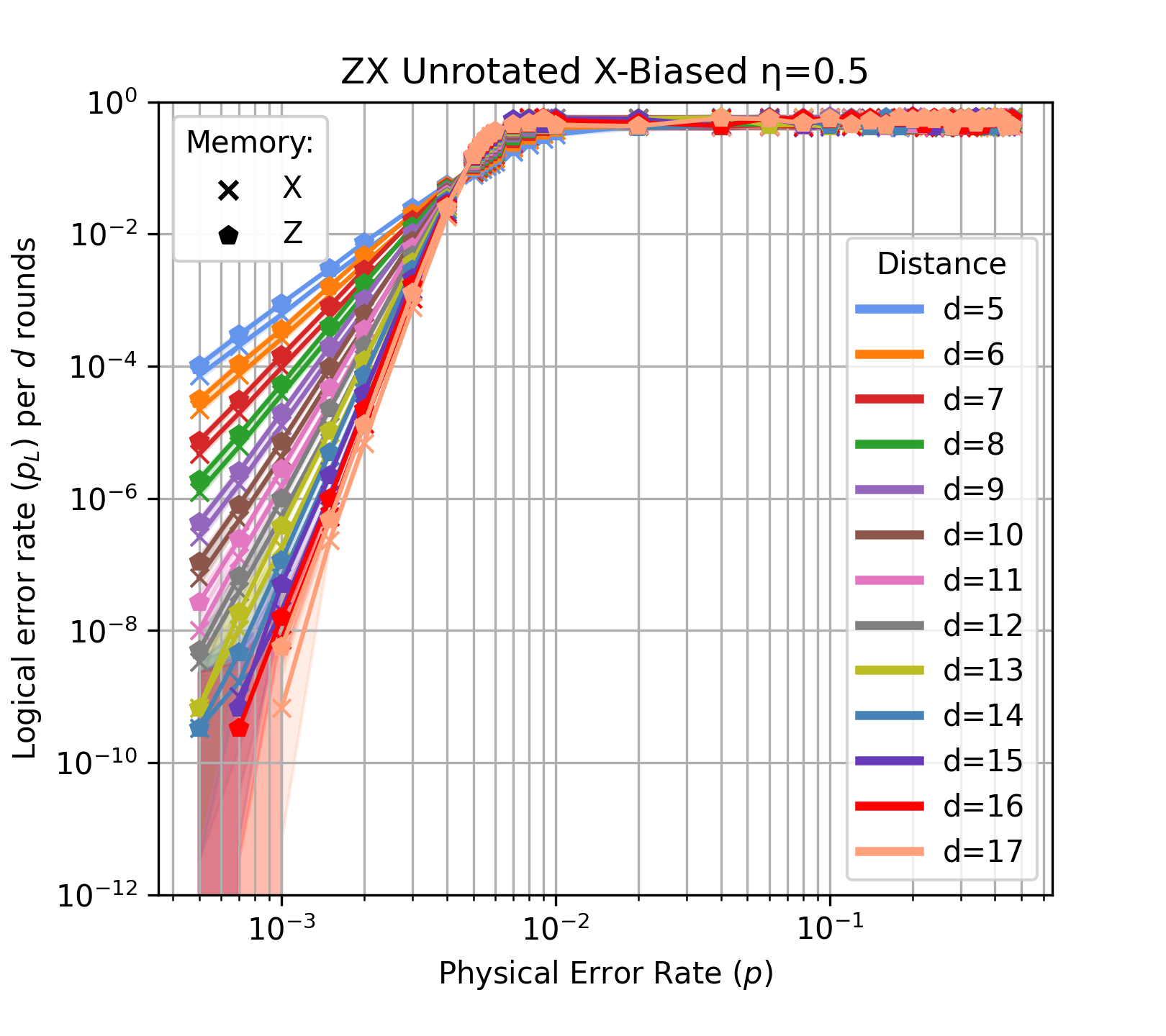}
        \caption{}
        \label{fig:zx_unrot_bi_a}
    \end{subfigure}
    \hfill
    \begin{subfigure}[t]{0.32\textwidth}
        \centering
        \includegraphics[width=\linewidth]{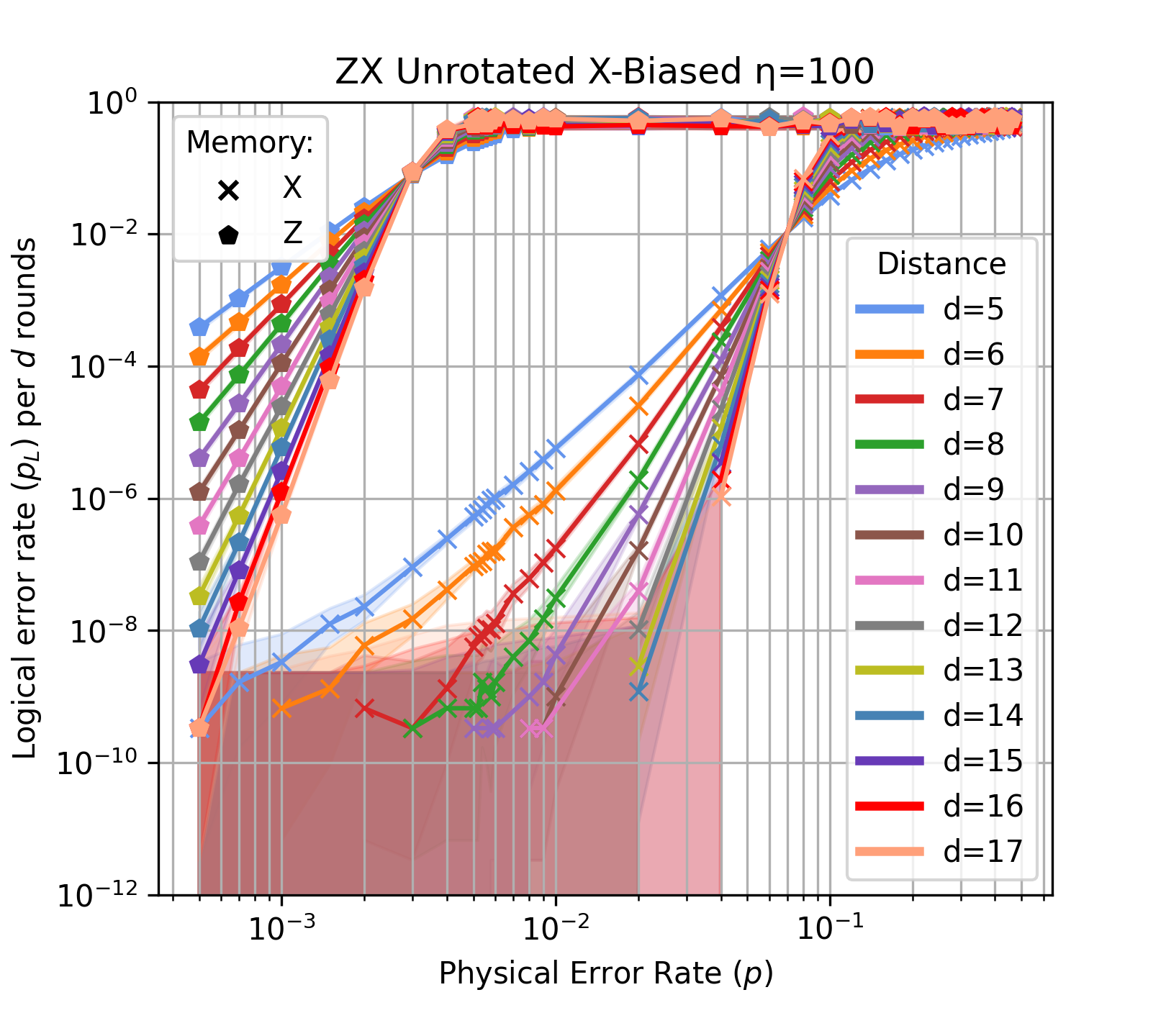}
        \caption{}
        \label{fig:zx_unrot_bi_b}
    \end{subfigure}
    \hfill
    \begin{subfigure}[t]{0.32\textwidth}
        \centering
        \includegraphics[width=\linewidth]{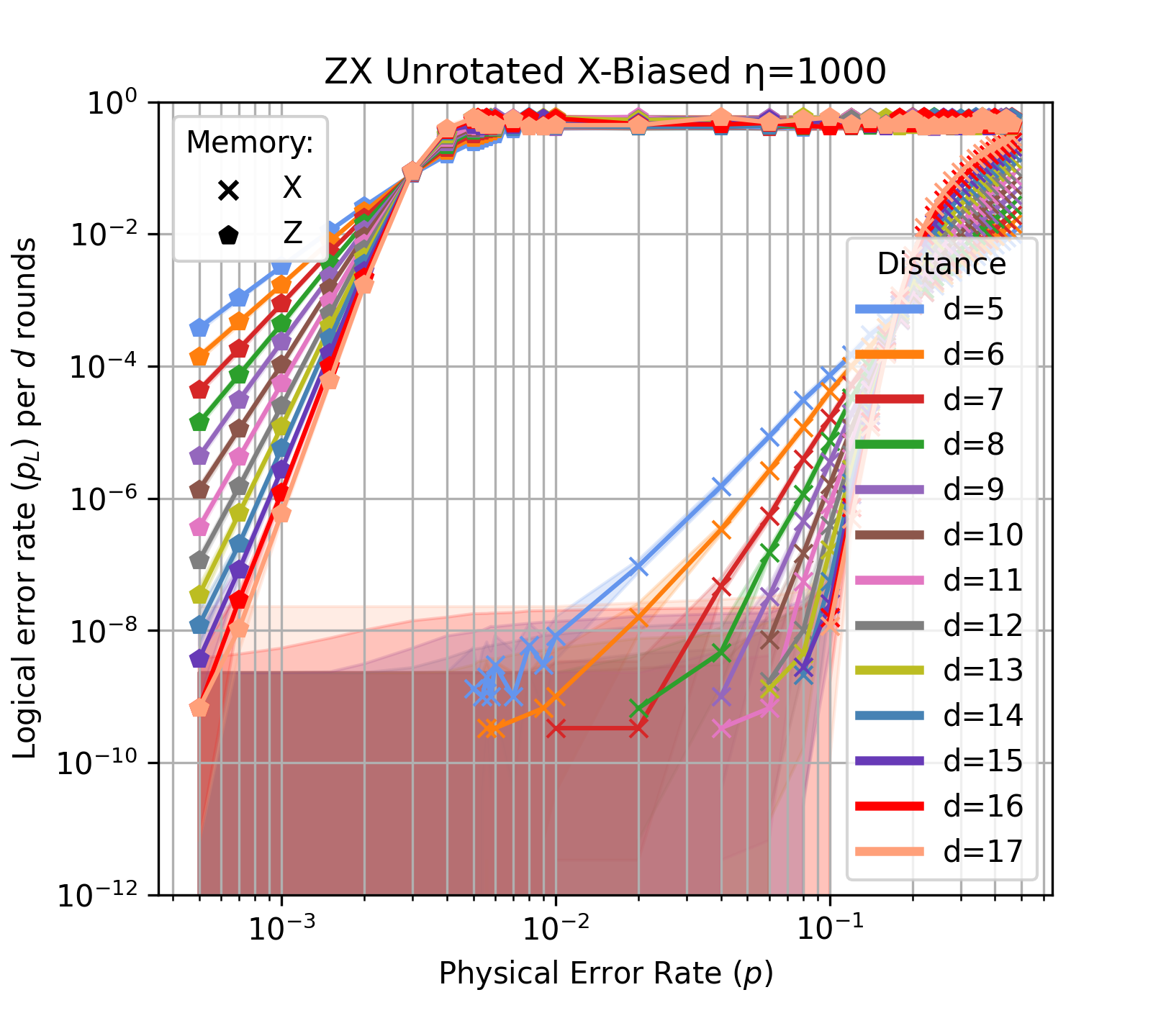}
        \caption{}
        \label{fig:zx_unrot_bi_c}
    \end{subfigure}

    \caption{Logical error rate per $d$ rounds of syndrome extraction $p_L$ versus physical error rate $p$ simulated on unrotated $ZX$ surface code with $X$-aligned CNOT order under Pauli-$X$ biased noise at $\eta = \{0.5, 100, 1000\}$.}

    \label{fig:zx_unrot_bi_selected}
\end{figure}

\begin{figure}[!htbp]
    \centering
    
    \begin{subfigure}[t]{0.32\textwidth}
        \centering
        \includegraphics[width=\linewidth]{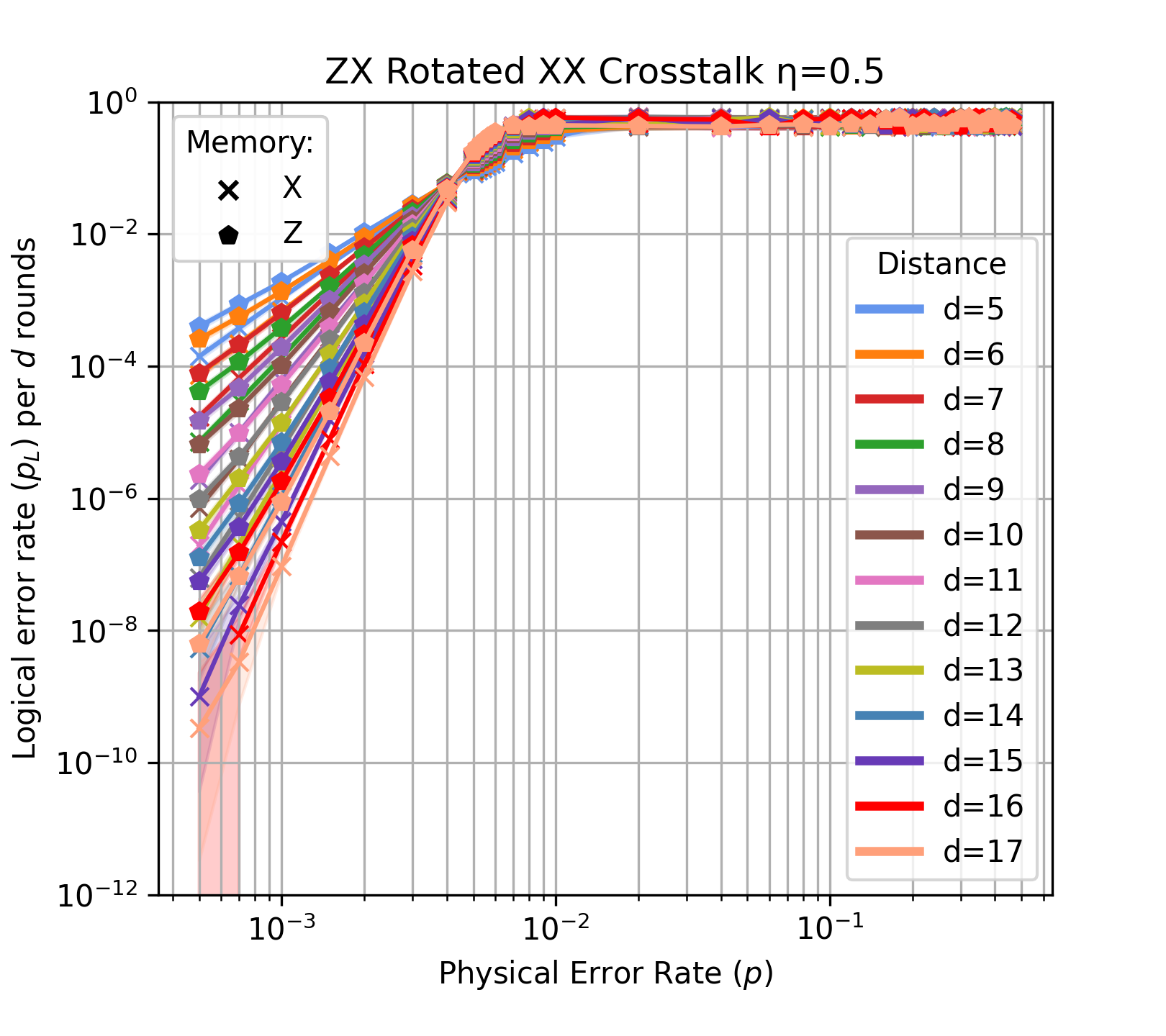}
        \caption{}
        \label{fig:zx_rot_wc_a}
    \end{subfigure}
    \hfill
    \begin{subfigure}[t]{0.32\textwidth}
        \centering
        \includegraphics[width=\linewidth]{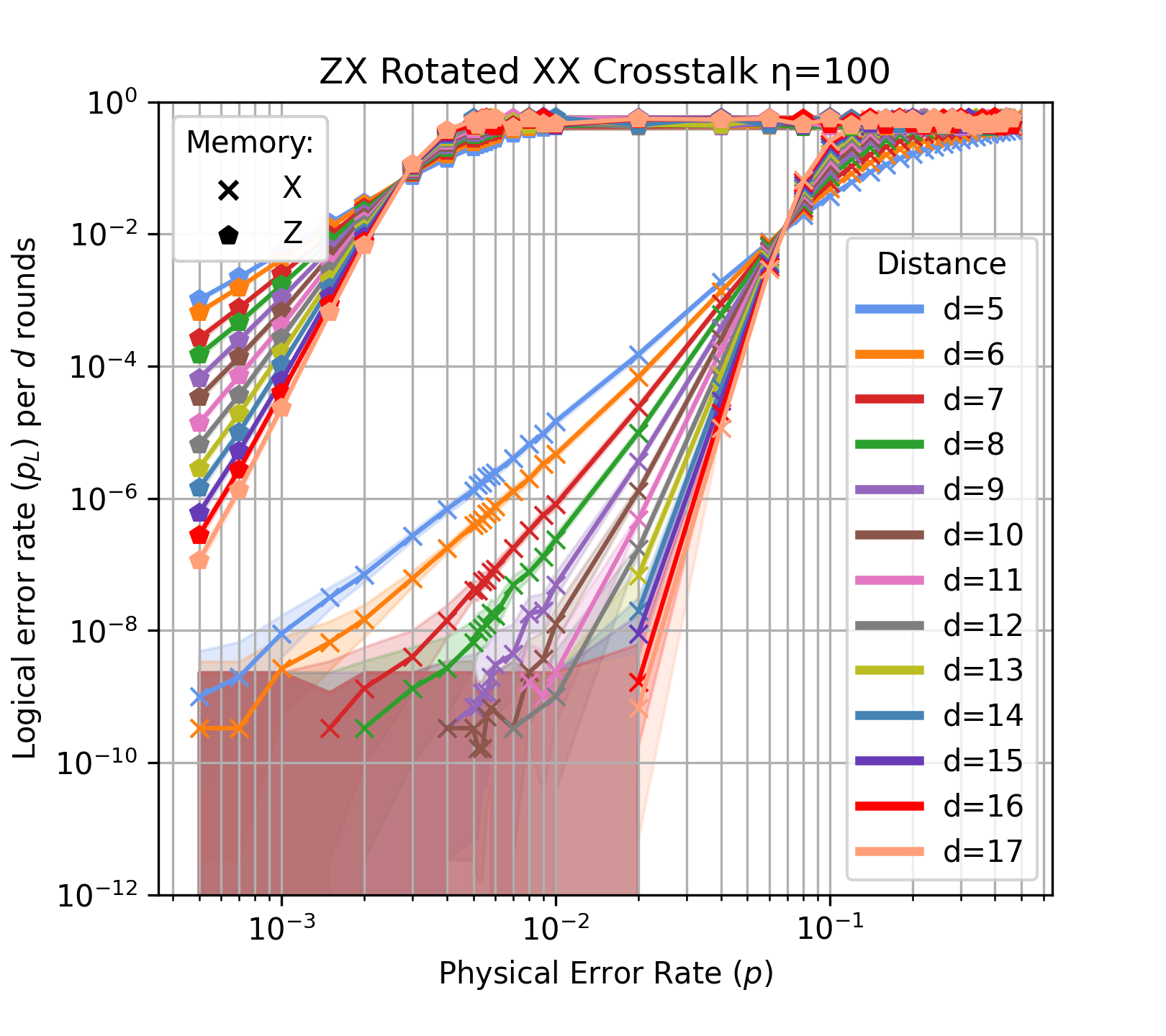}
        \caption{}
        \label{fig:zx_rot_wc_b}
    \end{subfigure}
    \hfill
    \begin{subfigure}[t]{0.32\textwidth}
        \centering
        \includegraphics[width=\linewidth]{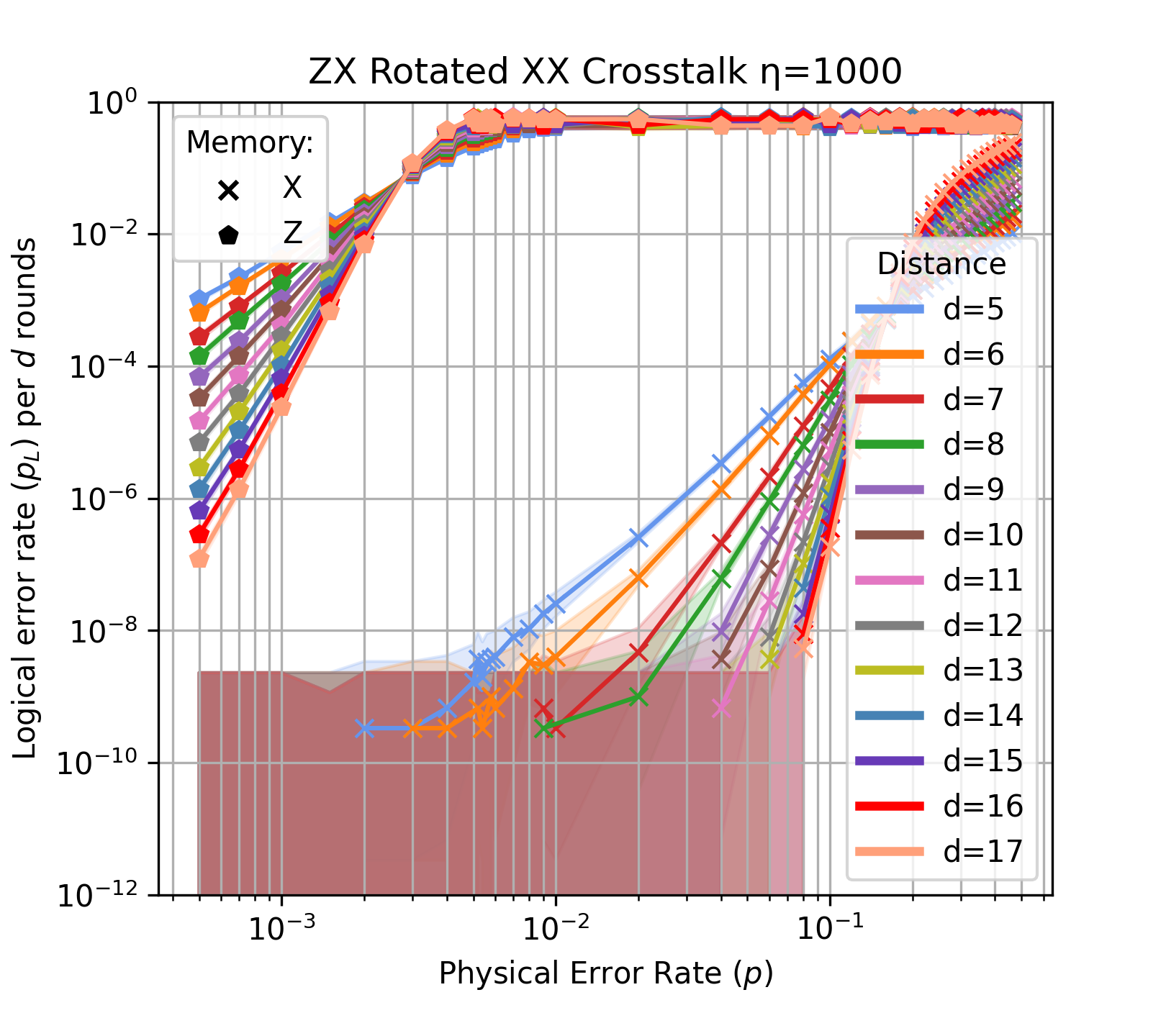}
        \caption{}
        \label{fig:zx_rot_wc_c}
    \end{subfigure}

    \caption{Logical error rate per $d$ rounds of syndrome extraction $p_L$ versus physical error rate $p$ simulated on rotated $ZX$ surface code with $X$-aligned CNOT order under Pauli-$X$ biased noise at $\eta = \{0.5, 100, 1000\}$ with an additional $XX$ crosstalk noise.}

    \label{fig:zx_rot_wc_selected}
\end{figure}

\begin{figure}[!htbp]
    \centering
    
    \begin{subfigure}[t]{0.32\textwidth}
        \centering
        \includegraphics[width=\linewidth]{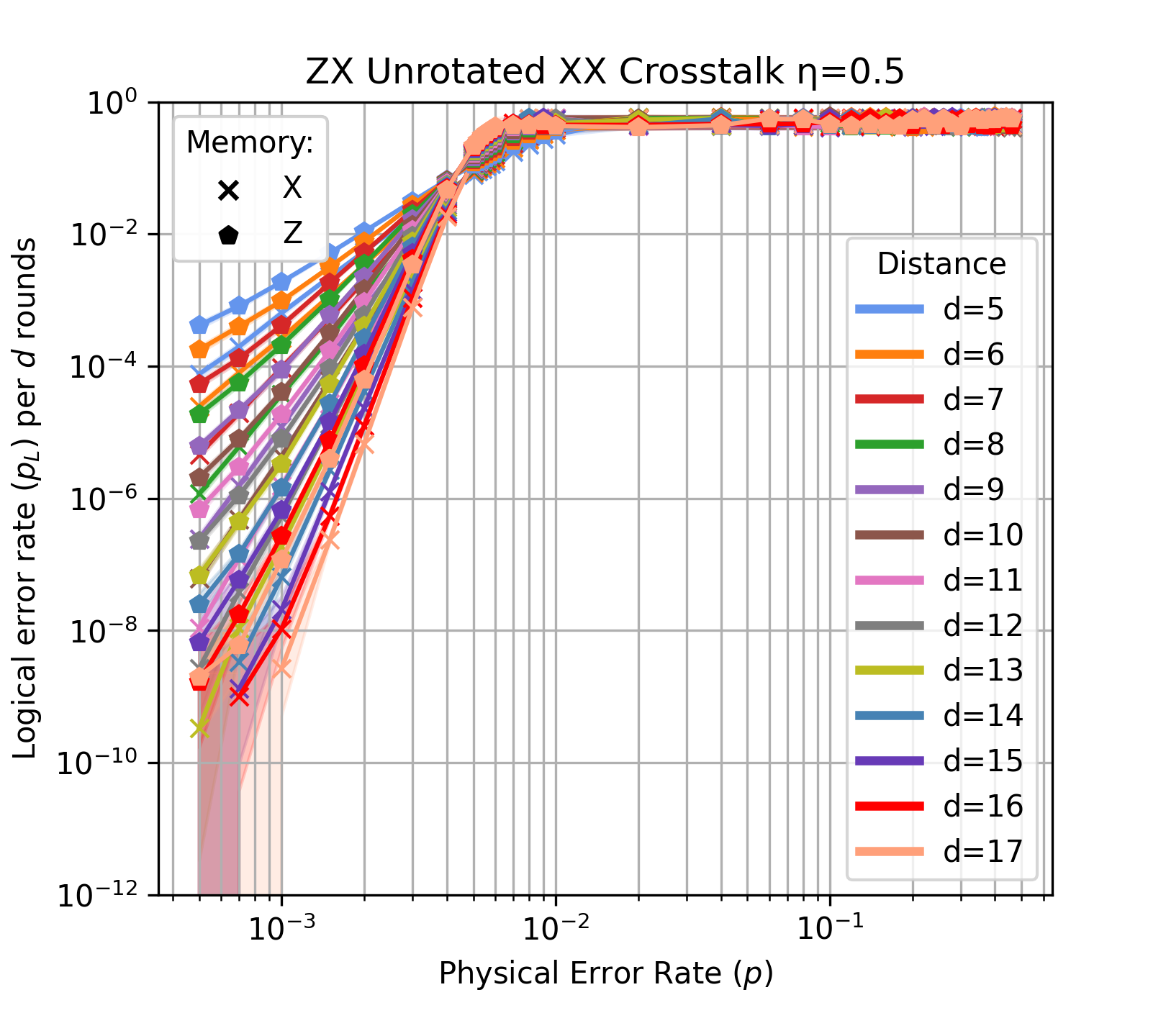}
        \caption{}
        \label{fig:zx_unrot_wc_a}
    \end{subfigure}
    \hfill
    \begin{subfigure}[t]{0.32\textwidth}
        \centering
        \includegraphics[width=\linewidth]{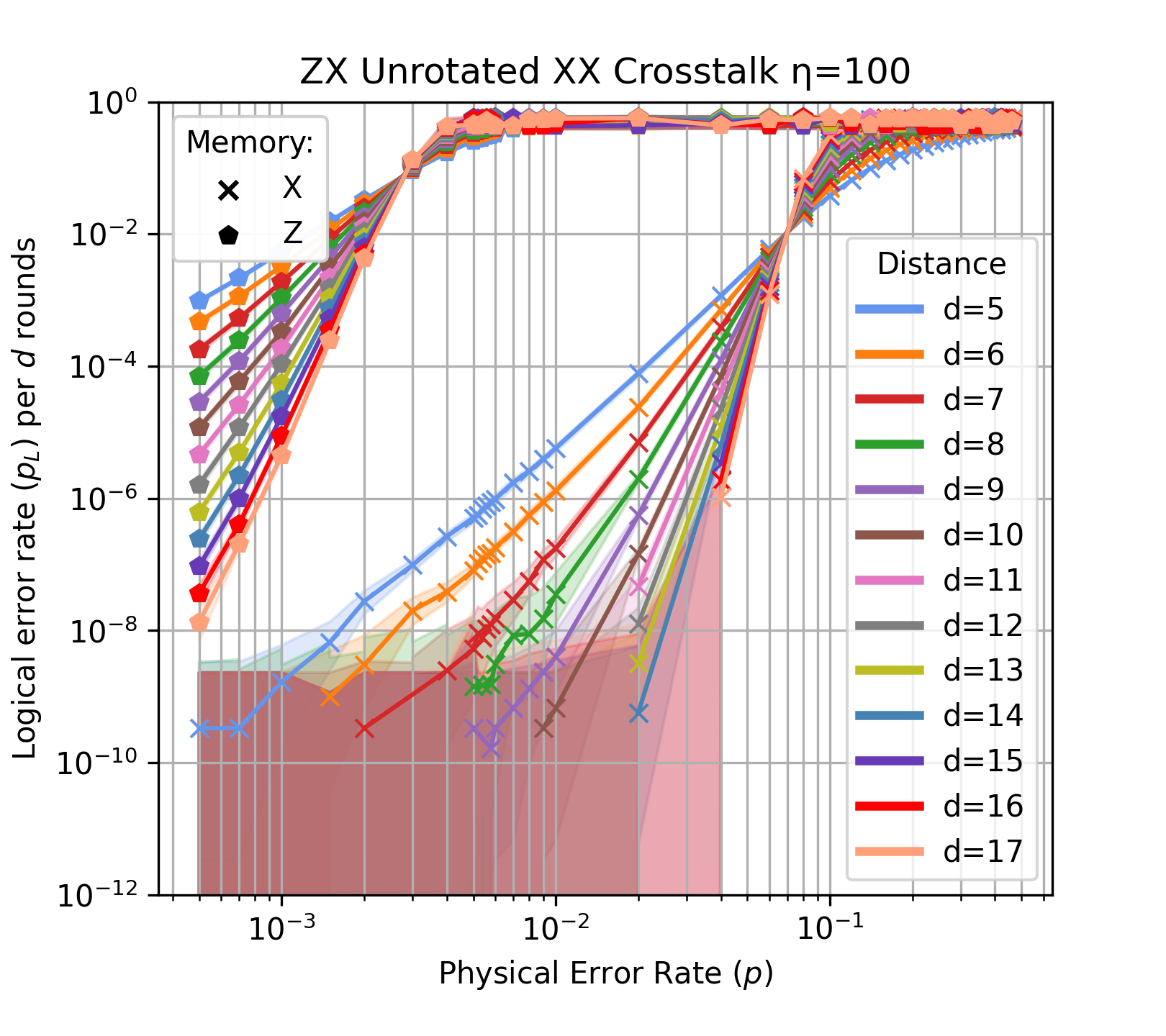}
        \caption{}
        \label{fig:zx_unrot_wc_b}
    \end{subfigure}
    \hfill
    \begin{subfigure}[t]{0.32\textwidth}
        \centering
        \includegraphics[width=\linewidth]{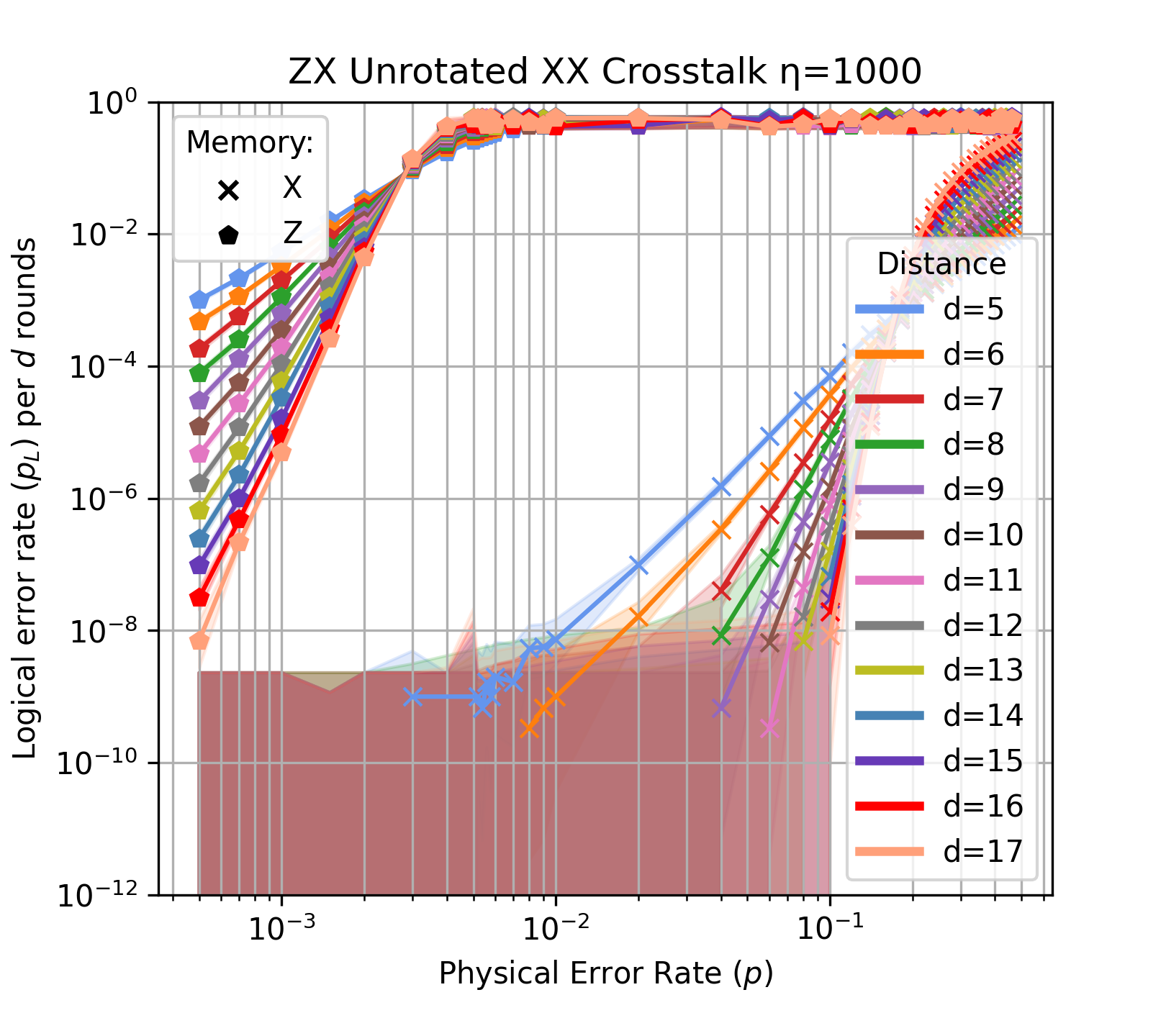}
        \caption{}
        \label{fig:zx_unrot_wc_c}
    \end{subfigure}

    \caption{Logical error rate per $d$ rounds of syndrome extraction $p_L$ versus physical error rate $p$ simulated on unrotated $ZX$ surface code with $X$-aligned CNOT order under Pauli-$X$ biased noise at $\eta = \{0.5, 100, 1000\}$ with an additional $XX$ crosstalk noise.}
    
    \label{fig:zx_unrot_wc_selected}
\end{figure}

In this section, we first present the results obtained by simulating the Pauli-$X$ biased noise on the rotated and unrotated variants of the standard $ZX$ surface code with $X$-aligned CNOT order (Table \ref{tab:zx_xbiased}, Figs.~\ref{fig:zx_rot_bi_selected}, \ref{fig:zx_unrot_bi_selected}). Then, we present the results obtained by simulating both Pauli-$X$ biased noise and gate-based $XX$ crosstalk noise on the same surface codes (Table \ref{tab:zx_crosstalk}, Figs.~\ref{fig:zx_rot_wc_selected}, \ref{fig:zx_unrot_wc_selected}).

\begin{table}[ht]
\centering
\caption{Threshold values for the $ZX$ surface code with $X$-aligned CNOT order, under Pauli-$X$ biased noise. Here, $\eta$ denotes bias parameter, $p_{\mathrm{th}}^X$ denotes $X$-memory threshold and $p_{\mathrm{th}}^Z$ denotes $Z$-memory threshold. The $Z$-memory thresholds carry a statistical fitting uncertainty $\sigma_{f} \approx 0.002 \times 10^{-2}$ at all $\eta$; the $X$-memory fit uncertainty increases with bias, from $\sigma_{f} \approx 0.002 \times 10^{-2}$ at low bias to $\sigma_{f} \approx 0.02 \times 10^{-2}$ for $\eta \geq 30$.}
\label{tab:zx_xbiased}
\small
\setlength{\tabcolsep}{12pt}
\begin{tabular}{c SS SS}
\toprule
& \multicolumn{2}{c}{Rotated Surface Code} 
& \multicolumn{2}{c}{Unrotated Surface Code} \\
\cmidrule(lr){2-3} \cmidrule(lr){4-5}
{$\eta$} &
{$p_{\mathrm{th}}^{X} (\times 10^{-2})$} &
{$p_{\mathrm{th}}^{Z} (\times 10^{-2})$} &
{$p_{\mathrm{th}}^{X} (\times 10^{-2})$} &
{$p_{\mathrm{th}}^{Z} (\times 10^{-2})$} \\
\midrule
0.5      & 0.428 & 0.428 & 0.462 & 0.458 \\
1        & 0.566 & 0.386 & 0.584 & 0.403 \\
3        & 0.990 & 0.321 & 1.005 & 0.349 \\
10       & 1.899 & 0.296 & 1.999 & 0.312 \\
30       & 3.738 & 0.284 & 3.839 & 0.302 \\
100      & 6.796 & 0.280 & 6.909 & 0.298 \\
300      & 11.108 & 0.278 & 11.432 & 0.296 \\
1000     & 16.365 & 0.277 & 17.148 & 0.295 \\
$\infty$ & {--}   & 0.276 & {--}   & 0.294 \\
\bottomrule
\end{tabular}
\end{table}

\begin{table}[!htbp]
\centering
\caption{Threshold values for the $ZX$ surface code with $X$-aligned CNOT order, under Pauli-$X$ biased noise and additional $XX$ crosstalk noise. Here, $\eta$ denotes bias parameter, $p_{\mathrm{th}}^X$ denotes $X$-memory threshold and $p_{\mathrm{th}}^Z$ denotes $Z$-memory threshold. The $Z$-memory thresholds carry a statistical fitting uncertainty $\sigma_{f} \approx 0.002 \times 10^{-2}$ at all $\eta$; the $X$-memory fit uncertainty increases with bias, from $\sigma_{f} \approx 0.002 \times 10^{-2}$ at low bias to $\sigma_{f} \approx 0.02 \times 10^{-2}$ for $\eta \geq 30$.}
\label{tab:zx_crosstalk}
\small
\setlength{\tabcolsep}{12pt}
\begin{tabular}{c SS SS}
\toprule
& \multicolumn{2}{c}{Rotated Surface Code} 
& \multicolumn{2}{c}{Unrotated Surface Code} \\
\cmidrule(lr){2-3} \cmidrule(lr){4-5}
{$\eta$} &
{$p_{\mathrm{th}}^{X} (\times 10^{-2})$} &
{$p_{\mathrm{th}}^{Z} (\times 10^{-2})$} &
{$p_{\mathrm{th}}^{X} (\times 10^{-2})$} &
{$p_{\mathrm{th}}^{Z} (\times 10^{-2})$} \\
\midrule
0.5      & 0.428  & 0.396  & 0.462  & 0.433 \\
1        & 0.566  & 0.373  & 0.584  & 0.385 \\
3        & 0.990  & 0.313  & 1.003  & 0.322 \\
10       & 1.899  & 0.284  & 1.999  & 0.290 \\
30       & 3.738  & 0.277  & 3.838  & 0.288 \\
100      & 6.795  & 0.277  & 6.907  & 0.287 \\
300      & 11.108 & 0.276  & 11.431 & 0.287 \\
1000     & 16.366 & 0.276  & 17.149 & 0.287 \\
$\infty$ & {--}   & 0.276  & {--}   & 0.287 \\
\bottomrule
\end{tabular}
\end{table}

For each bias parameter $\eta$, the logical error rates $p_L$ were plotted as a function of the physical error rates $p$ for both $X$- and $Z$-memory experiments. Thus, we have separate plots for each $\eta$. In the figures of this paper, we only show the plots for three $\eta$ values. The rest of the plots can be found in the github repository link provided in section~\ref{sec:code}. As expected, we clearly see the curves for different distances of a given memory experiment cross at a common point, which we call the threshold point. Below the threshold point, increasing the distance $d$ of the surface code suppresses the logical error rate $p_L$. 

For the standard depolarizing case with $\eta=0.5$, the $X$ memory threshold $p_{\mathrm{th}}^X$ and $Z$ memory threshold $p_{\mathrm{th}}^Z$ are equal to within the reported uncertainty for the rotated surface code (Table~\ref{tab:zx_xbiased}), while there is a difference of $0.004 \times 10^{-2}$ between the $X$ and $Z$ memory thresholds for the unrotated surface code, which is comparable to the combined fitting uncertainty. This separation is seen more clearly in the full $p_L$ curves of Fig.~\ref{fig:zx_unrot_bi_a}, compared to the curves of Fig.~\ref{fig:4a}.

Under Pauli-$X$- biased noise (Table~\ref{tab:zx_xbiased}), the $X$-memory threshold $p_{\mathrm{th}}^{X}$ increases monotonically with the bias parameter $\eta$ for both rotated and unrotated surface codes (Figs.~\ref{fig:zx_rot_bi_selected}, \ref{fig:zx_unrot_bi_selected}). In the standard depolarizing case ($\eta=0.5$), the threshold value $p_{\mathrm{th}}^{X}$ is approximately $0.428 \times 10^{-2}$ for the rotated code and $0.462 \times 10^{-2}$ for the unrotated code. It increases steadily with increasing bias and at $\eta=1000$, it reaches approximately $16.4 \times 10^{-2}$ for the rotated code and $17.1 \times 10^{-2}$ for the unrotated code. Eventually, the threshold point flows out of the observed window $(p \leq 0.46)$ for pure $X$-biased noise ($\eta=\infty$), so no $X$-memory threshold is reported for $\eta = \infty$. The results in Table~\ref{tab:zx_xbiased} show that the unrotated surface code exhibits a numerically higher $X$-memory threshold $p_{\mathrm{th}}^{X}$ than the rotated surface code at every bias value. This difference exceeds the combined fitting uncertainty for all $\eta$ except $\eta=3$, where the separation ($0.015 \times 10^{-2}$) is comparable to the combined uncertainty and is therefore not resolved; from $\eta=10$ onward the gap is clearly resolved and widens as the bias increases.

In contrast, under the same noise model, the $Z$-memory threshold $p_{\mathrm{th}}^{Z}$ decreases as $\eta$ increases and rapidly approaches a saturation value. In the rotated surface code, $p_{\mathrm{th}}^{Z}$ saturates near $0.276 \times 10^{-2} $, and in the unrotated surface code, it saturates near $0.294\times 10^{-2}$. Across all bias parameters considered, the unrotated geometry shows a slightly higher $p_{\mathrm{th}}^{Z}$ compared to the rotated geometry for the Pauli-$X$ biased noise.

When gate-based $XX$ crosstalk noise is included (Table~\ref{tab:zx_crosstalk}), the qualitative trend with $\eta$ remains unchanged. As the bias parameter $\eta$ increases, the $X$-memory threshold $p_{\mathrm{th}}^{X}$ continues to increase monotonically, and eventually flows out of the observed window for $\eta = \infty$. But the $Z$-memory threshold $p_{\mathrm{th}}^{Z}$ decreases and saturates for higher $\eta$. The effect of adding crosstalk noise on $X$-memory threshold falls within the combined fitting uncertainty across all $\eta$, for both rotated and unrotated surface codes. However, the addition of crosstalk noise causes a decrease in the $Z$-memory thresholds $p_{\mathrm{th}}^{Z}$ for this noise model compared to the only $X$-biased noise case. For the unrotated surface code, this reduction is significant across all bias values as the saturation point for $p_{\mathrm{th}}^{Z}$ shifts from $0.294 \times 10^{-2}$ under $X$-biased noise to $0.287\times 10^{-2}$ under the additional crosstalk noise. For the rotated surface code, the reduction in $p_{\mathrm{th}}^{Z}$ is significant at low-to-moderate bias ($\eta \lesssim 30$), but the saturation point at high $\eta$ converges to $0.276 \times 10^{-2}$ in both noise models, which is within the combined fitting uncertainty.

An additional remark that we want to make here is that, although the error model and simulation pipeline are exactly the same for both rotated and unrotated surface codes, there is a visible separation between the performance of memory $X$ and $Z$ even for the standard depolarizing case $(\eta = 0.5)$ for the unrotated surface code with $X$-biased noise (Fig.~\ref{fig:zx_unrot_bi_a}), while there is no separation in memory $X$ and $Z$ performance in the rotated surface code (Fig.~\ref{fig:zx_unrot_bi_a}). Though the separation is not seen in the rotated surface code under $X$-biased noise, the separation in performance of $X$ memory and $Z$ memory is visible in the rotated surface code under crosstalk noise. We can see the separate memory $X$ and $Z$ curves in the plots in Fig.~\ref{fig:zx_unrot_bi_a} and Fig.~\ref{fig:zx_rot_wc_a}, but the separation can be seen even more prominently for the unrotated surface code under crosstalk noise (Fig.~\ref{fig:zx_unrot_wc_a}), where the gap between $p_{\mathrm{th}}^{X}$ and $p_{\mathrm{th}}^{Z}$ at $\eta =0.5$ is nearly an order of magnitude (a factor of $\sim7$) larger than under $X$-biased noise alone.

\subsection{ZX Surface Code with Z-aligned CNOT Order}
\label{sec:zx_z-align_results}

\begin{figure}[!htbp]
    \centering
    
    \begin{subfigure}[t]{0.32\textwidth}
        \centering
        \includegraphics[width=\linewidth]{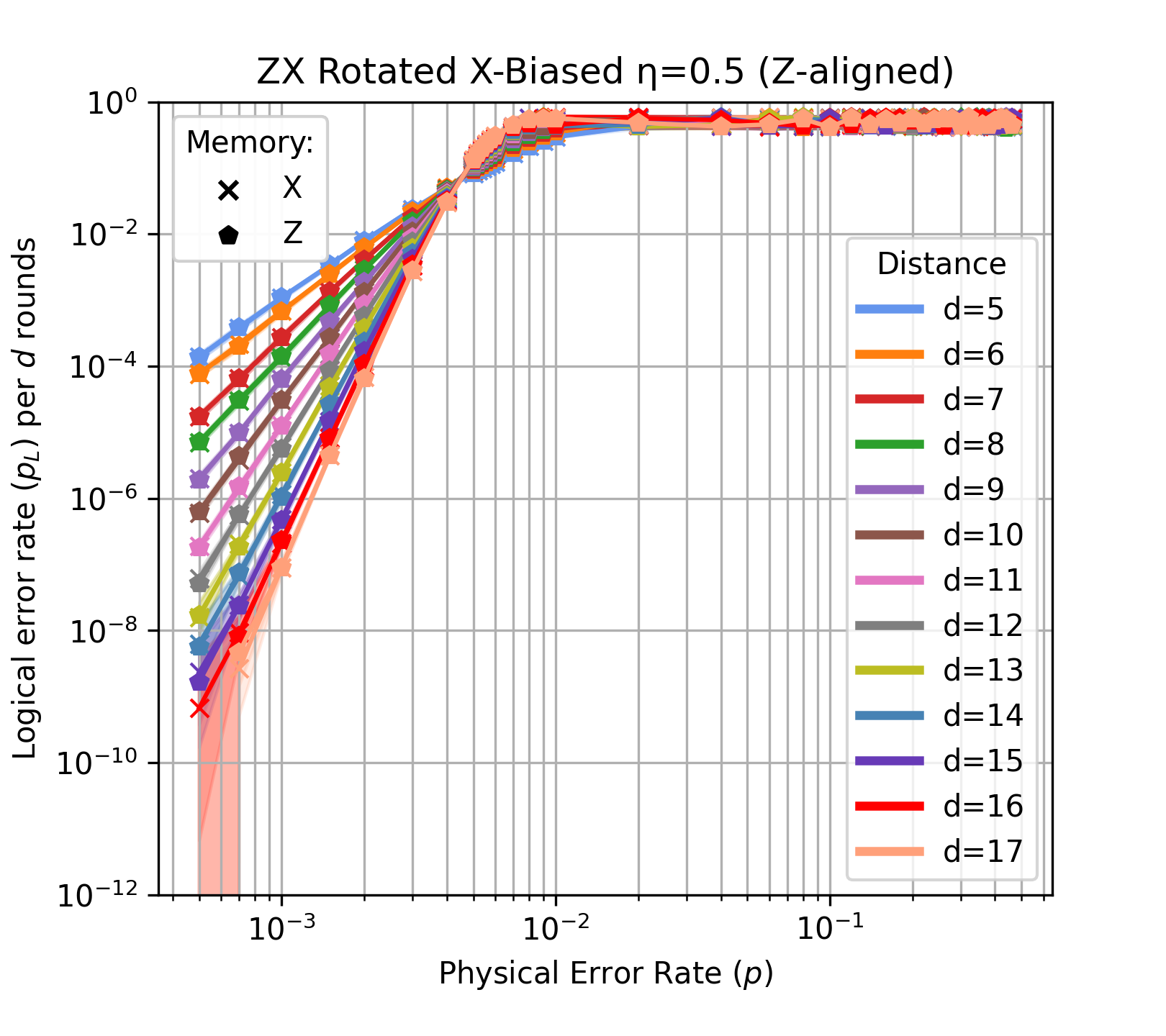}
        \caption{}
        \label{fig:zx_zl_rot_bi_a}
    \end{subfigure}
    \hfill
    \begin{subfigure}[t]{0.32\textwidth}
        \centering
        \includegraphics[width=\linewidth]{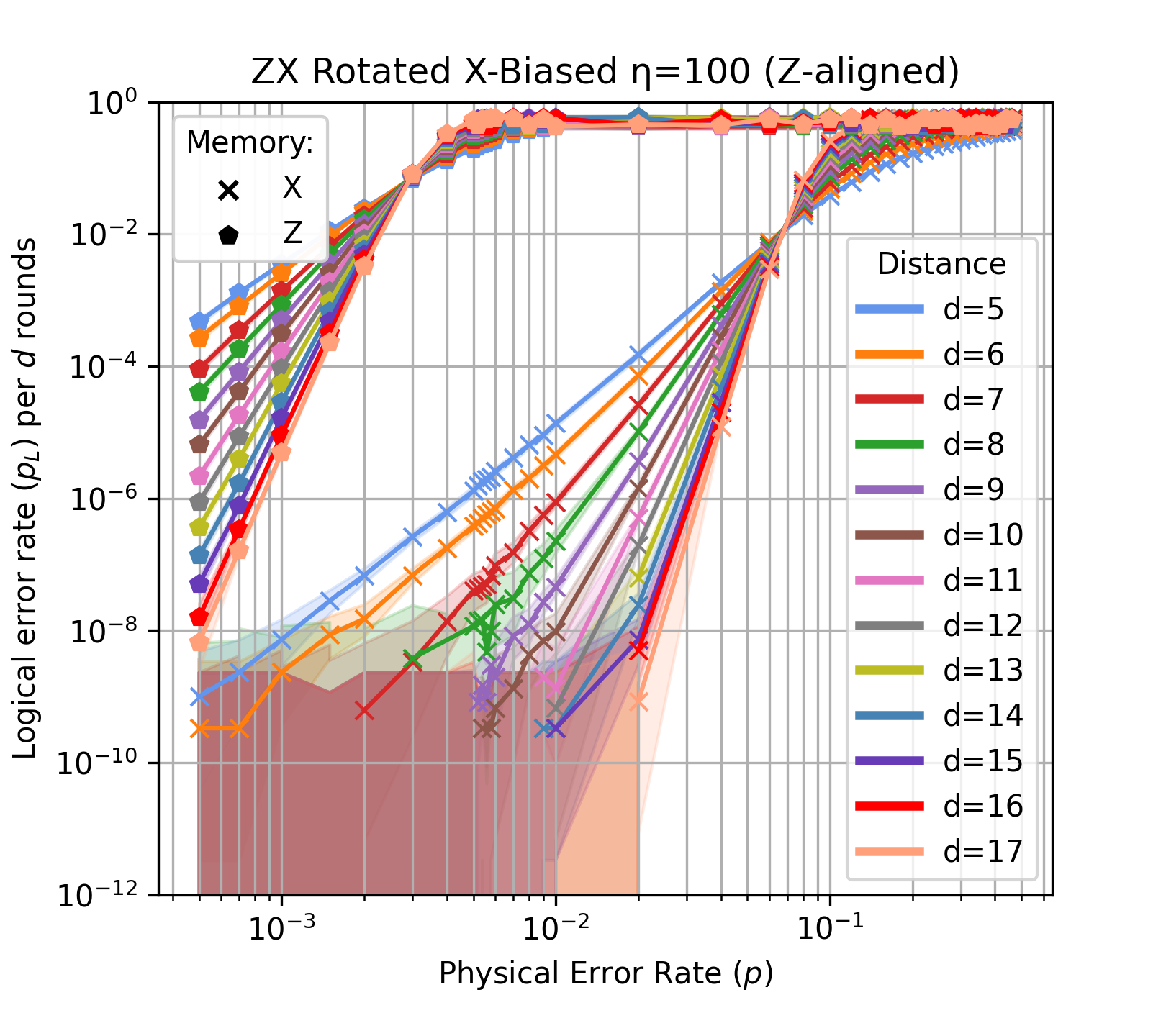}
        \caption{}
        \label{fig:zx_zl_rot_bi_b}
    \end{subfigure}
    \hfill
    \begin{subfigure}[t]{0.32\textwidth}
        \centering
        \includegraphics[width=\linewidth]{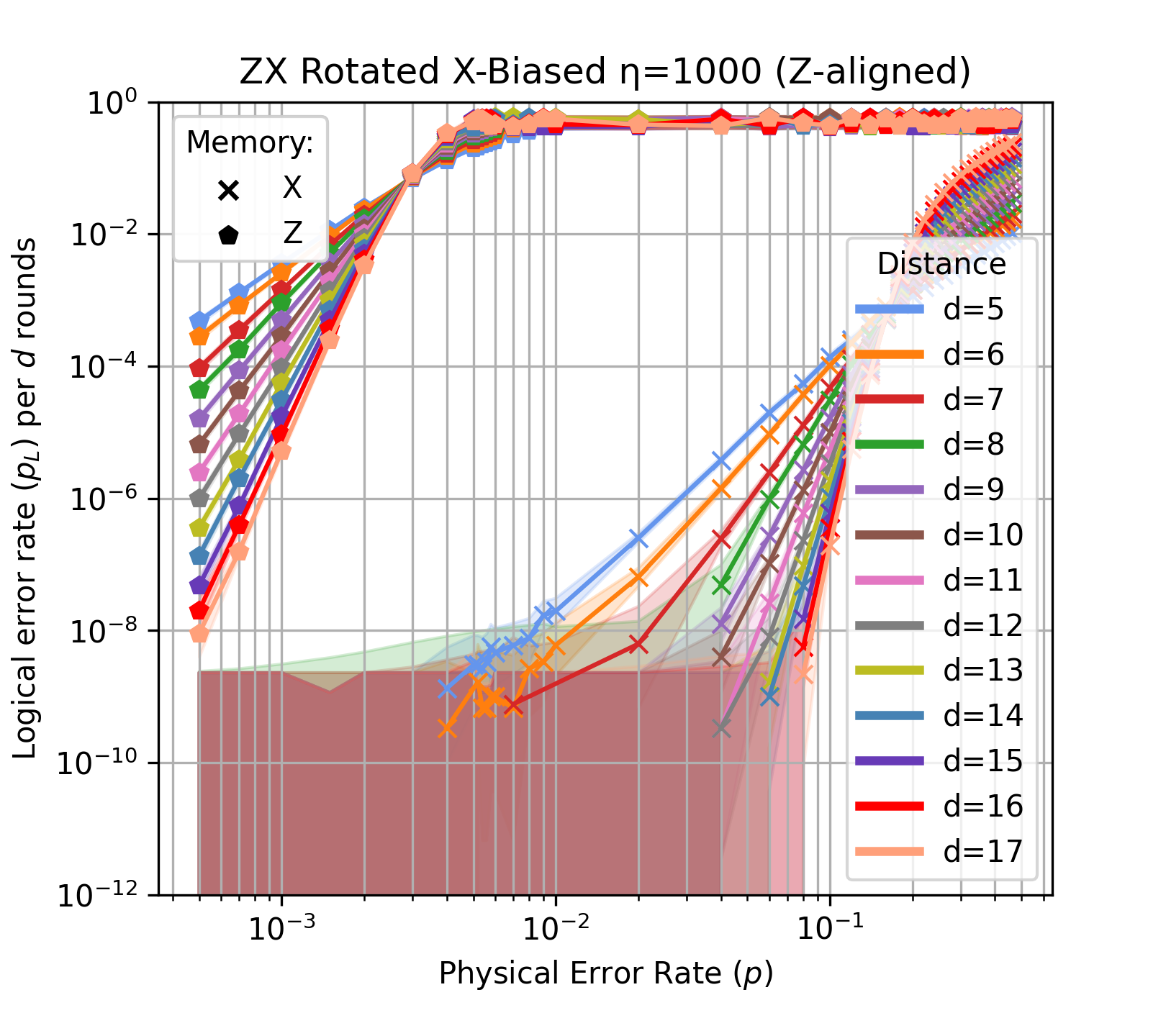}
        \caption{}
        \label{fig:zx_zl_rot_bi_c}
    \end{subfigure}

    \caption{Logical error rate per $d$ rounds of syndrome extraction $p_L$ versus physical error rate $p$ simulated on rotated $ZX$ surface code with $Z$-aligned CNOT order under Pauli-$X$ biased noise at $\eta = \{0.5, 100, 1000\}$.}

    \label{fig:zx_zl_rot_bi}
\end{figure}

\begin{figure}[!htbp]
    \centering
    
    \begin{subfigure}[t]{0.32\textwidth}
        \centering
        \includegraphics[width=\linewidth]{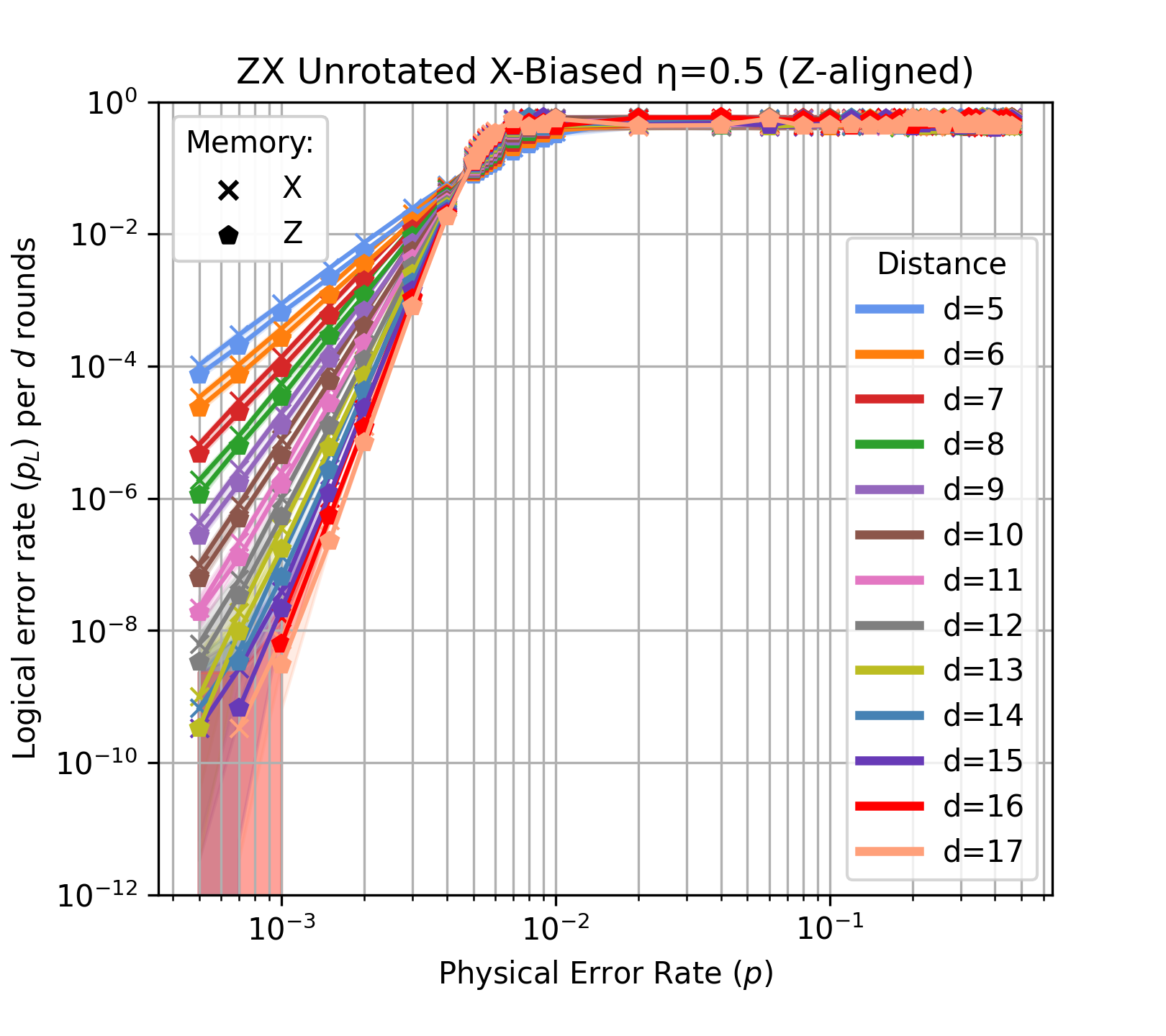}
        \caption{}
        \label{fig:zx_zl_unrot_bi_a}
    \end{subfigure}
    \hfill
    \begin{subfigure}[t]{0.32\textwidth}
        \centering
        \includegraphics[width=\linewidth]{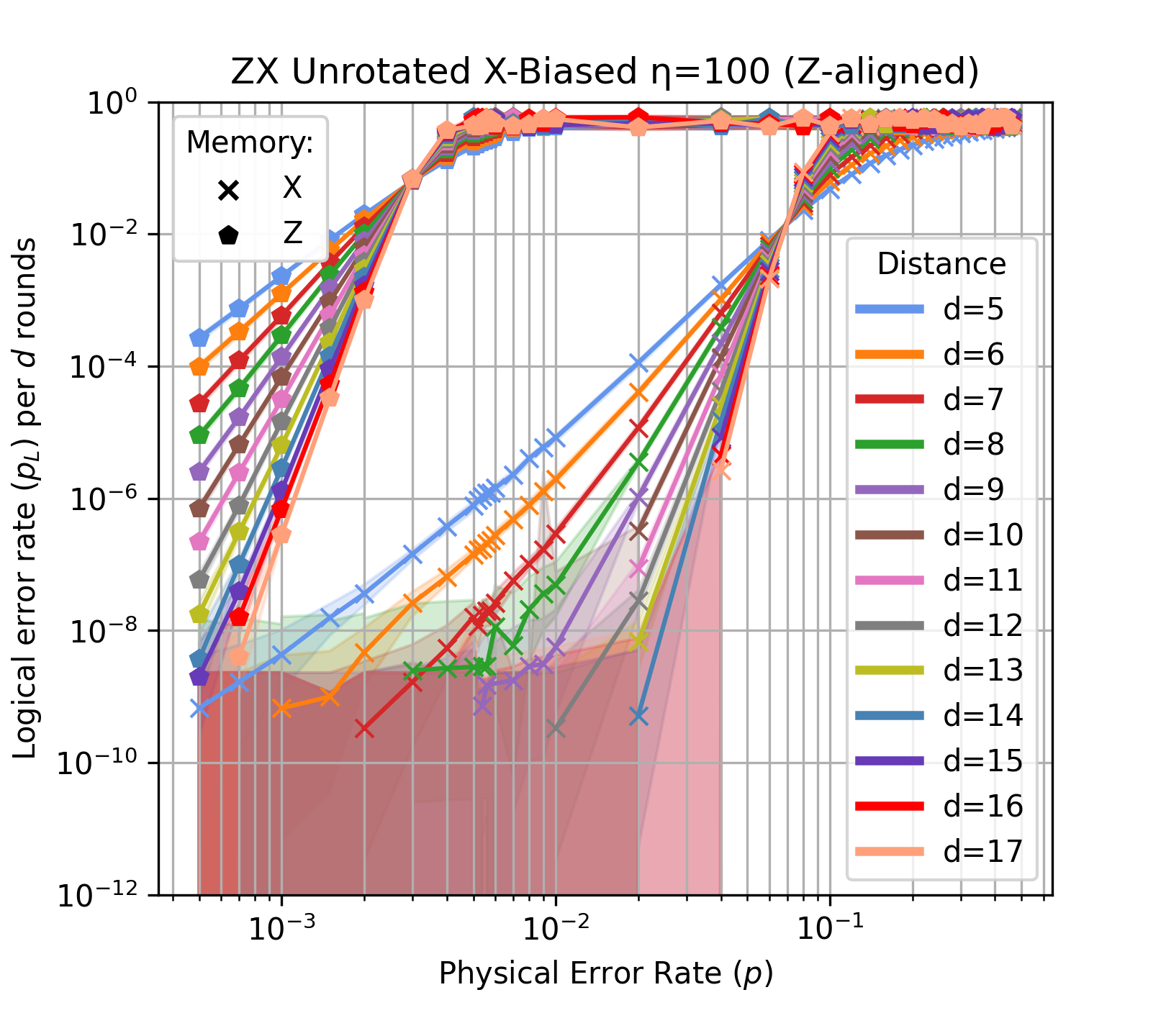}
        \caption{}
        \label{fig:zx_zl_unrot_bi_b}
    \end{subfigure}
    \hfill
    \begin{subfigure}[t]{0.32\textwidth}
        \centering
        \includegraphics[width=\linewidth]{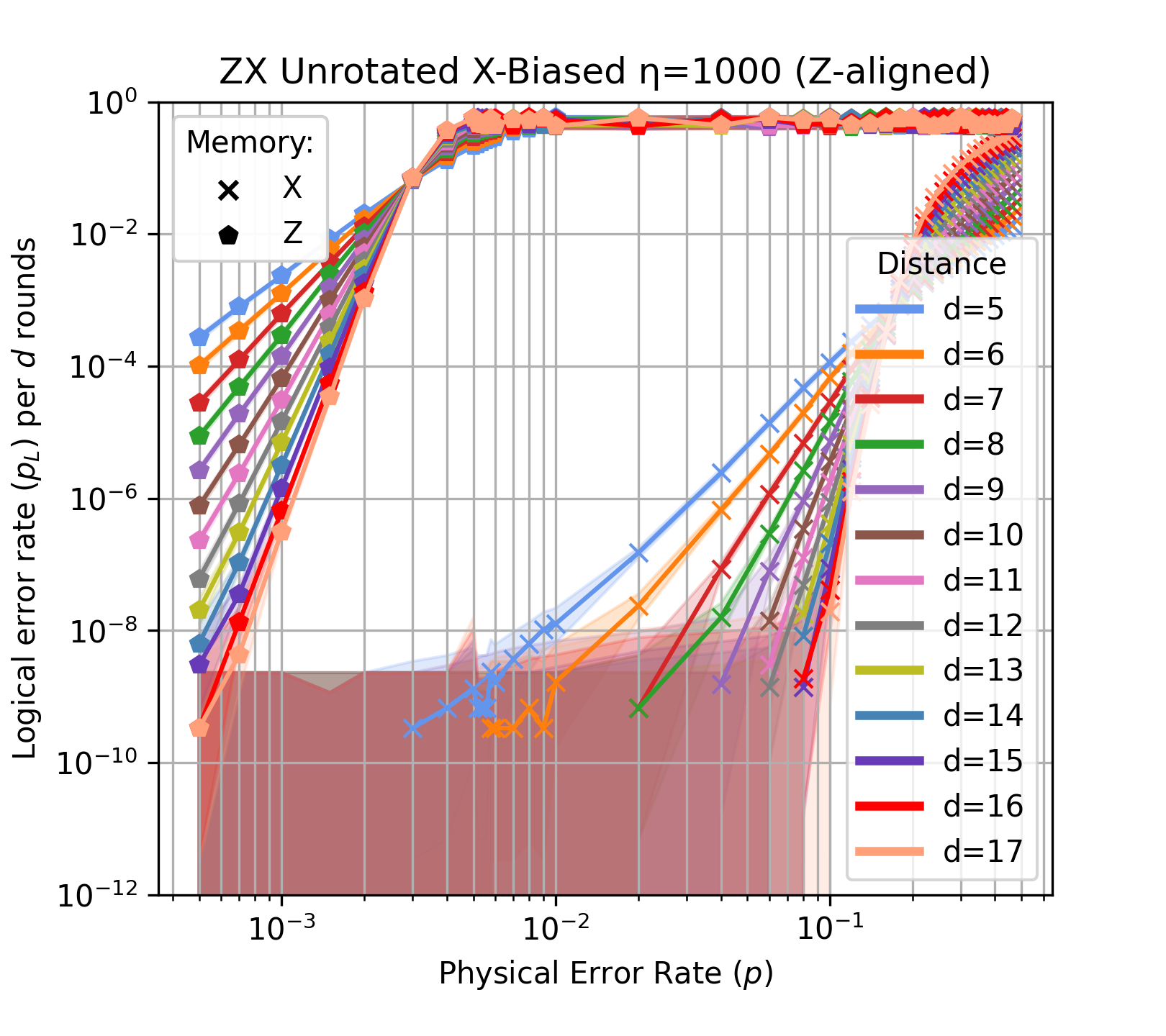}
        \caption{}
        \label{fig:zx_zl_unrot_bi_c}
    \end{subfigure}

    \caption{Logical error rate per $d$ rounds of syndrome extraction $p_L$ versus physical error rate $p$ simulated on unrotated $ZX$ surface code with $Z$-aligned CNOT order under Pauli-$X$ biased noise at $\eta = \{0.5, 100, 1000\}$.} 
    
    \label{fig:zx_zl_unrot_bi}
\end{figure}

\begin{figure}[ht]
    \centering
    
    \begin{subfigure}[t]{0.32\textwidth}
        \centering
        \includegraphics[width=\linewidth]{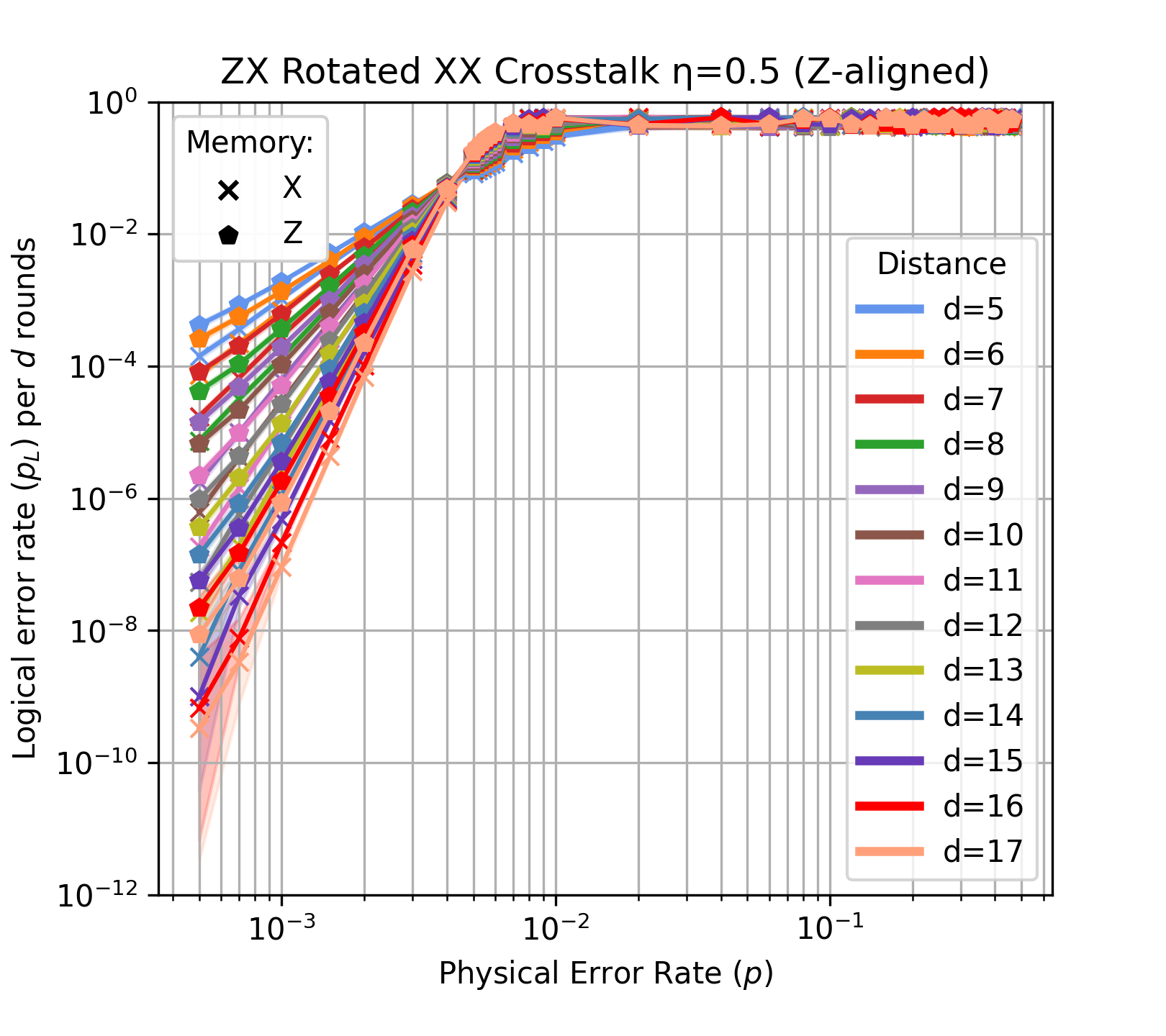}
        \caption{}
        \label{fig:zx_zl_rot_wc_a}
    \end{subfigure}
    \hfill
    \begin{subfigure}[t]{0.32\textwidth}
        \centering
        \includegraphics[width=\linewidth]{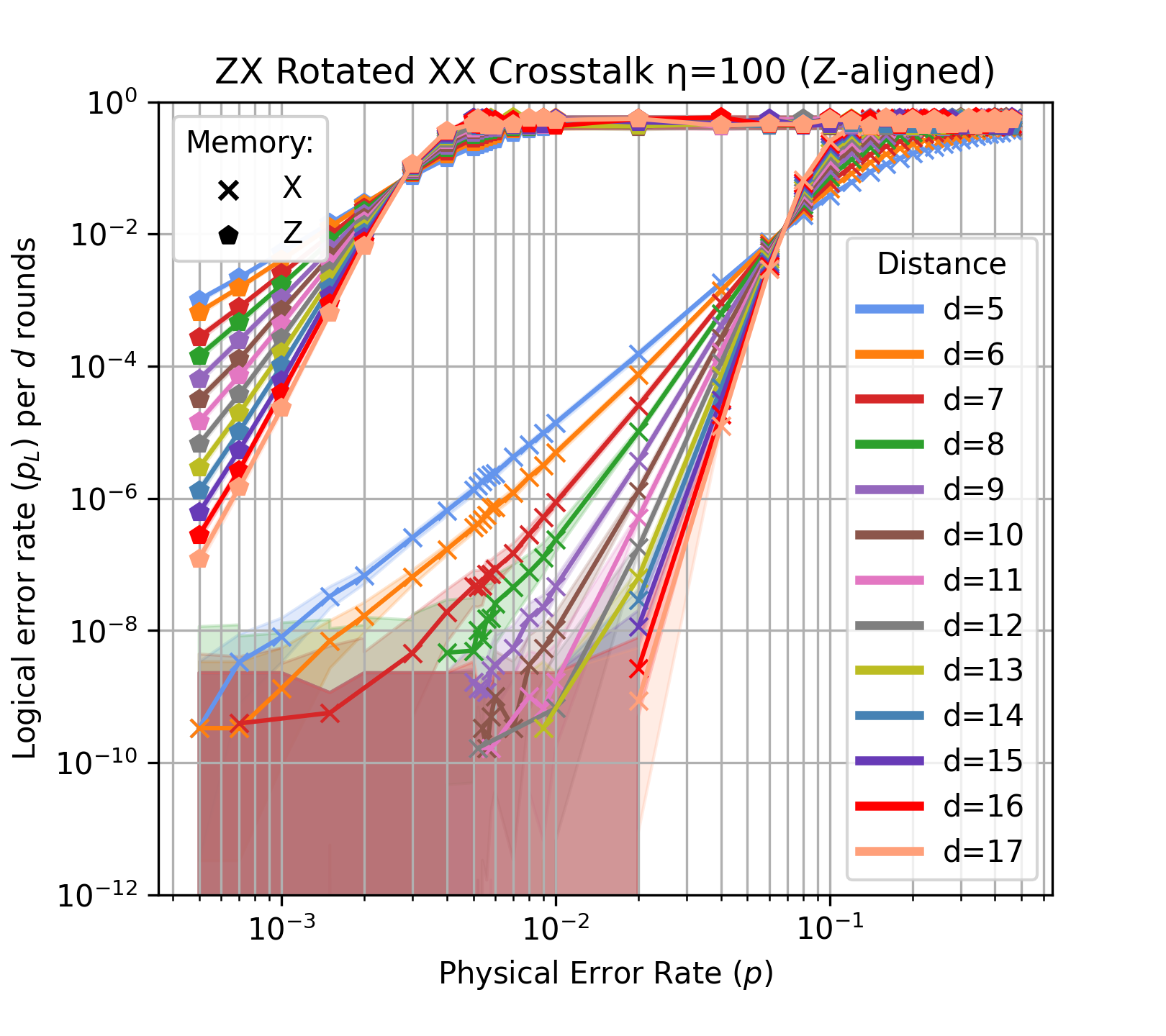}
        \caption{}
        \label{fig:zx_zl_rot_wc_b}
    \end{subfigure}
    \hfill
    \begin{subfigure}[t]{0.32\textwidth}
        \centering
        \includegraphics[width=\linewidth]{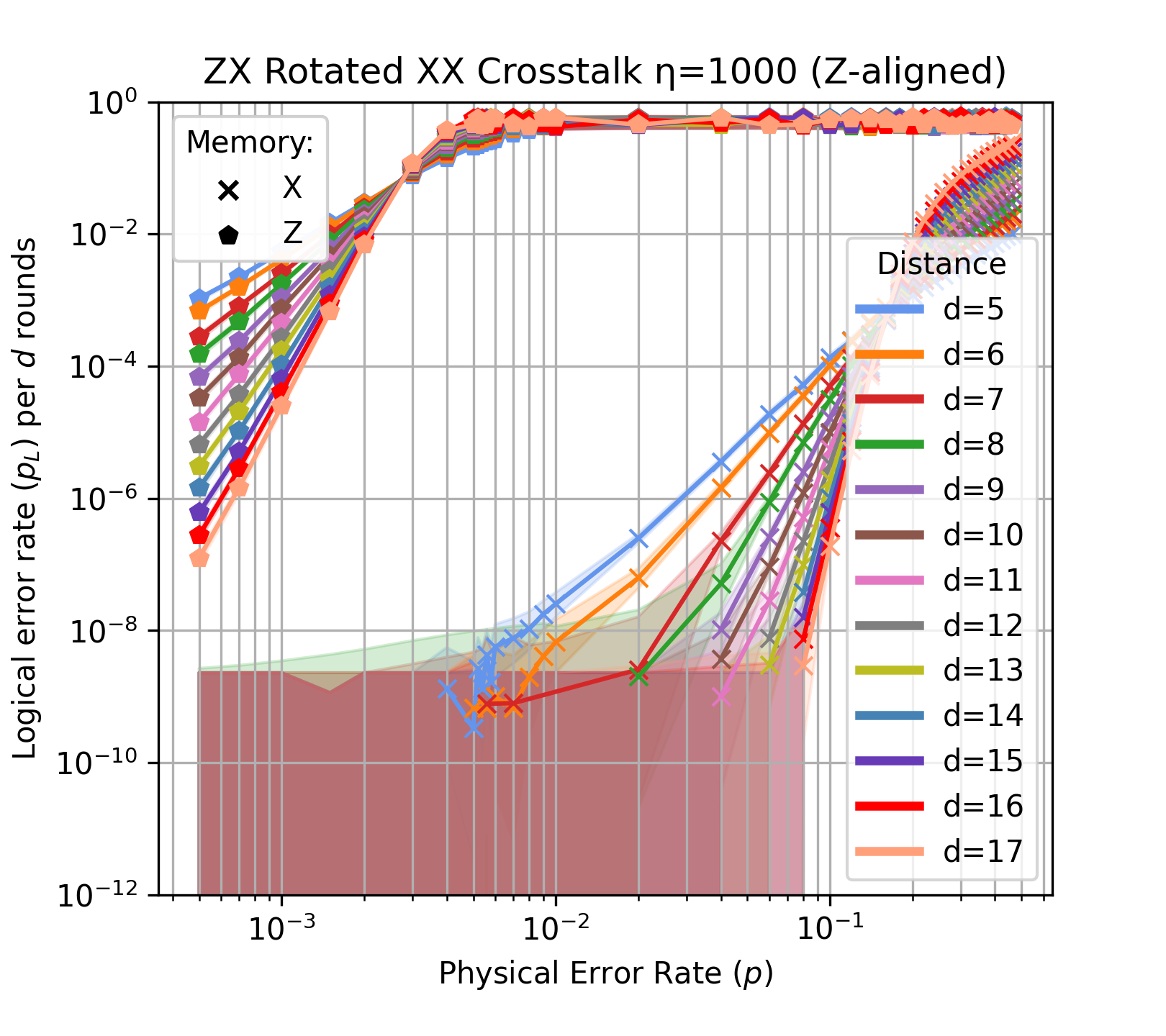}
        \caption{}
        \label{fig:zx_zl_rot_wc_c}
    \end{subfigure}

    \caption{Logical error rate per $d$ rounds of syndrome extraction $p_L$ versus physical error rate $p$ simulated on rotated $ZX$ surface code with $Z$-aligned CNOT order under Pauli-$X$ biased noise at $\eta = \{0.5, 100, 1000\}$ with an additional $XX$ crosstalk noise.}
    
    \label{fig:zx_zl_rot_wc}
\end{figure}

\begin{figure}[ht]
    \centering
    
    \begin{subfigure}[t]{0.32\textwidth}
        \centering
        \includegraphics[width=\linewidth]{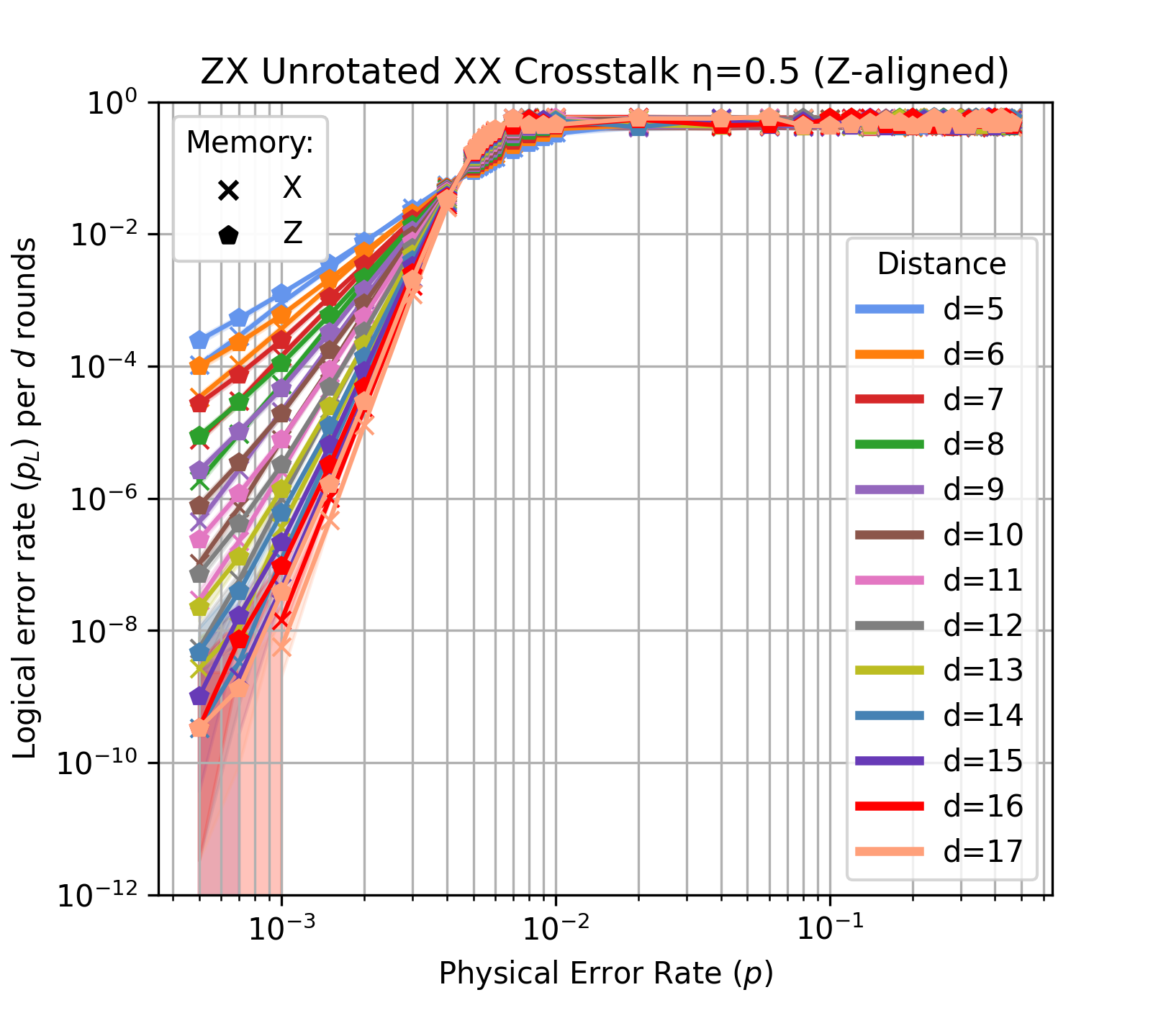}
        \caption{}
        \label{fig:zx_zl_unrot_wc_a}
    \end{subfigure}
    \hfill
    \begin{subfigure}[t]{0.32\textwidth}
        \centering
        \includegraphics[width=\linewidth]{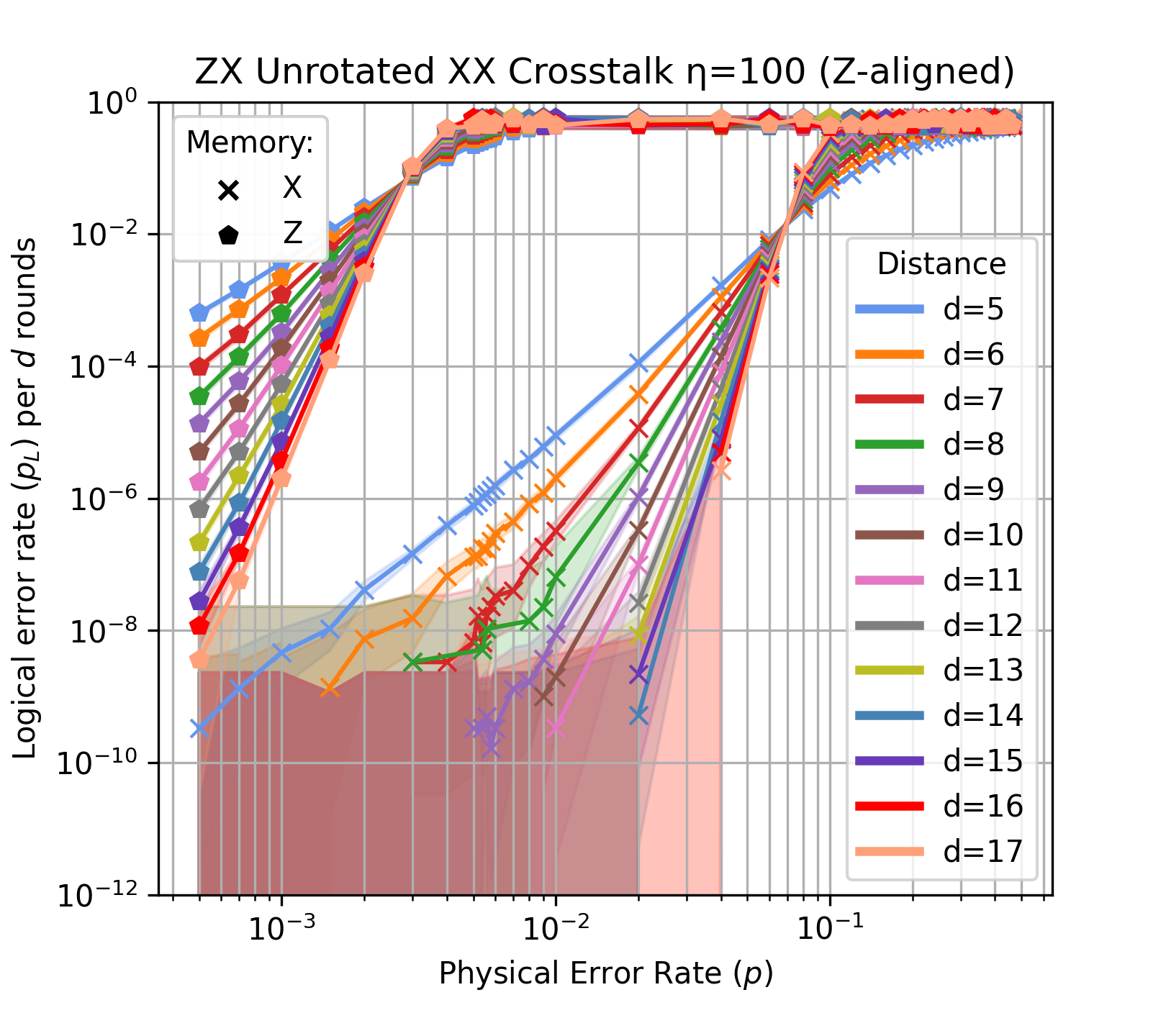}
        \caption{}
        \label{fig:zx_zl_unrot_wc_b}
    \end{subfigure}
    \hfill
    \begin{subfigure}[t]{0.32\textwidth}
        \centering
        \includegraphics[width=\linewidth]{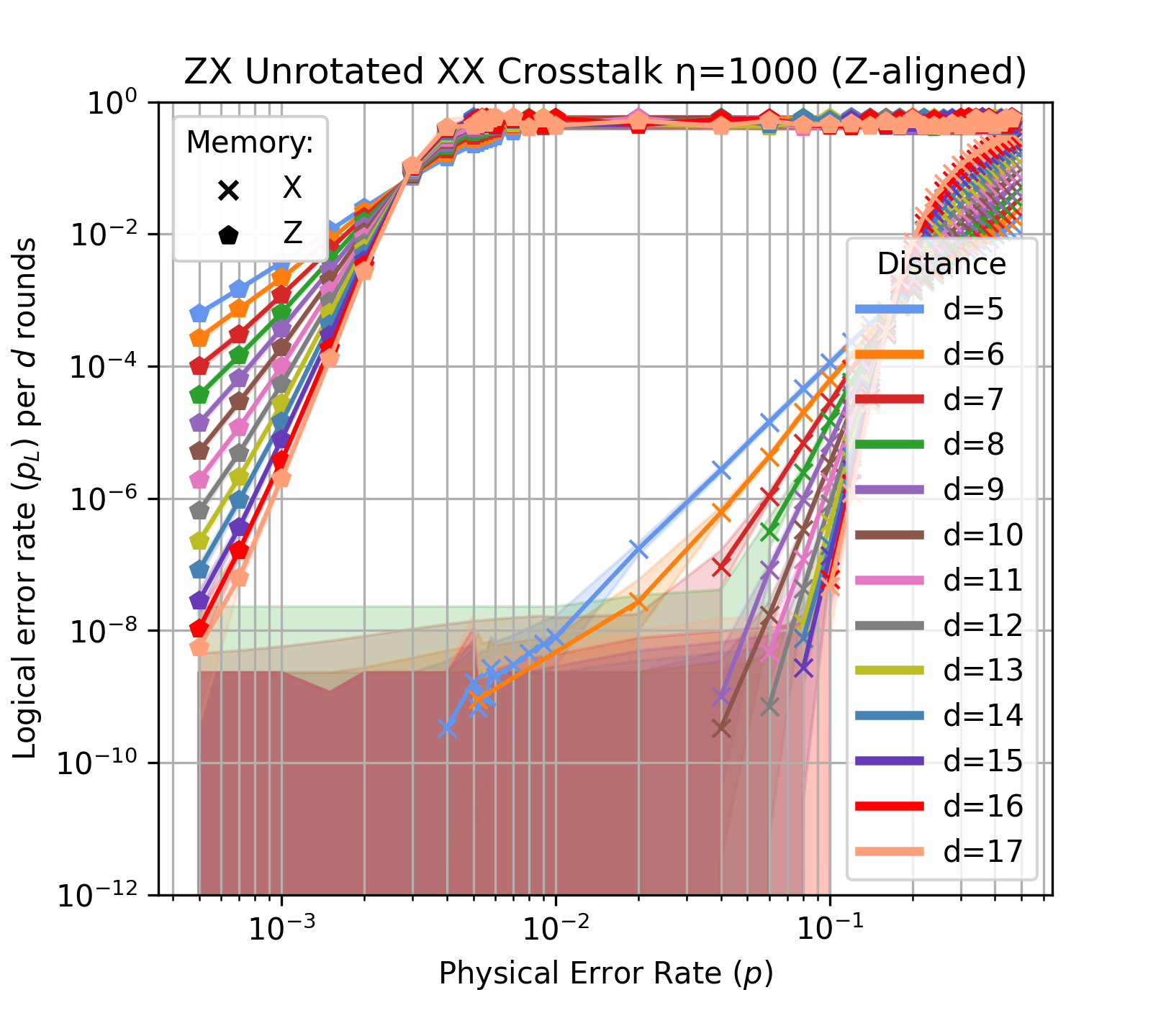}
        \caption{}
        \label{fig:zx_zl_unrot_wc_c}
    \end{subfigure}

    \caption{Logical error rate per $d$ rounds of syndrome extraction $p_L$ versus physical error rate $p$ simulated on unrotated $ZX$ surface code with $Z$-aligned CNOT order under Pauli-$X$ biased noise at $\eta = \{0.5, 100, 1000\}$ with an additional $XX$ crosstalk noise.}
    
    \label{fig:zx_zl_unrot_wc}
\end{figure}

In this section, we first present the results obtained by simulating Pauli $X$-biased noise on the rotated and unrotated variants of the $ZX$ surface code with $Z$-aligned CNOT order (Table \ref{tab:zx_biased_zalign}, Figs.~\ref{fig:zx_zl_rot_bi}, \ref{fig:zx_zl_unrot_bi}). Then, we present the result obtained by simulating the additional gate-based $XX$ crosstalk noise on the same surface codes (Table \ref{tab:zx_crosstalk_zalign}, Figs.~\ref{fig:zx_zl_rot_wc}, \ref{fig:zx_zl_unrot_wc}). As discussed earlier, for each bias parameter $\eta$, the logical error rates $p_L$ are plotted as a function of the physical error rates $p$ for both $X$ and $Z$ memory experiments, resulting in separate plots for each $\eta$. A clear threshold is observed for both memory experiments at each $\eta$. 

\begin{table}[!htbp]
\centering
\caption{Threshold values $p_{\mathrm{th}}$ for the $ZX$ surface code with $Z$-aligned CNOT order, under Pauli-$X$ biased noise. Here, $\eta$ denotes bias parameter, $p_{\mathrm{th}}^X$ denotes $X$-memory threshold and $p_{\mathrm{th}}^Z$ denotes $Z$-memory threshold. The $Z$-memory thresholds carry a statistical fitting uncertainty $\sigma_{f} \approx 0.002 \times 10^{-2}$ at all $\eta$; the $X$-memory fit uncertainty increases with bias, from $\sigma_{f} \approx 0.002 \times 10^{-2}$ at low bias to $\sigma_{f} \approx 0.02 \times 10^{-2}$ for $\eta \geq 30$.}
\label{tab:zx_biased_zalign}
\small
\setlength{\tabcolsep}{12pt}
\begin{tabular}{c SS SS}
\toprule
& \multicolumn{2}{c}{Rotated Surface Code} 
& \multicolumn{2}{c}{Unrotated Surface Code} \\
\cmidrule(lr){2-3} \cmidrule(lr){4-5}
{$\eta$} &
{$p_{\mathrm{th}}^{X} (\times 10^{-2})$} &
{$p_{\mathrm{th}}^{Z} (\times 10^{-2})$} &
{$p_{\mathrm{th}}^{X} (\times 10^{-2})$} &
{$p_{\mathrm{th}}^{Z} (\times 10^{-2})$} \\
\midrule
0.5      & 0.428  & 0.428 & 0.461  & 0.469 \\
1        & 0.561  & 0.386 & 0.584  & 0.41  \\
3        & 0.99   & 0.334 & 1.004  & 0.356 \\
10       & 1.888  & 0.307 & 1.998  & 0.322 \\
30       & 3.733  & 0.287 & 3.838  & 0.308 \\
100      & 6.789  & 0.28  & 6.841  & 0.301 \\
300      & 11.103 & 0.278 & 11.165 & 0.299 \\
1000     & 16.357 & 0.277 & 16.881 & 0.299 \\
$\infty$ & {--}   & 0.276 & {--}   & 0.299 \\
\bottomrule
\end{tabular}
\end{table}

\begin{table}[!htbp]
\centering
\caption{Threshold values $p_{\mathrm{th}}$ for the $ZX$ surface code with $Z$-aligned CNOT order, under Pauli-$X$ biased noise and additional $XX$ crosstalk noise. Here, $\eta$ denotes bias parameter, $p_{\mathrm{th}}^X$ denotes $X$-memory threshold and $p_{\mathrm{th}}^Z$ denotes $Z$-memory threshold. The $Z$-memory thresholds carry a statistical fitting uncertainty $\sigma_{f} \approx 0.002 \times 10^{-2}$ at all $\eta$; the $X$-memory fit uncertainty increases with bias, from $\sigma_{f} \approx 0.002 \times 10^{-2}$ at low bias to $\sigma_{f} \approx 0.02 \times 10^{-2}$ for $\eta \geq 30$.}
\label{tab:zx_crosstalk_zalign}
\small
\setlength{\tabcolsep}{12pt}
\begin{tabular}{c SS SS}
\toprule
& \multicolumn{2}{c}{Rotated Surface Code} 
& \multicolumn{2}{c}{Unrotated Surface Code} \\
\cmidrule(lr){2-3} \cmidrule(lr){4-5}
{$\eta$} &
{$p_{\mathrm{th}}^{X} (\times 10^{-2})$} &
{$p_{\mathrm{th}}^{Z} (\times 10^{-2})$} &
{$p_{\mathrm{th}}^{X} (\times 10^{-2})$} &
{$p_{\mathrm{th}}^{Z} (\times 10^{-2})$} \\
\midrule
0.5      & 0.428  & 0.396 & 0.432  & 0.441 \\
1        & 0.565  & 0.373 & 0.582  & 0.389 \\
3        & 0.990  & 0.312 & 1.001  & 0.326 \\
10       & 1.899  & 0.284 & 1.970  & 0.293 \\
30       & 3.738  & 0.277 & 3.828  & 0.289 \\
100      & 6.795  & 0.276 & 6.812  & 0.288 \\
300      & 11.104 & 0.276 & 11.105 & 0.288 \\
1000     & 16.362 & 0.276 & 16.871 & 0.288 \\
$\infty$ & {--}   & 0.276 & {--}   & 0.288 \\

\bottomrule
\end{tabular}
\end{table}

Similar to the $ZX$ surface code with $X$-aligned CNOT order (section \ref{sec:zx_x-align_result}), under Pauli $X$-biased noise (Table \ref{tab:zx_biased_zalign}), the $X$-memory threshold $p_{\mathrm{th}}^{X}$ for the $ZX$ surface code with $Z$-aligned CNOT order increases monotonically as $\eta$ increases for both rotated and unrotated surface codes. Also, following the same qualitative trend as section \ref{sec:zx_x-align_result}, the $Z$-memory threshold $p_{\mathrm{th}}^{Z}$ decreases as $\eta$ increases until it reaches a saturation point.

For the unrotated surface code, $p_{\mathrm{th}}^{Z}$ under $Z$-aligned CNOT order (Table~\ref{tab:zx_biased_zalign}) is higher than the $p_{\mathrm{th}}^{Z}$ under $X$-aligned CNOT order (Table~\ref{tab:zx_xbiased}) at all bias values. The saturation point of $p_{\mathrm{th}}^{Z}$ for the unrotated surface code increases from $0.294 \times 10^{-2}$ in the $X$-aligned schedule (Table~\ref{tab:zx_xbiased}) to $0.299 \times 10^{-2}$ in the $Z$-aligned schedule (Table~\ref{tab:zx_biased_zalign}). For the rotated surface code, the increase in $p_{\mathrm{th}}^{Z}$ under the $Z$-aligned CNOT order relative to the $X$-aligned order exceeds the combined fitting uncertainty only at moderate bias ($\eta \approx 3$--$10$).

In contrast to the behavior of $p_{\mathrm{th}}^{Z}$, the $X$-memory threshold $p_{\mathrm{th}}^{X}$ does not increase for the $Z$-aligned CNOT order compared to the $X$-aligned CNOT order for either geometry (Tables~\ref{tab:zx_biased_zalign}, \ref{tab:zx_xbiased}). For the rotated surface code, the reduction between the $X$-aligned and $Z$-aligned $p_{\mathrm{th}}^{X}$ is small, at most $\approx 0.01 \times 10^{-2}$ in magnitude. Because $\sigma_f$ for the $X$-memory fits grows with bias, this reduction does not exceed twice the combined fitting uncertainty at any bias value. At low bias it is comparable to the combined fitting uncertainty, and at higher bias it falls within the combined fitting uncertainty. So, we treat the two CNOT orders as giving statistically indistinguishable $p_{\mathrm{th}}^{X}$ for the rotated code. Similarly, for the unrotated surface code, the two CNOT orders give statistically indistinguishable $p_{\mathrm{th}}^{X}$ for $\eta \leq 30$. At $\eta = 100$, the $Z$-aligned order gives a lower $p_{\mathrm{th}}^{X}$, but the reduction ($\approx 0.07 \times 10^{-2}$) is only about twice the combined fitting uncertainty and is thus marginally resolved. At $\eta = 300$ and $\eta = 1000$, the reduction grows to $\approx 0.27 \times 10^{-2}$, well beyond the combined fitting uncertainty, so the $Z$-aligned order gives a clearly lower $p_{\mathrm{th}}^{X}$ at these bias values.

When gate-based $XX$ crosstalk noise is included (Table~\ref{tab:zx_crosstalk_zalign}), the qualitative trend with $\eta$ remains unchanged as $p_{\mathrm{th}}^{X}$ increases monotonically with bias and $p_{\mathrm{th}}^{Z}$ decreases and saturates. The addition of crosstalk reduces $p_{\mathrm{th}}^{Z}$ at all finite bias values. This reduction is largest at low-to-moderate bias, where it exceeds the combined fitting uncertainty. For example, in the rotated code at $\eta=0.5$, $p_{\mathrm{th}}^{Z}$ decreases from $0.428\times 10^{-2}$ under Pauli $X$-biased noise to $0.396\times 10^{-2}$ when gate-based $XX$ crosstalk is included. The effect of crosstalk on $p_{\mathrm{th}}^{X}$ differs between the rotated and unrotated codes. For the rotated surface code, crosstalk does not lower $p_{\mathrm{th}}^{X}$, instead there is a slight increase. Wherever the difference is resolvable, it is a small increase that reaches approximately four times the combined fitting uncertainty. For the unrotated surface code, crosstalk instead reduces $p_{\mathrm{th}}^{X}$ at every $\eta$, by amounts that exceed the combined fitting uncertainty except at $\eta=1$ and $\eta=3$.

When gate-based $XX$ crosstalk noise is included (Table~\ref{tab:zx_crosstalk_zalign}), the qualitative trend with $\eta$ remains unchanged as $p_{\mathrm{th}}^{X}$ increases monotonically with bias and $p_{\mathrm{th}}^{Z}$ decreases and saturates. The addition of crosstalk reduces $p_{\mathrm{th}}^{Z}$ at all finite bias values. This reduction is largest at low-to-moderate bias, where it exceeds the combined fitting uncertainty. For example, in the rotated code at $\eta=0.5$, $p_{\mathrm{th}}^{Z}$ decreases from $0.428\times 10^{-2}$ under Pauli $X$-biased noise to $0.396\times 10^{-2}$ when gate-based $XX$ crosstalk is included. The effect of crosstalk on $p_{\mathrm{th}}^{X}$ differs between the rotated and unrotated codes. For the rotated surface code, crosstalk does not lower $p_{\mathrm{th}}^{X}$. The observed differences are small positive shifts of at most $\approx 0.01 \times 10^{-2}$. Because the $X$-memory fit uncertainty grows with bias, these shifts remain within the combined fitting uncertainty at every $\eta$. So, we do not treat them as a resolved enhancement. For the unrotated surface code, crosstalk lowers $p_{\mathrm{th}}^{X}$ at every $\eta$, but this reduction is resolved only at $\eta = 0.5$, where it reaches $\approx 0.03 \times 10^{-2}$. It is marginal at $\eta = 10$ and $\eta = 300$, and falls within the combined fitting uncertainty at the remaining bias values.

Additionally, following the remark made at the end of section \ref{sec:zx_x-align_result}, looking at the results obtained by simulating Pauli $X$-biased noise with $\eta=0.5$, we see a clear separation in the performance of memory $X$ and memory $Z$ of the unrotated surface code (Fig.~\ref{fig:zx_zl_unrot_bi_a}), which is not seen in the rotated surface code (Fig.~\ref{fig:zx_zl_rot_bi_a}). The separation is, however, visible in all cases under crosstalk noise (Figs.~\ref{fig:zx_zl_rot_wc_a}, \ref{fig:zx_zl_unrot_wc_a}). 

We want to make one final remark. We compare the unbiased depolarizing case ($\eta=0.5$) between the two CNOT orderings. In the $X$-aligned CNOT order (Section~\ref{sec:zx_x-align_result}), the logical $X$ memory exhibits slightly lower logical error rates than logical $Z$ in the unrotated geometry. In contrast, in the $Z$-aligned CNOT ordering, this behavior reverses as we see logical $Z$ memory having lower logical error rates than logical $X$ (Figures~\ref{fig:zx_zl_unrot_bi_a} and~\ref{fig:zx_zl_unrot_wc_a}). 

\subsection{ZY Surface Code}
\label{sec:zy_results}

\begin{figure}[!htbp]
    \centering
    
    \begin{subfigure}[t]{0.32\textwidth}
        \centering
        \includegraphics[width=\linewidth]{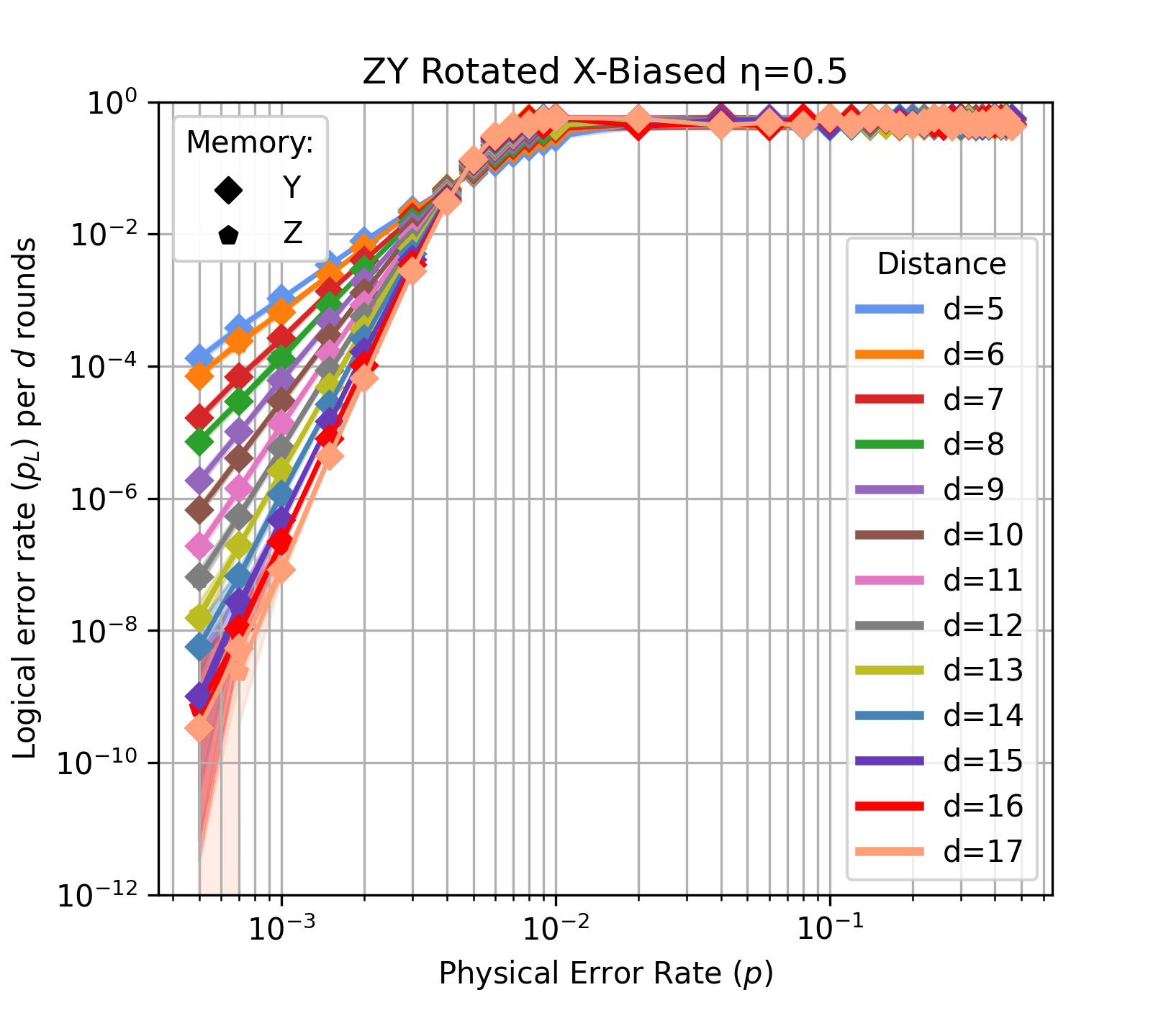}
        \caption{}
        \label{fig:zy_rot_bi (a) }
    \end{subfigure}
    \hfill
    \begin{subfigure}[t]{0.32\textwidth}
        \centering
        \includegraphics[width=\linewidth]{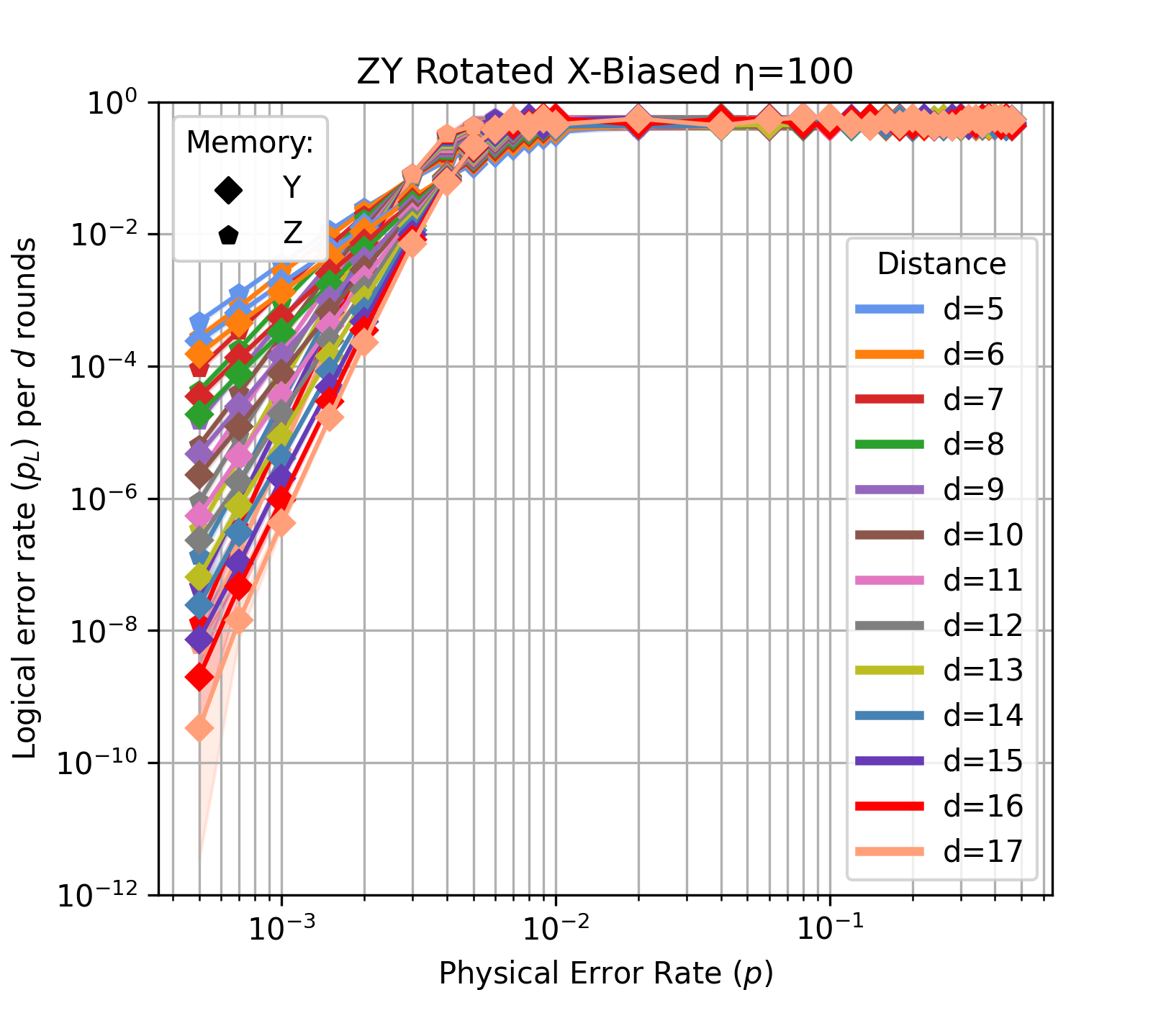}
        \caption{}
        \label{fig:zy_rot_bi (b) }
    \end{subfigure}
    \hfill
    \begin{subfigure}[t]{0.32\textwidth}
        \centering
        \includegraphics[width=\linewidth]{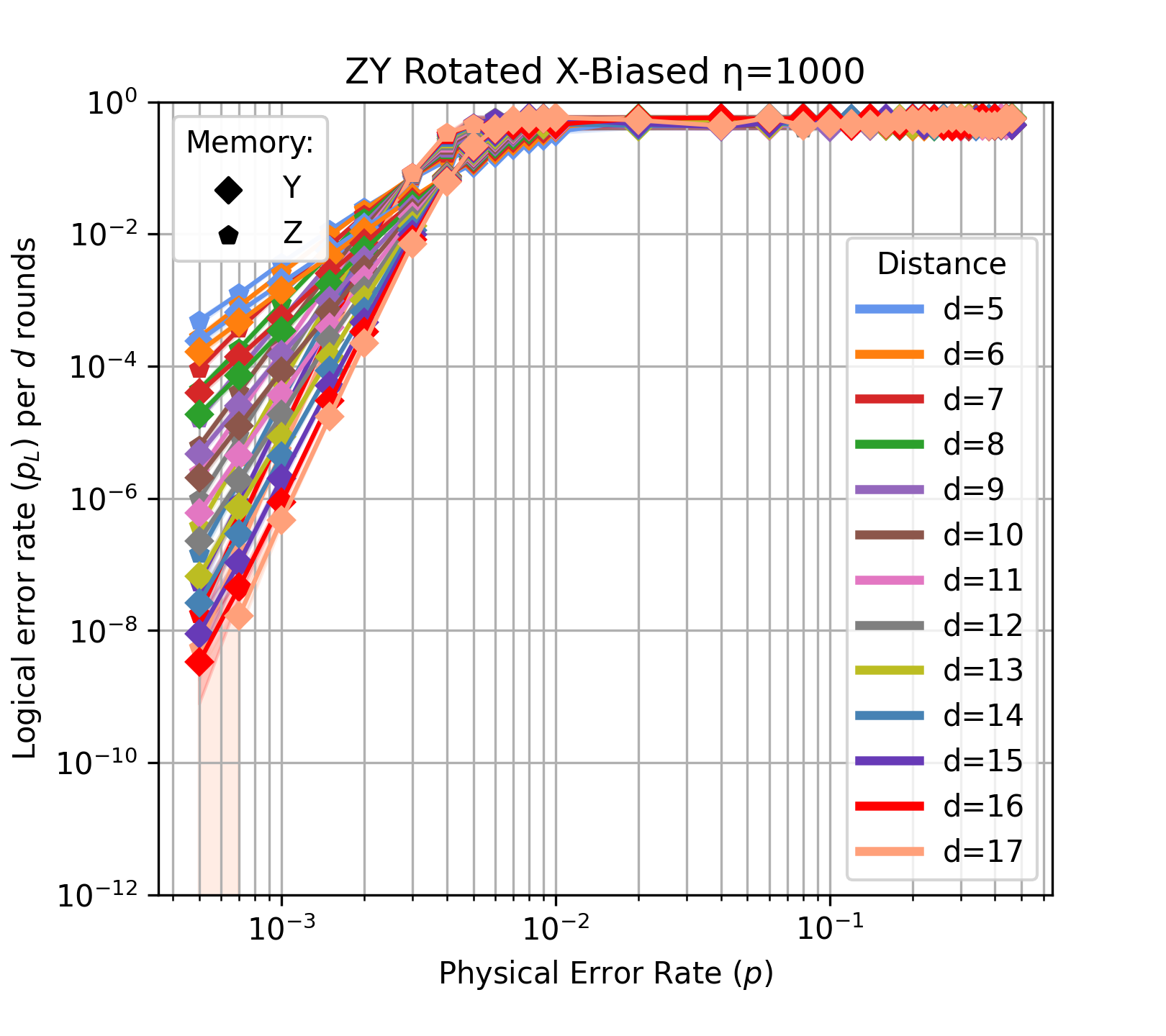}
        \caption{}
        \label{fig:zy_rot_bi (c) }
    \end{subfigure}

    \caption{Logical error rate per $d$ rounds of syndrome extraction $p_L$ versus physical error rate $p$ simulated on rotated $ZY$ surface code with $X$-aligned CNOT order under Pauli-$X$ biased noise at $\eta = \{0.5, 100, 1000\}$.}

    \label{fig:zy_rot_bi}
\end{figure}

\begin{figure}[!htbp]
    \centering
    
    \begin{subfigure}[t]{0.32\textwidth}
        \centering
        \includegraphics[width=\linewidth]{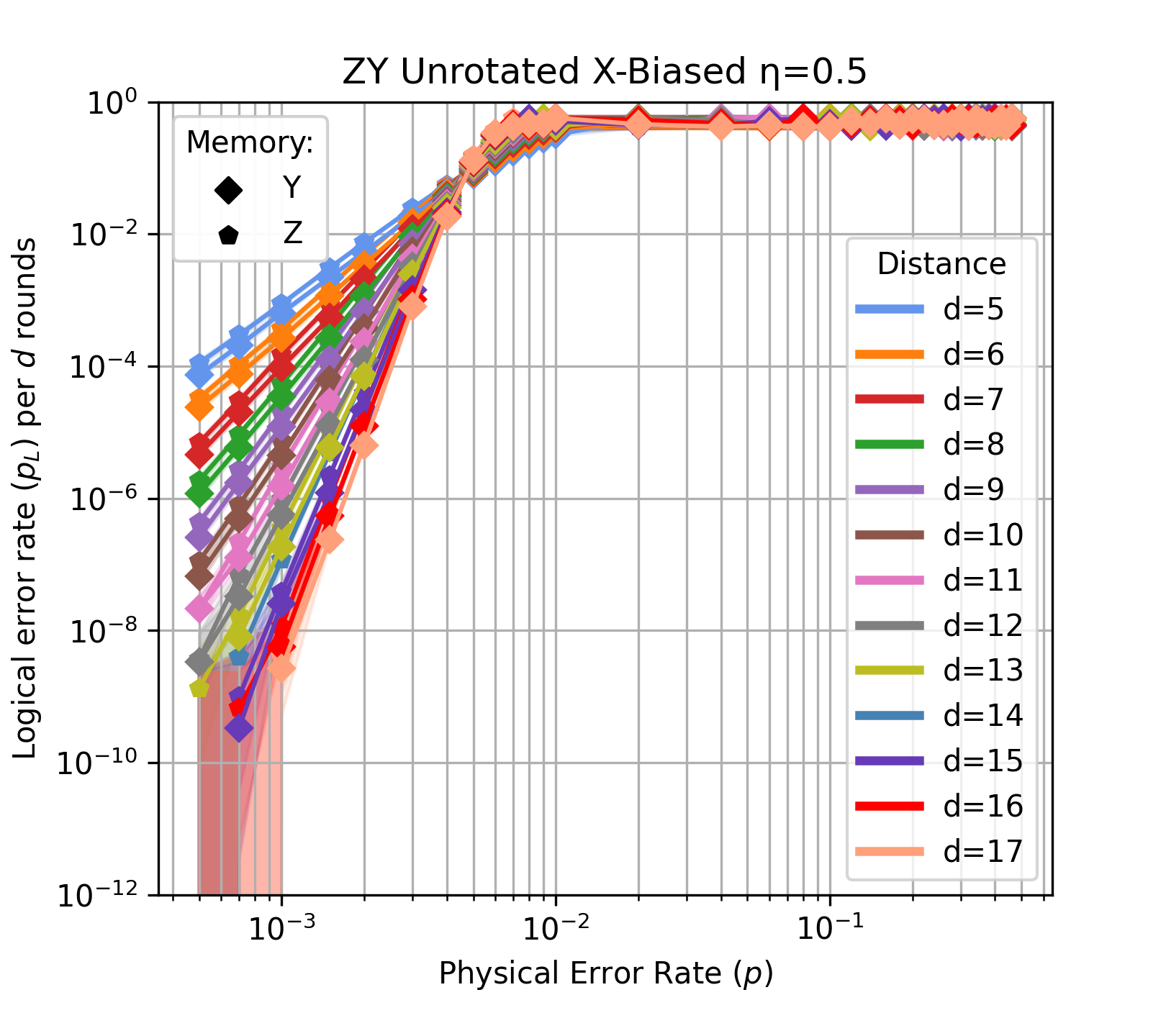}
        \caption{}
        \label{fig:zy_unrot_bi (a)}
    \end{subfigure}
    \hfill
    \begin{subfigure}[t]{0.32\textwidth}
        \centering
        \includegraphics[width=\linewidth]{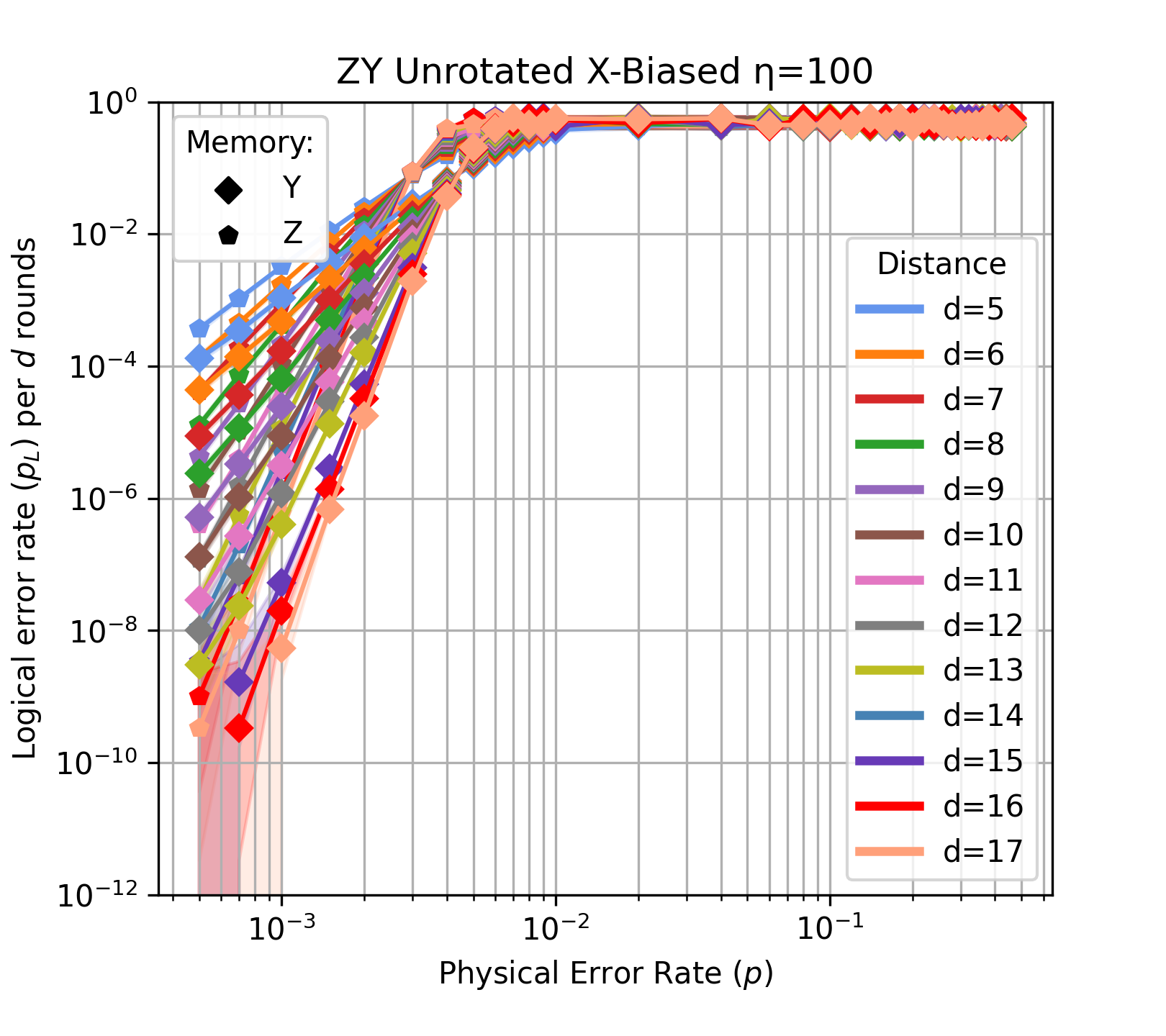}
        \caption{}
        \label{fig:zy_unrot_bi (b)}
    \end{subfigure}
    \hfill
    \begin{subfigure}[t]{0.32\textwidth}
        \centering
        \includegraphics[width=\linewidth]{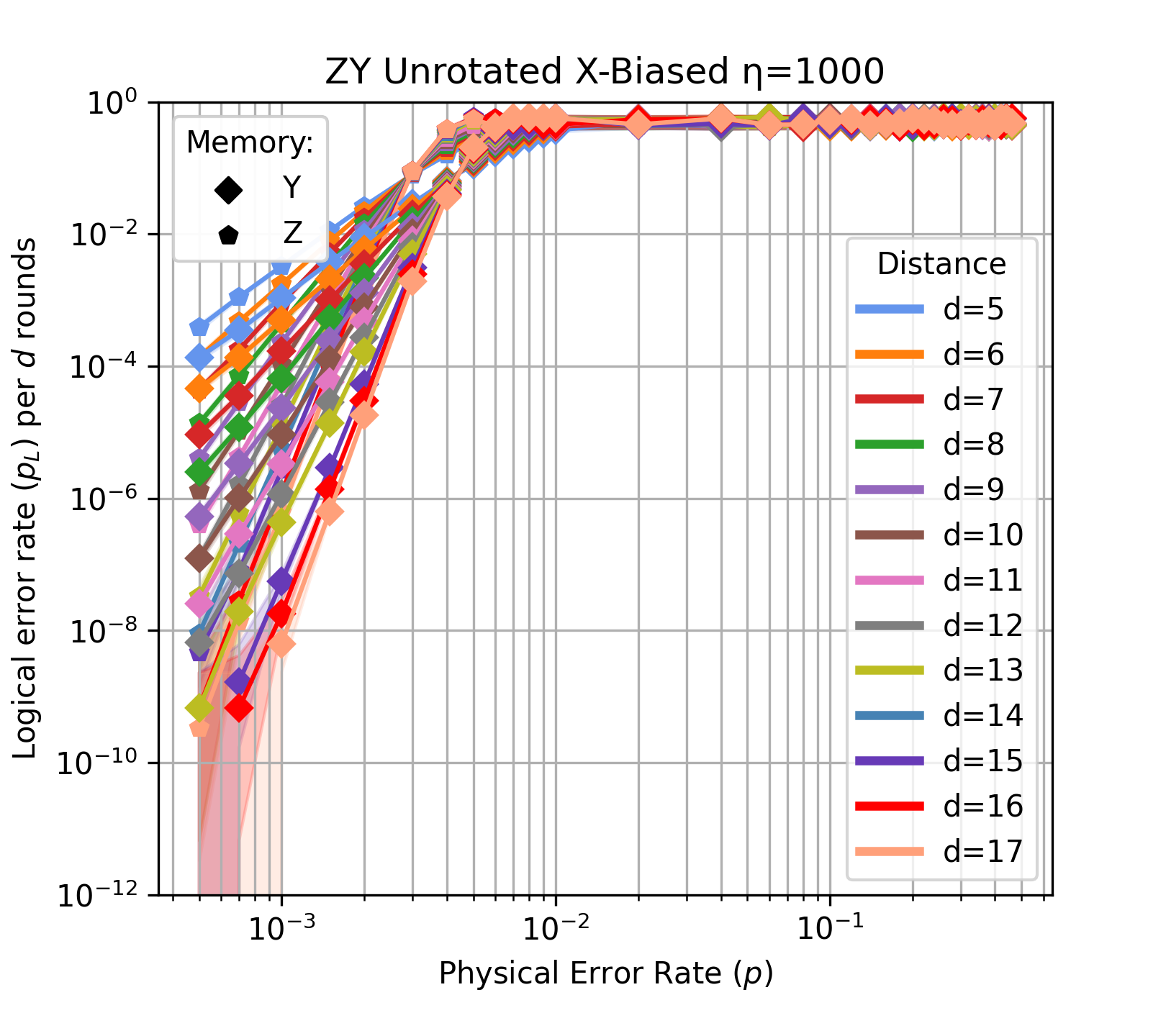}
        \caption{}
        \label{fig:zy_unrot_bi (c)}
    \end{subfigure}

    \caption{Logical error rate per $d$ rounds of syndrome extraction $p_L$ versus physical error rate $p$ simulated on unrotated $ZY$ surface code with $X$-aligned CNOT order under Pauli-$X$ biased noise at $\eta = \{0.5, 100, 1000\}$.} 
    
    \label{fig:zy_unrot_bi}
\end{figure}

\begin{figure}[ht]
    \centering
    
    \begin{subfigure}[t]{0.32\textwidth}
        \centering
        \includegraphics[width=\linewidth]{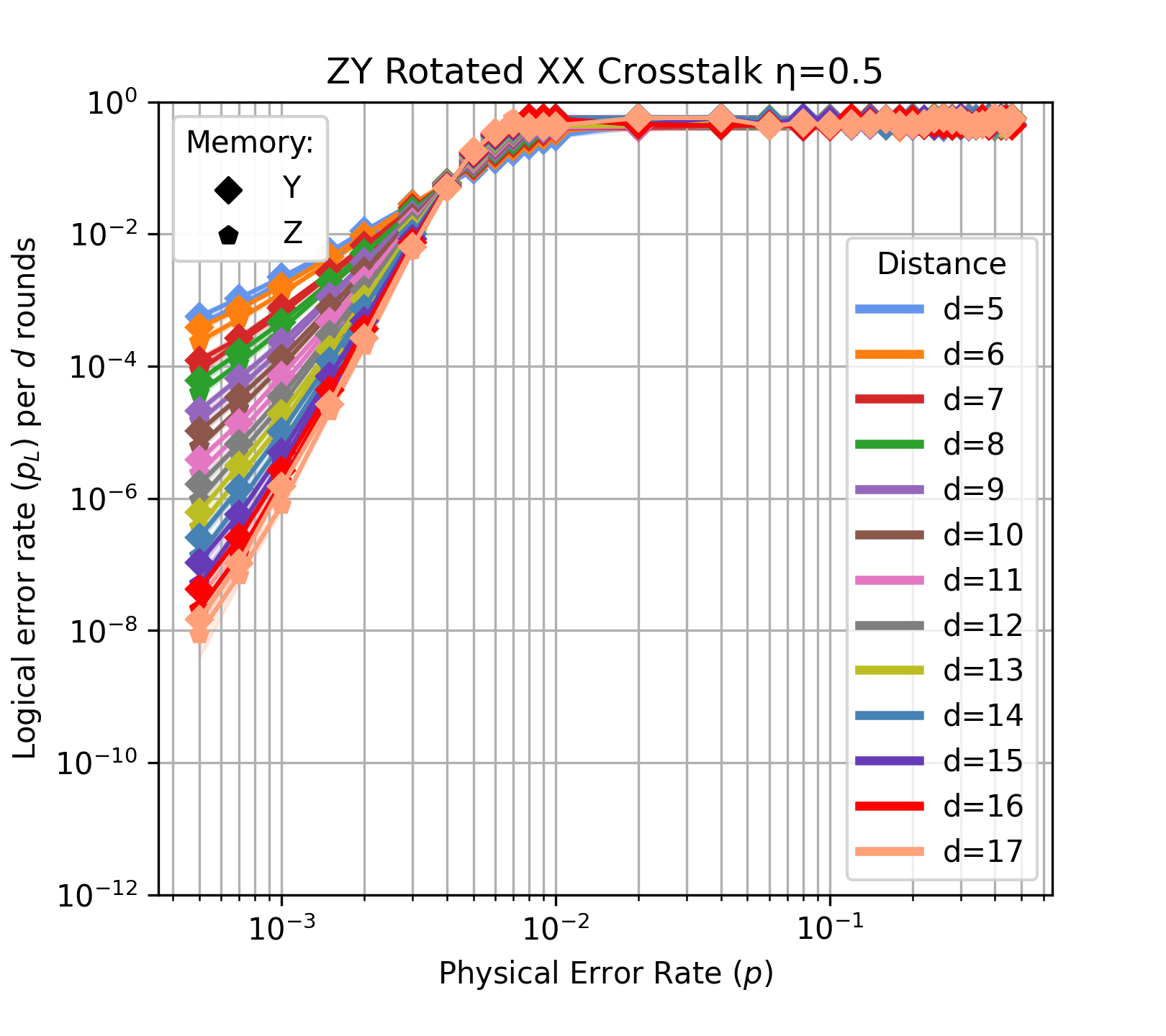}
        \caption{}
        \label{fig:zy_rot_wc (a)}
    \end{subfigure}
    \hfill
    \begin{subfigure}[t]{0.32\textwidth}
        \centering
        \includegraphics[width=\linewidth]{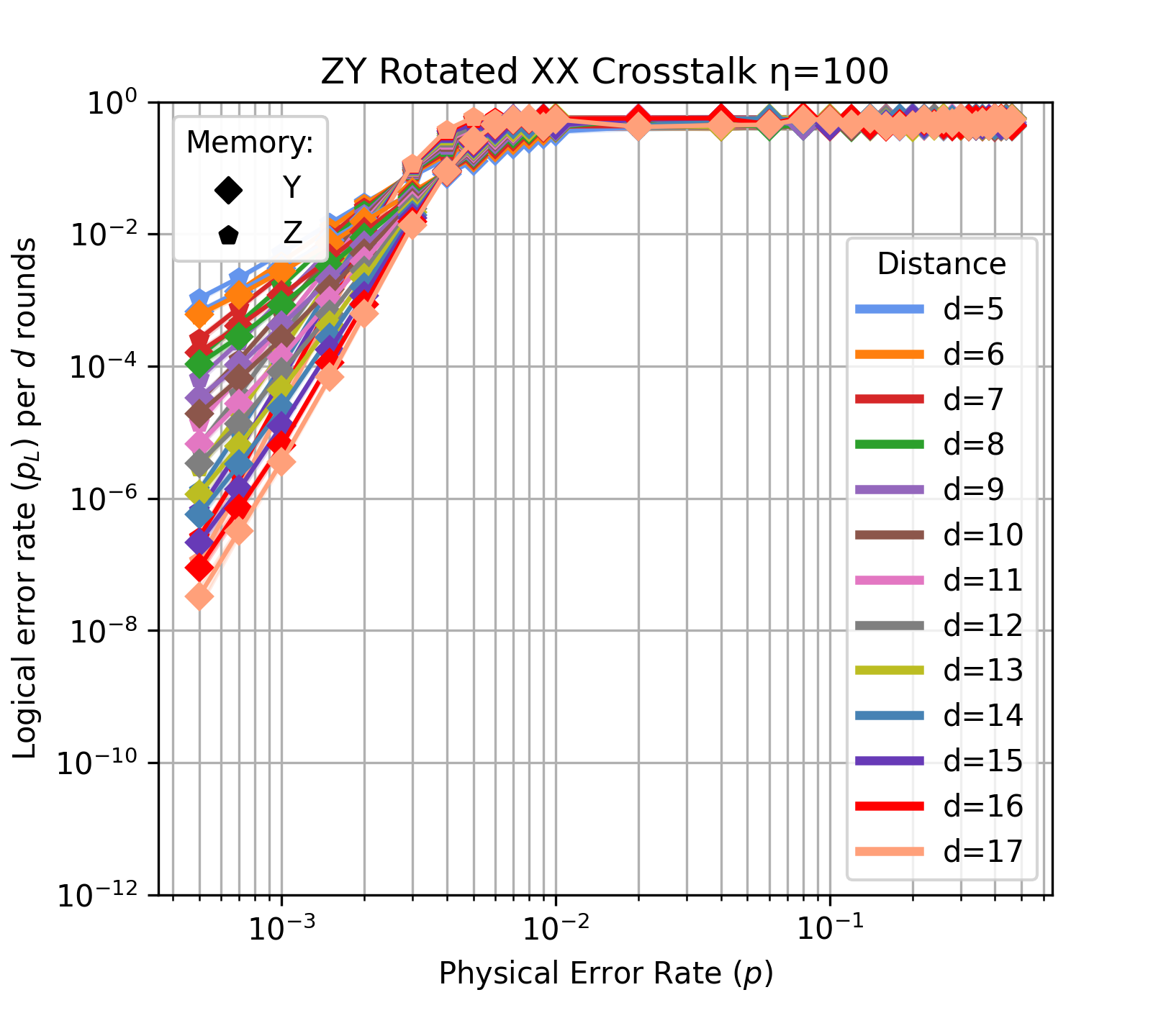}
        \caption{}
        \label{fig:zy_rot_wc (b)}
    \end{subfigure}
    \hfill
    \begin{subfigure}[t]{0.32\textwidth}
        \centering
        \includegraphics[width=\linewidth]{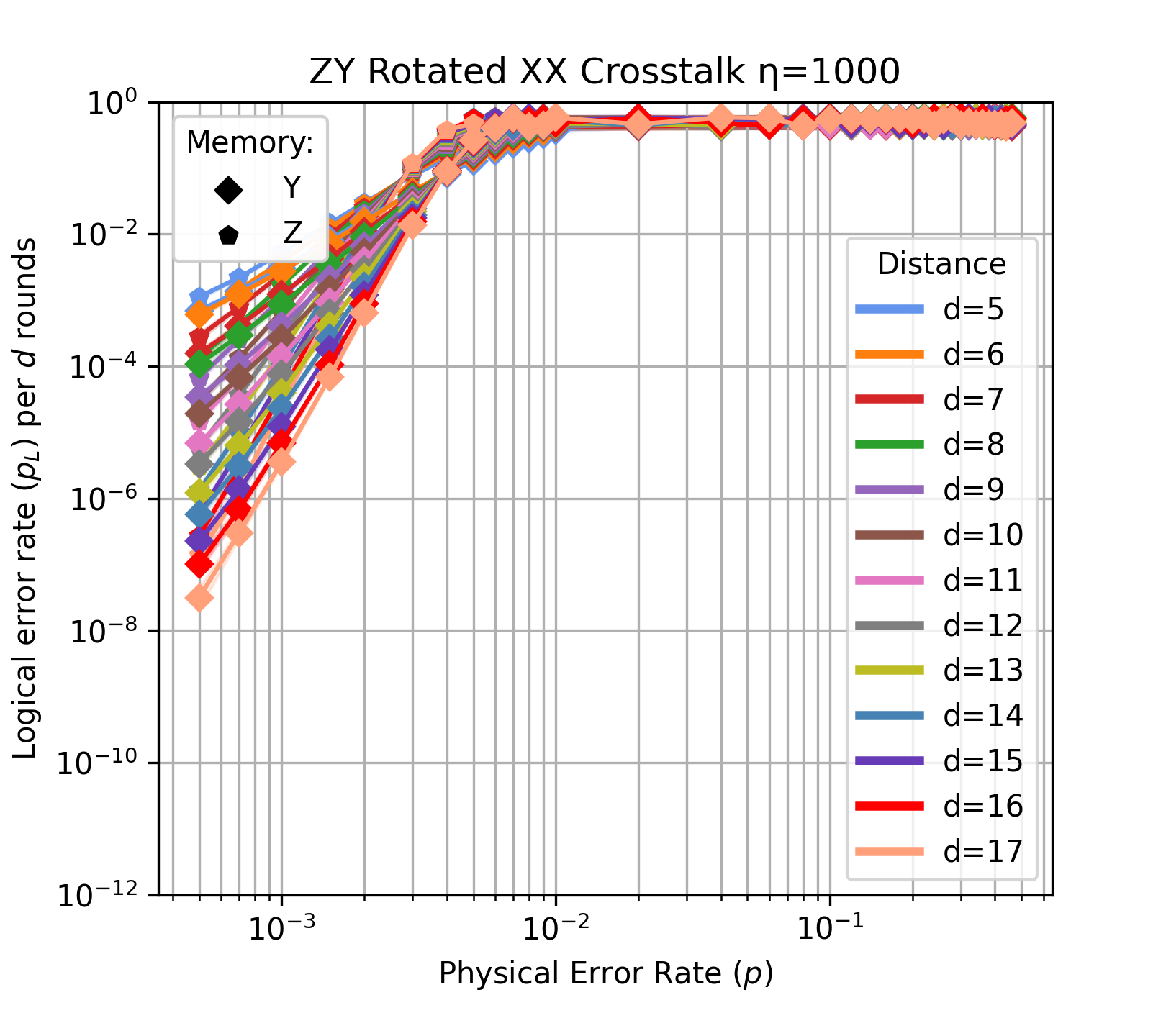}
        \caption{}
        \label{fig:zy_rot_wc (c)}
    \end{subfigure}

    \caption{Logical error rate per $d$ rounds of syndrome extraction $p_L$ versus physical error rate $p$ simulated on rotated $ZY$ surface code with $X$-aligned CNOT order under Pauli-$X$ biased noise at $\eta = \{0.5, 100, 1000\}$ with an additional $XX$ crosstalk noise.}
    
    \label{fig:zy_rot_wc}
\end{figure}

\begin{figure}[ht]
    \centering
    
    \begin{subfigure}[t]{0.32\textwidth}
        \centering
        \includegraphics[width=\linewidth]{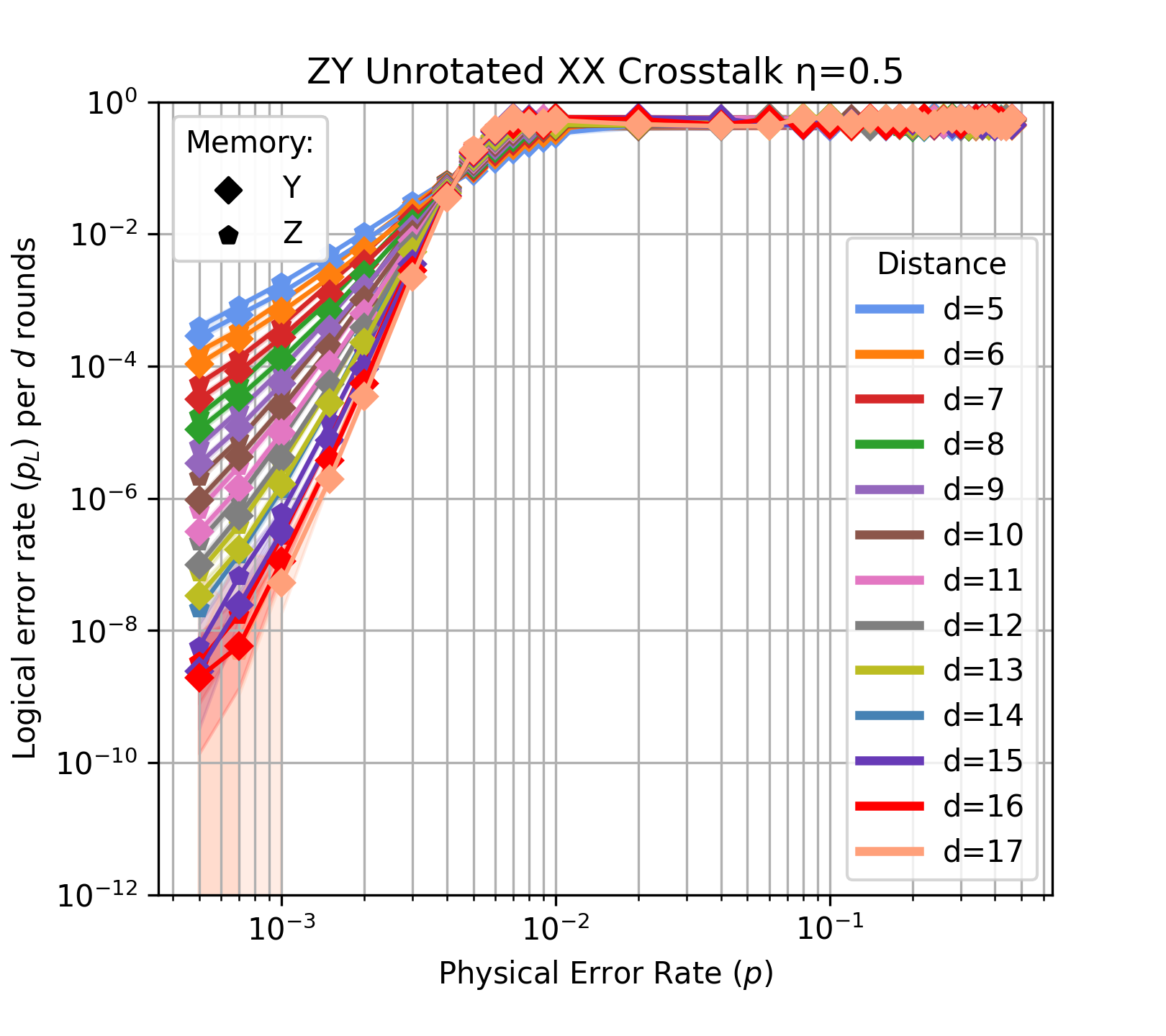}
        \caption{}
        \label{fig:zy_unrot_wc (a)}
    \end{subfigure}
    \hfill
    \begin{subfigure}[t]{0.32\textwidth}
        \centering
        \includegraphics[width=\linewidth]{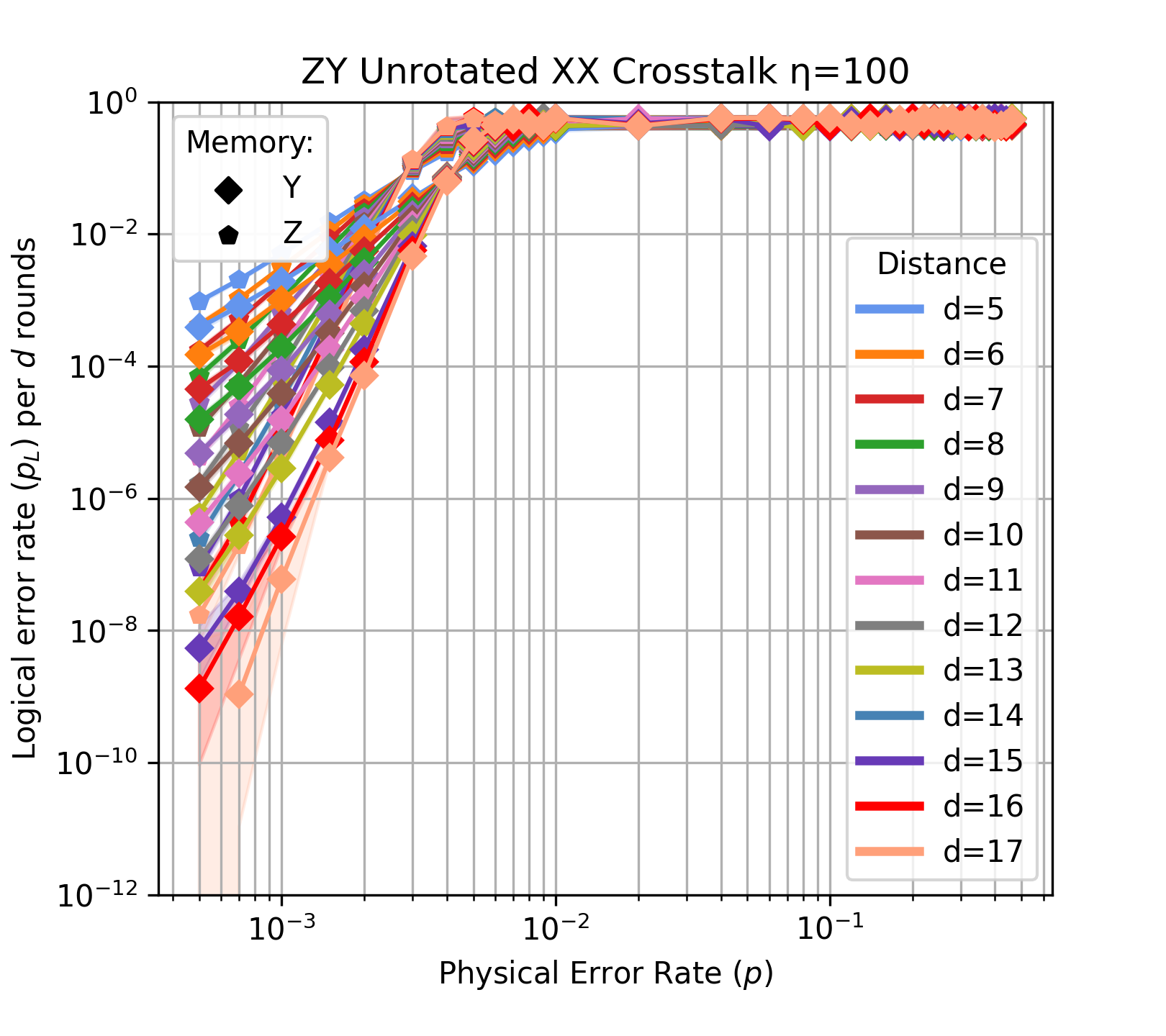}
        \caption{}
        \label{fig:zy_unrot_wc (b)}
    \end{subfigure}
    \hfill
    \begin{subfigure}[t]{0.32\textwidth}
        \centering
        \includegraphics[width=\linewidth]{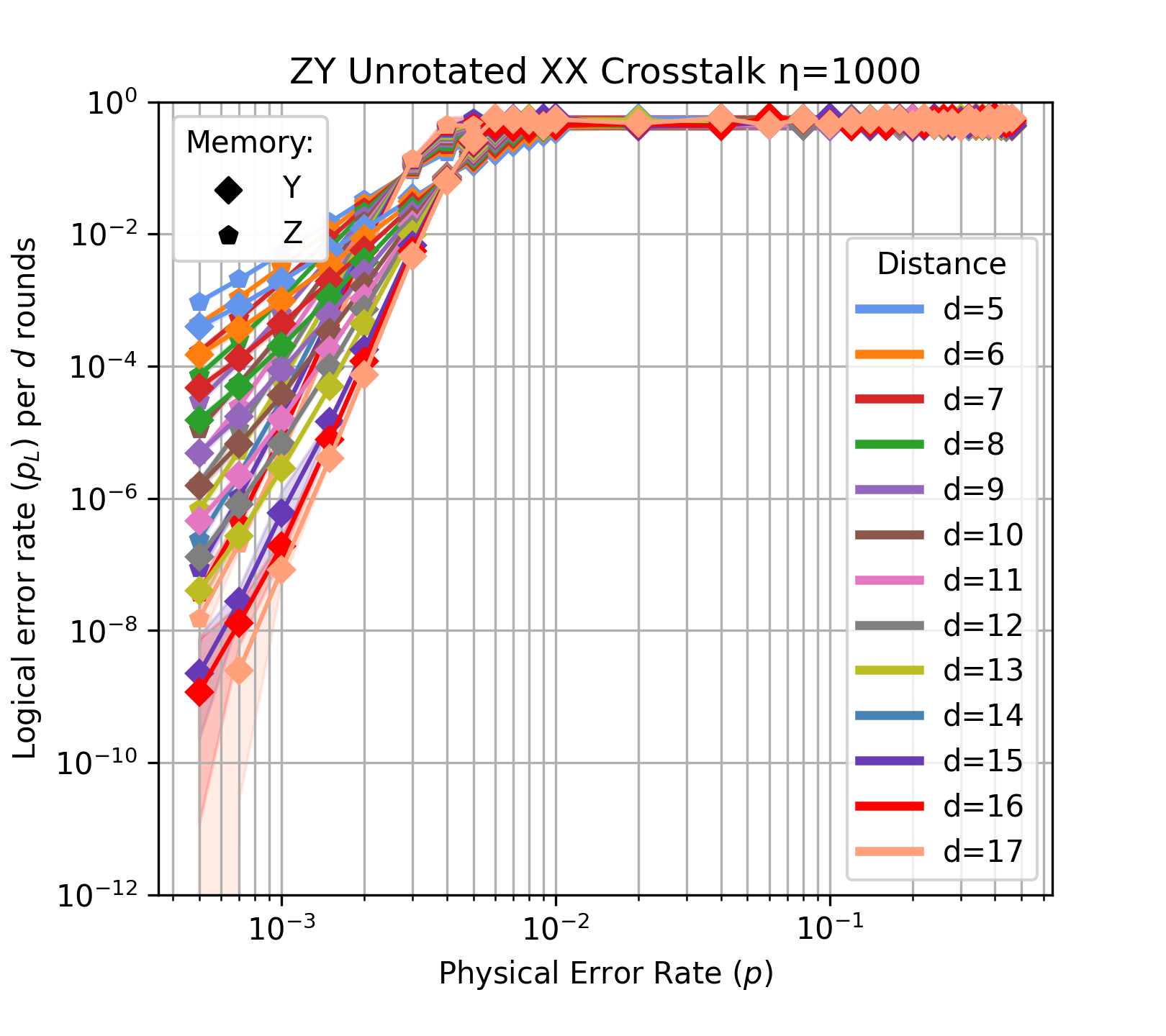}
        \caption{}
        \label{fig:zy_unrot_wc (c)}
    \end{subfigure}

    \caption{Logical error rate per $d$ rounds of syndrome extraction $p_L$ versus physical error rate $p$ simulated on unrotated $ZY$ surface code with $X$-aligned CNOT order under Pauli-$X$ biased noise at $\eta = \{0.5, 100, 1000\}$ with an additional $XX$ crosstalk noise.}
    
    \label{fig:zy_unrot_wc}
\end{figure}

Finally, in this section, we present the results obtained by simulating the two noise models on the $ZY$ surface code with $Z$-type and $Y$-type stabilizers. Here, we first present the results obtained by simulating Pauli $X$-biased noise on the $ZY$ surface code (Table \ref{tab:zy_xbiased}, Figs.~\ref{fig:zy_rot_bi}, \ref{fig:zy_unrot_bi}), and then we present the results obtained by simulating an additional gate-based $XX$ crosstalk noise on the $ZY$ surface code (Table \ref{tab:zy_xbiased_crosstalk}, Figs.~\ref{fig:zy_rot_wc}, \ref{fig:zy_unrot_wc}). For the $ZY$ surface code, we only simulated the $10231203$ CNOT order, which we call the $X$-aligned CNOT order. The meaning and motivation for the choice of this order is explained \ref{sec:cnot_order}, Fig.~\ref{fig:cnot_order}.

\begin{table}[ht]
\centering
\caption{Threshold values $p_{\mathrm{th}}$ for the $ZY$ surface code under Pauli-$X$ biased noise. Here, $\eta$ denotes bias parameter, $p_{\mathrm{th}}^Y$ denotes $Y$-memory threshold and $p_{\mathrm{th}}^Z$ denotes $Z$-memory threshold. All values carry a statistical fitting uncertainty of order $\sigma_{f} \approx 0.002 \times 10^{-2}$.}
\label{tab:zy_xbiased}
\small
\setlength{\tabcolsep}{12pt}
\begin{tabular}{c SS SS}
\toprule
& \multicolumn{2}{c}{Rotated Surface Code} 
& \multicolumn{2}{c}{Unrotated Surface Code} \\
\cmidrule(lr){2-3} \cmidrule(lr){4-5}
{$\eta$} &
{$p_{\mathrm{th}}^{Y} (\times 10^{-2})$} &
{$p_{\mathrm{th}}^{Z} (\times 10^{-2})$} &
{$p_{\mathrm{th}}^{Y} (\times 10^{-2})$} &
{$p_{\mathrm{th}}^{Z} (\times 10^{-2})$} \\
\midrule
0.5      & 0.428 & 0.428 & 0.460 & 0.458 \\
1        & 0.423 & 0.386 & 0.449 & 0.403 \\
3        & 0.415 & 0.321 & 0.439 & 0.350 \\
10       & 0.411 & 0.296 & 0.437 & 0.313 \\
30       & 0.410 & 0.284 & 0.435 & 0.302 \\
100      & 0.404 & 0.279 & 0.435 & 0.299 \\
300      & 0.403 & 0.278 & 0.434 & 0.296 \\
1000     & 0.402 & 0.276 & 0.434 & 0.295 \\
$\infty$ & 0.402 & 0.275 & 0.434 & 0.294 \\
\bottomrule
\end{tabular}
\end{table}

\begin{table}[ht]
\centering
\caption{Threshold values $p_{\mathrm{th}}$ for the $ZY$ surface code under Pauli-$X$ biased noise and additional $XX$ crosstalk noise. Here, $\eta$ denotes bias parameter, $p_{\mathrm{th}}^Y$ denotes $Y$-memory threshold and $p_{\mathrm{th}}^Z$ denotes $Z$-memory threshold. All values carry a statistical fitting uncertainty of order $\sigma_{f} \approx 0.002 \times 10^{-2}$.}
\label{tab:zy_xbiased_crosstalk}
\small
\setlength{\tabcolsep}{12pt}
\begin{tabular}{c SS SS}
\toprule
& \multicolumn{2}{c}{Rotated Surface Code} 
& \multicolumn{2}{c}{Unrotated Surface Code} \\
\cmidrule(lr){2-3} \cmidrule(lr){4-5}
{$\eta$} &
{$p_{\mathrm{th}}^{Y} (\times 10^{-2})$} &
{$p_{\mathrm{th}}^{Z} (\times 10^{-2})$} &
{$p_{\mathrm{th}}^{Y} (\times 10^{-2})$} &
{$p_{\mathrm{th}}^{Z} (\times 10^{-2})$} \\
\midrule
0.5      & 0.396 & 0.396 & 0.435 & 0.433 \\
1        & 0.394 & 0.373 & 0.431 & 0.385 \\
3        & 0.391 & 0.313 & 0.414 & 0.325 \\
10       & 0.390 & 0.284 & 0.415 & 0.290 \\
30       & 0.384 & 0.277 & 0.414 & 0.288 \\
100      & 0.383 & 0.277 & 0.412 & 0.287 \\
300      & 0.382 & 0.276 & 0.411 & 0.287 \\
1000     & 0.382 & 0.276 & 0.411 & 0.287 \\
$\infty$ & 0.380 & 0.276 & 0.411 & 0.287 \\
\bottomrule
\end{tabular}
\end{table}

For this surface code as well, we simulated both the rotated and unrotated surface codes. For each bias parameter $\eta$, we calculated the logical error rates $p_L$ and plotted them as a function of physical error rates $p$. Similar to what we observed in $ZX$ surface codes (section \ref{sec:zx_x-align_result}, \ref{sec:zx_z-align_results}), for each bias parameter $\eta$, we observed a clear threshold for both memory $Y$ and memory $Z$. However, the qualitative behavior of the thresholds in the $ZY$ surface code differs significantly from that observed in the $ZX$ surface codes.

Under the Pauli $X$-biased noise, in the standard depolarizing case $(\eta=0.5)$, we find that the thresholds for the rotated $ZY$ surface code (Table \ref{tab:zy_xbiased}) are exactly the same as the thresholds for rotated $ZX$ surface code (Table \ref{tab:zx_xbiased}), with the role of memory $Y$ and memory $X$ swapped. In the unrotated surface code as well, at $\eta=0.5$ the performance of $Z$ memory is identical in both $ZX$ and $ZY$ surface codes. But $Y$ memory threshold $p_{\mathrm{th}}^Y$ for the unrotated $ZY$ surface code at $\eta = 0.5$ is found to be slightly lower than the $X$ memory threshold $p_{\mathrm{th}}^X$ of the unrotated $ZX$ surface code at the same $\eta$. The difference is $0.002 \times 10^{-2}$ which is within the combined fitting uncertainty.

Similarly, under Pauli-$X$-biased noise (Table~\ref{tab:zy_xbiased}), $p_{\mathrm{th}}^{Y}$ shows only a weak dependence on the bias parameter $\eta$ in both rotated and unrotated surface codes, in sharp contrast to the $X$-memory threshold of the $ZX$ surface code. For example, in the rotated code, $p_{\mathrm{th}}^{Y}$ decreases slowly from $0.428 \times 10^{-2}$ at $\eta=0.5$ to $0.402 \times 10^{-2}$ at $\eta=\infty$. A similar weak dependence is observed in the unrotated surface code, where the threshold $p_{\mathrm{th}}^{Y}$ decreases from $0.460 \times 10^{-2} $ to $0.434\times 10^{-2}$. This is the central qualitative difference compared to the $ZX$ surface code. Over the same bias range the $ZX$ $X$-memory threshold rises steeply, from $0.428\times10^{-2}$ at $\eta=0.5$ to $16.4\times10^{-2}$ at $\eta=1000$ (rotated; Table~\ref{tab:zx_xbiased}), which is a factor of ${\sim}38$, whereas the $ZY$ $Y$-memory threshold instead decreases slightly. Therefore, increasing $X$-bias does not enhance the $Y$-memory threshold of the $ZY$ code.

On the other hand, the $Z$-memory threshold $p_{\mathrm{th}}^{Z}$ decreases with increasing $\eta$ and rapidly approaches a saturation value, similar to the trend observed in the $ZX$ surface codes. In the rotated code, $p_{\mathrm{th}}^{Z}$ decreases from approximately $0.428\times 10^{-2}$ at $\eta=0.5$ to $0.275\times 10^{-2}$ at $\eta=\infty$, while in the unrotated code, it decreases from $0.458\times 10^{-2}$ to $0.294\times 10^{-2}$. In fact, the $Z$-memory thresholds of the $ZX$ and $ZY$ surface codes agree to within the combined fitting uncertainty at every $\eta$, not only at the saturation point (Tables~\ref{tab:zy_xbiased}, \ref{tab:zx_xbiased}). This result is consistent with the fact that the two codes share identical $Z$-type stabilizers since the $ZX$ to $ZY$ modification changes only the complementary stabilizer type, which primarily affects the $X$/$Y$ memory rather than the $Z$ memory. The saturation values are comparable to those observed in the $ZX$ surface codes. Comparing rotated and unrotated codes, we again observe that the unrotated code consistently gives slightly higher threshold values for both $Y$- and $Z$-memory experiments across all bias parameters, and these differences exceed the combined fitting uncertainty.

When gate-based $XX$ crosstalk noise is added on top of the Pauli-$X$ biased noise (Table~\ref{tab:zy_xbiased_crosstalk}), the qualitative $\eta$-dependence of the thresholds in $ZY$ surface code is preserved, exactly as it was for the $ZX$ codes. The $Y$-memory threshold stays nearly flat in $\eta$, which is the distinctive feature of the $ZY$ code as noted above, while the $Z$-memory threshold decreases and saturates. Beyond preserving these trends, the addition of crosstalk decreases the thresholds for both $Y$ and $Z$ memories for the $ZY$ surface code, in contrast to the $X$-aligned $ZX$ surface code (\ref{sec:zx_x-align_result}), where crosstalk left the $X$-memory threshold within the fit uncertainty (Table~\ref{tab:zx_crosstalk}). For the $Y$-memory, the reduction induced by additional crosstalk noise exceeds the combined fit uncertainty across all $\eta$ in both geometries and stays roughly constant, of order $0.02$--$0.03\times10^{-2}$. For the $Z$-memory, the reduction exceeds the combined fit uncertainty at all $\eta$ for the unrotated code but only at low-to-moderate bias for the rotated code, where it falls within the fit uncertainty for $\eta \geq 100$; it is most prominent at low bias and diminishes as $\eta$ increases. For instance, in the rotated geometry at $\eta=0.5$, both $p_{\mathrm{th}}^{Y}$ and $p_{\mathrm{th}}^{Z}$ decrease from $0.428\times10^{-2}$ without crosstalk to $0.396\times10^{-2}$ with crosstalk.

\section{Discussion}
\label{sec:discussion}

In this section, we interpret the results obtained from our simulations and discuss the behavior of our tailored surface codes under Pauli-$X$ biased noise and an additional gate-based $XX$ crosstalk noise with practical minimum-weight perfect matching decoders (MWPM). 

\subsection{Threshold Behavior of ZX Surface Code}
\label{sec:zx_discussion}

We observed that the threshold behavior of the $ZX$ surface code, both under Pauli-$X$ biased noise and under additional gate-based $XX$ crosstalk noise follows a clear pattern. As the bias $\eta$ increases, the $X$ memory threshold $p_{\mathrm{th}}^X$ increases monotonically, and eventually flows out of the observed window, while the $Z$-memory threshold $p_{\mathrm{th}}^Z$ decreases and saturates (Table~\ref{tab:zx_xbiased}, \ref{tab:zx_crosstalk}). This asymmetry occurs because of the commutation relations between the dominant error and the logical operators. Since the dominant $X$ errors commute with $X_L$, they cannot flip the $X$-basis logical observable and therefore cannot cause $X$-memory failures. An $X$-memory failure instead requires a logical $Z_L$, built from $Z$- or $Y$-type errors, which by construction of our biased noise model (Section~\ref{sec:error model}) become increasingly rare as $\eta$ tends to  $\infty$. Consequently, $p_{\mathrm{th}}^{X}$ increases monotonically with $\eta$. For the $Z$ memory, by contrast, a failure is caused by a logical $X_L$ built from $X$-error chains. These dominant $X$ errors are detected only by the $Z$-type stabilizers, since the $X$-type stabilizers commute with $X$ errors and thus are blind to them. As $\eta$ tends to $\infty$ the noise becomes essentially pure $X$ noise, thus the effective bit-flip problem seen by the $Z$ memory stops changing and $p_{\mathrm{th}}^{Z}$ saturates at a low value. 

This qualitative behavior is consistent across both rotated and unrotated variants of the surface code, both noise models, and both CNOT orders. Across all cases, the unrotated code yields higher thresholds than the rotated code for both $X$- and $Z$-memory. For the $Z$ memory, the difference in the threshold values between the rotated and unrotated codes remain small at all $\eta$, of order $0.01$--$0.04\times10^{-2}$. For the $X$ memory these differences are comparably small at low bias but grow substantially with $\eta$, reaching $\approx 0.8\times10^{-2}$ at $\eta=1000$.

The effect of crosstalk differs between the two memories and, for the $X$ memory, between the two CNOT orders. For the $X$-aligned CNOT order, adding crosstalk changes the $X$-memory threshold by less than the combined fitting uncertainty across all $\eta$ in both geometries. For the $Z$-aligned order, the crosstalk-induced shifts in the $X$-memory threshold are larger in magnitude but, once the bias dependence of the fit uncertainty is taken into account, remain mostly within the combined fitting uncertainty. Crosstalk reduces the unrotated $X$-memory threshold at every $\eta$, but this reduction is clearly resolved only at $\eta = 0.5$. It is marginal at $\eta = 10$ and $\eta = 300$, and remains within the combined uncertainty at the other bias values. We return to this CNOT-order dependence in Section~\ref{sec:cnot_order_discussion}.

The $Z$-memory threshold, by contrast, is reduced by crosstalk in every configuration. For the unrotated geometry this reduction exceeds the combined fitting uncertainty at all $\eta$ in both CNOT orders. For the rotated geometry it exceeds the combined fitting uncertainty up to $\eta \approx 30$ for both CNOT orders, then diminishes as the $Z$-memory threshold approaches saturation. This is consistent with the correlated $XX$ crosstalk introducing additional $X$-type errors, which make the $Z$ memory harder to protect. The $X$ memory, whose failures require $Z$- or $Y$-type errors, is largely insensitive to this extra $X$ noise; the residual shifts under the $Z$-aligned schedule are small and, apart from a few low- and moderate-bias points, lie within the combined fitting uncertainty.

The choice of CNOT order introduces a further redistribution of threshold performance. Under $Z$-aligned CNOT order, the $Z$-memory threshold is enhanced relative to the $X$-aligned case, particularly in the unrotated geometry, while the $X$-memory threshold is reduced at high bias. This indicates that the CNOT scheduling influences which logical operator is better protected, without uniformly improving or degrading overall performance.

\subsection{Threshold Behavior of ZY Surface Code}

The $ZY$ surface code was constructed by replacing the $X$-type stabilizers with $Y$-type stabilizers, so that both the $Y$-type and $Z$-type stabilizers detect $X$ errors. This doubles the number of stabilizers that detect the dominant $X$ errors and thus give more syndrome information for error correction, following the same principle as the tailored code of Ref.~\cite{Tuckett2018Ultrahigh}. Here, we interpret the results reported in section \ref{sec:zy_results} in two ways.

First, the $Y$-memory threshold $p_{\mathrm{th}}^Y$ of the $ZY$ code remains approximately constant as $\eta$ increases, decreasing only slightly from $0.428 \times 10^{-2}$ to $0.402 \times 10^{-2}$ in the rotated geometry (Table~\ref{tab:zy_xbiased}). This contrasts sharply with the $ZX$ code, where $p_{\mathrm{th}}^X$ rises from $0.428 \times 10^{-2}$ to $16.4\times 10^{-2}$ (Table~\ref{tab:zx_xbiased}) over the same $\eta$ range. The reason is structural. In the $ZY$ code the logical operator protected by the $Y$-memory is a $Y$-type string. Since a single-qubit $X$ error anticommutes with $Y$ on that qubit, a chain of the dominant $X$ errors running along the support of this logical operator flips the logical $Y$ parity and causes a $Y$-memory failure. The dominant $X$ errors therefore attack the $Y$-memory directly, and as $\eta$ increases this error type only becomes more prevalent. This is the opposite of the $ZX$ case, where the $X$-memory logical is a pure-$X$ string that commutes with the dominant $X$ errors, giving the passive protection responsible for the steep rise of $p_{\mathrm{th}}^{X}$.

Second, the $Z$-memory threshold $p_{\mathrm{th}}^{Z}$ is statistically indistinguishable between the $ZX$ and $ZY$ codes. For example, at $\eta=1000$ in the rotated geometry, $p_{\mathrm{th}}^{Z} = 0.277\times10^{-2}$ for the $ZX$ code (Table~\ref{tab:zx_xbiased}) and $0.276\times10^{-2}$ for the $ZY$ code (Table~\ref{tab:zy_xbiased}), a difference of $0.001\times10^{-2}$, within the combined fitting uncertainty. In fact, the two agree to within the combined fitting uncertainty at every bias value (Tables~\ref{tab:zx_xbiased}, \ref{tab:zy_xbiased}), not only at saturation. The indistinguishable thresholds show that PyMatching 2.3.1 with correlated matching does not translate the additional $Y$ stabilizer syndrome into an improved threshold. This result is consistent with Ref.~\cite{Tuckett2018Ultrahigh} that minimum-weight matching fails for their tailored code. Additionally, this extends their observation under code-capacity noise to circuit-level noise.

The $ZY$ code provides more syndrome information than the $ZX$ but does not change how error strings propagate through the lattice. The $X$ errors still form two-dimensional error chains in the usual way. To benefit from the doubled syndrome, a decoder would need to jointly reason over both $Y$- and $Z$-type syndrome data to infer that both stabilizer types were triggered by the same physical $X$ error. Our results show that the correlated matching extension of PyMatching 2.3.1 does not perform this type of joint reasoning. This suggests that for matching-based decoders, code modifications that simplify the error geometry may be more effective than code modifications that enrich the syndrome structure.

\subsection{Effect of CNOT Ordering}
\label{sec:cnot_order_discussion}

Comparing $X$-aligned and $Z$-aligned CNOT orders in the $ZX$ code, we observe a redistribution of threshold values between the $X$- and $Z$-memories (Tables~\ref{tab:zx_xbiased}, \ref{tab:zx_biased_zalign}). For the unrotated surface code, the $Z$-aligned order gives a higher $p_{\mathrm{th}}^{Z}$ at all bias values. This increase exceeds the combined fitting uncertainty up to moderate bias but the increase narrows to within the combined fitting uncertainty as $p_{\mathrm{th}}^{Z}$ saturates at high bias ($\eta \geq 100$). The $Z$-aligned order also gives a lower $p_{\mathrm{th}}^{X}$ at high bias. The reduction is marginal at $\eta = 100$ and it is clearly resolved at $\eta = 300$ and $\eta = 1000$. For the rotated surface code, the increase in $p_{\mathrm{th}}^{Z}$ exceeds the combined fitting uncertainty only at moderate bias ($\eta \approx 3$--$10$), while the changes in $p_{\mathrm{th}}^{X}$ remain within the combined fitting uncertainty at all bias values and are therefore not resolved. This asymmetry suggests that the unrotated surface code is more sensitive to CNOT scheduling than the rotated code, likely because of its different boundary structure and larger number of physical qubits. The redistribution of $X$- and $Z$-memory performance we observe is consistent with the CNOT-order sensitivity reported by \cite{orourke2024comparethepair}. More importantly, these results suggest that CNOT ordering can serve as a tunable parameter. Depending on which logical memory is more critical for a given application, one can choose the ordering that preferentially protects it. We have reported the effect of the different CNOT ordering here, but a detailed analysis of the underlying mechanism that causes this difference is out of the scope of this work.

\section{Outlook}
\label{sec:outlook}

In this work, we have simulated independent and correlated noise models on multiple surface-code structures (rotated and unrotated; $X$-aligned and $Z$-aligned CNOT order; $ZX$ and $ZY$), decoding throughout with the MWPM decoder PyMatching v2.3.1. A limitation of this approach is that it cannot fully exploit the syndrome structure of the $ZY$ code. Recall that the $ZY$ code replaces the $X$-type stabilizers with $Y$-type stabilizers (Section~\ref{sec:surface_code}), which is allowed because the $Y$-type stabilizers commute with the $Z$-type stabilizers. Since both the $Y$- and $Z$-type stabilizers detect $X$ errors, the $ZY$ code carries twice as many syndrome bits about the dominant $X$ errors compared to the $ZX$ code. This mirrors the doubling of useful syndrome information exploited by the tailored $XY$ code of Ref.~\cite{Tuckett2018Ultrahigh}. However, this additional information is correlated, and it is not matchable by MWPM decoders \cite{higgott2023improved}. A single $X$ error triggers both its adjacent $Z$- and $Y$-type stabilizers, so the $X$-error syndrome forms a hypergraph rather than the graph on which matching decoders operate. A matching decoder must therefore decode on a matchable subgraph and cannot access the extra $Y$-stabilizer information, which is consistent with our finding that the $Z$-memory thresholds of the $ZX$ and $ZY$ codes are indistinguishable. To address these shortcomings of MWPM decoder, we also implemented the BeliefPropagation, BP-OSD, and BeliefFind decoders, but found that they were computationally expensive and failed to converge for the $d \geq 5$ codes. Further exploration of these decoders is beyond the scope of this work. Studying the performance of our construction on other decoders is a natural next step.

These observations point to a concrete open problem. The non-matchable, correlated syndrome structure of the $ZY$ code requires decoders that exploit this information without the matchability and convergence limitations of matching- and BP-based methods. The most direct route is tensor-network (TN) decoding, which was used to reach the ultrahigh thresholds of the tailored $XY$ code of Ref.~\cite{Tuckett2018Ultrahigh}. General TN decoders for 2D Pauli codes~\cite{chubb2021general} apply directly to the $ZY$ code. Learning-based approaches, such as graph-neural-network or reinforcement-learning decoders trained on the $ZY$ syndrome structure, offer a complementary, longer-term direction. Developing such a decoder and characterizing the resulting threshold improvements over MWPM, under both code-capacity and circuit-level noise, would be a valuable extension of the present work.

\section{Data Availability}
All the data that support the plots and values in the tables within this paper are available from the authors upon request.

\section{Code Availability}
\label{sec:code}
The codes used to generate STIM circuits, run Sinter sampling processes, and analyze error rates are available at: \textit{https://github.com/Pritesh402/zxzy-qec}

\section{Acknowledgement}
This research is supported by the College of Computational, Mathematical, and Physical Sciences at Brigham Young University. We extend our sincere thanks to the BYU Quantum Group for their valuable insights and meaningful discussions. We also thank the high-performance computing cluster access provided by the BYU Office of Research Computing where we performed all the surface code simulations for large distances using billions of Monte Carlo shots. 


\bibliographystyle{quantum}
\bibliography{mybibliography}

\end{document}